\documentclass[letterpaper,12pt,titlepage,oneside,final]{book}
 
\pdfoutput=1

\newcommand{\href}[1]{#1} % does nothing, but defines the command so the
\usepackage{ifthen}
\newboolean{PrintVersion}
\setboolean{PrintVersion}{false} 
\usepackage{amsmath,amssymb,amstext} % Lots of math symbols and environments
\usepackage[pdftex]{graphicx} % For including graphics N.B. pdftex graphics driver 

\usepackage[pdftex,letterpaper=true,pagebackref=false]{hyperref} % with basic options
\usepackage{rotating}
\hypersetup{
    plainpages=false,       % needed if Roman numbers in frontpages
    pdfpagelabels=true,     % adds page number as label in Acrobat's page count
    bookmarks=true,         % show bookmarks bar?
    unicode=false,          % non-Latin characters in Acrobat’s bookmarks
    pdftoolbar=true,        % show Acrobat’s toolbar?
    pdfmenubar=true,        % show Acrobat’s menu?
    pdffitwindow=false,     % window fit to page when opened
    pdfstartview={FitH},    % fits the width of the page to the window
    pdftitle={uWaterloo\ LaTeX\ Thesis\ Template},    % title: CHANGE THIS TEXT!
    pdfnewwindow=true,      % links in new window
    colorlinks=true,        % false: boxed links; true: colored links
    linkcolor=blue,         % color of internal links
    citecolor=green,        % color of links to bibliography
    filecolor=magenta,      % color of file links
    urlcolor=cyan           % color of external links
}
\ifthenelse{\boolean{PrintVersion}}{   % for improved print quality, change some hyperref options
\hypersetup{	% override some previously defined hyperref options
    citecolor=black,%
    filecolor=black,%
    linkcolor=black,%
    urlcolor=black}
}{} % end of ifthenelse (no else)

\let\origdoublepage\cleardoublepage
\newcommand{\clearemptydoublepage}{%
  \clearpage{\pagestyle{empty}\origdoublepage}}
\let\cleardoublepage\clearemptydoublepage

\newcommand{\R}{\mathbb{R}}
\newcommand{\C}{\mathbb{C}}
\newcommand{\h}{\mathbb{H}}
\newcommand{\CH}{\mathbb{C}\otimes\mathbb{H}}
\newcommand{\CO}{\mathbb{C}\otimes\mathbb{O}}

\newcommand{\RCHO}{\mathbb{R}\otimes\mathbb{C}\otimes\mathbb{H}\otimes\mathbb{O}}
\newcommand{\COOO}{\mathbb{C}\otimes\overleftarrow{\mathbb{O}}}
\newcommand{\RCHOOO}{\mathbb{R}\otimes\mathbb{C}\otimes\mathbb{H}\otimes\overleftarrow{\mathbb{O}}}

\newcommand{\CLseven}{\mathbb{C}l(7)}
\newcommand{\CLsix}{\mathbb{C}l(6)}
\newcommand{\CLtwo}{\mathbb{C}l(2)}
\newcommand{\CLfour}{\mathbb{C}l(4)}
\newcommand{\CLn}{\mathbb{C}l(n)}
\newcommand{\CLtwelve}{\mathbb{C}l(12)}
\newcommand{\CLten}{\mathbb{C}l(10)}

\newcommand{\oot}{\omega \omega^{\dagger}}
\newcommand{\oto}{\omega ^{\dagger}\omega }

\newcommand{\spacer}{0.2cm}

\begin{document}

% For a large document, it is a good idea to divide your thesis
% into several files, each one containing one chapter.
% To illustrate this idea, the "front pages" (i.e., title page,
% declaration, borrowers' page, abstract, acknowledgements,
% dedication, table of contents, list of tables, list of figures,
% nomenclature) are contained within the file "uw-ethesis-frontpgs.tex" which is
% included into the document by the following statement.
%----------------------------------------------------------------------
% FRONT MATERIAL
%----------------------------------------------------------------------
% T I T L E   P A G E
% -------------------
% Last updated May 24, 2011, by Stephen Carr, IST-Client Services
% The title page is counted as page `i' but we need to suppress the
% page number.  We also don't want any headers or footers.
\pagestyle{empty}
\pagenumbering{roman}

% The contents of the title page are specified in the "titlepage"
% environment.
\begin{titlepage}
        \begin{center}
        \vspace*{1.0cm}

        \Huge
        {\bf  Standard model physics from an algebra? }

        \vspace*{1.0cm}

        \normalsize
        by \\

        \vspace*{1.0cm}

        \Large
        C. Furey \\

        \vspace*{3.0cm}

        \normalsize
        A thesis \\
        presented to the University of Waterloo \\ 
        in fulfillment of the \\
        thesis requirement for the degree of \\
        Doctor of Philosophy \\
        in \\
        Physics \\

        \vspace*{2.0cm}

        Waterloo, Ontario, Canada, 2015 \\

        \vspace*{1.0cm}

        \copyright\ C. Furey 2015 \\
        \end{center}
\end{titlepage}

% The rest of the front pages should contain no headers and be numbered using Roman numerals starting with `ii'
\pagestyle{plain}
\setcounter{page}{2}

\cleardoublepage % Ends the current page and causes all figures and tables that have so far appeared in the input to be printed.
% In a two-sided printing style, it also makes the next page a right-hand (odd-numbered) page, producing a blank page if necessary.

% D E C L A R A T I O N   P A G E
% -------------------------------
  % The following is the sample Delaration Page as provided by the GSO
  % December 13th, 2006.  It is designed for an electronic thesis.

%\section{Author's declaration}
%  \begin{center}\textbf{Author's Declaration}\end{center}

%  \noindent
%I hereby declare that I am the sole author of this thesis.  I understand that my thesis may be made electronically available to the public.

\cleardoublepage
%\newpage

% A B S T R A C T
% ---------------

%\begin{center}\textbf{Abstract}\end{center}
\section{Abstract}

This thesis constitutes a first attempt to derive aspects of standard  model particle physics from little more than an algebra.  Here, we argue that physical concepts such as particles, causality, and irreversible time may result from the algebra acting on itself.

We then focus on a special case by considering the algebra $\RCHO$, the tensor product of the only four normed division algebras over the real numbers.  Using nothing more than $\RCHO$ acting on itself, we set out to find standard model particle representations:  a task which occupies the remainder of this text.

From the $\CH$ portion of the algebra, we find generalized ideals, and show that they describe concisely all of the Lorentz representations of the standard model.

From just the $\CO$ portion of the algebra, we find minimal left ideals, and show that they mirror the behaviour of a generation of  quarks and leptons under $su(3)_c$ and $u(1)_{em}$.  These unbroken symmetries, $su(3)_c$ and $u(1)_{em}$, appear \it uniquely \rm in this model as particular symmetries of  the algebra's ladder operators.  Electric charge, here, is seen to be simply a number operator for the system.

We then combine the $\CH$ and $\CO$ portions of $\RCHO$, and focus on a leptonic subspace, so as to demonstrate a rudimentary electroweak model.  Here, the underlying ladder operators are found to have a symmetry generated \it uniquely \rm by $su(2)_L$ and $u(1)_Y$.  Furthermore, we find that this model yields a straightforward explanation as to why $SU(2)_L$ acts only on left-handed states.

We then make progress towards a three-generation model.  The action of $\CO$ on itself can be seen to generate a 64-complex-dimensional algebra, wherein we are able to identify two sets of generators for $SU(3)_c$.  We  apply these generators to the rest of the space, and find that it breaks down into the $SU(3)_c$ representations of exactly three generations of quarks and leptons.  Furthermore, we show that these three-generation results can be extended, so as to include all 48 fermionic $U(1)_{em}$ charges.

 \cleardoublepage
%\newpage

% A C K N O W L E D G E M E N T S
% -------------------------------

%\begin{center}\textbf{Acknowledgements}\end{center}

\section{Acknowledgements}

This research was supported by NSERC, the University of Waterloo, the Templeton Foundation, and by the Perimeter Institute for Theoretical Physics.  Research at Perimeter Institute is supported by the Government of Canada through Industry Canada and by the Province of Ontario through the Ministry of Research and Innovation.

A special thanks to my committee members:  L. Boyle, F. Cachazo, F. Girelli, A. Kempf, R. Mann, and P. Ramond.   I am especially grateful to L. Boyle,   A. Kempf, and R. Mann for giving me a hand when I needed one, and feel very fortunate to have had A. Kempf as a supervisor during these final stages of my degree.  He has been an outstanding mentor.

My time in Waterloo has been both wonderful and trying, and I will be forever indebted to those who were there with me through the days that were difficult.    Alex, Carey, Calida, Deb, Steve, Luke, Sheri, Hilary, Gaetano, George, Vito, Steph, Josh, and of course, Mike, Dianna, Shona, and Kelly:  please know how much I value your friendship.

\cleardoublepage
%\newpage

% D E D I C A T I O N
% -------------------

\begin{center}\textbf{ }\end{center}

\vspace{5cm}

And I want to know the same thing.  We all want to know, how's it going to end?

- \it Brennan and Waits \rm
\cleardoublepage
%\newpage

% T A B L E   O F   C O N T E N T S
% ---------------------------------
\renewcommand\contentsname{Table of Contents}
\tableofcontents
\cleardoublepage
\phantomsection
%\newpage

% L I S T   O F   T A B L E S
% ---------------------------

%%\addcontentsline{toc}{chapter}{List of Tables}
%%\listoftables
%%\cleardoublepage
%%\phantomsection		% allows hyperref to link to the correct page

%\newpage

% L I S T   O F   F I G U R E S
% -----------------------------

%%\addcontentsline{toc}{chapter}{List of Figures}
%%\listoffigures
%%\cleardoublepage
%%\phantomsection		% allows hyperref to link to the correct page

%\newpage

% L I S T   O F   S Y M B O L S
% -----------------------------
% To include a Nomenclature section
% \addcontentsline{toc}{chapter}{\textbf{Nomenclature}}
% \renewcommand{\nomname}{Nomenclature}
% \printglossary
% \cleardoublepage
% \phantomsection % allows hyperref to link to the correct page
% \newpage

% Change page numbering back to Arabic numerals
\pagenumbering{arabic}

%----------------------------------------------------------------------
% MAIN BODY
%----------------------------------------------------------------------
% Because this is a short document, and to reduce the number of files
% needed for this template, the chapters are not separate
% documents as suggested above, but you get the idea. If they were
% separate documents, they would each start with the \chapter command, i.e, 
% do not contain \documentclass or \begin{document} and \end{document} commands.
%======================================================================
\chapter{Introduction}
%======================================================================
\section{Towards conceptual parsimony}

% what this thesis is

The safest conceivable theory is one which implements the fewest initial assumptions possible.
%Beyond satisfying a certain sense of aesthetics, the assumption that  \it nature is simple \rm  has the advantage of being a conservative guess.  
That is, any model based on elaborate input will by definition specify more detail than would a simple one, and hence will be more likely to be in conflict with reality.  

Our current theories of fundamental physics are often described in terms of an assortment of objects such as manifolds, metric tensor fields, matter fields, gauge fields, Fock spaces, etc. One might wonder if it could be possible to choose our starting point more conservatively.

This thesis may be thought of as an experiment of sorts.  It is a first attempt to see just how close one can get to standard model predictions, while using as little theoretical input as possible.  The original proposal of this idea was put forward in a Part III research essay for a Master's degree in 2006, \cite{cm}, \cite{cmnotes}.  Certainly, this project can be seen to be far from complete.  However, it does demonstrate how certain carefully-chosen, low-dimensional, mathematical objects can parallel a considerable amount of the standard model's structure.

% principles of unification and simplification, which have driven much of the theoretical work since the 70s

% casts everything into a single algebra, but doesn't unify the gauge groups in the usual sense.  instead unifies bosons and fermions into the same algebra, but without supersymmetry

Over the next 85 pages, we will work towards developing a model which is based on the same principles of unification and simplification, which have driven much of theoretical physics since the 1970s.  All objects in our model will be pieced together from the same algebra, although, we will not unify the gauge groups of the standard model in the same sense as do grand unified theories.  We will find that both fermions and bosons should arise together from the same algebra.  Having said that, we will not implement supersymmetry at this stage.  Although the road to unification has been an arduous one, we maintain that it is an idea worthy of staying the course.

\section{Outline}

In Chapter~\ref{m1} of this thesis, we give a brief sketch of our experiment, which motivates the algebraic model being proposed.  In Chapter~\ref{CH}, we introduce the algebra of the complex quaternions, $\CH$, and with them, demonstrate all of the Lorentz representations necessary to describe  the standard model.  Then, in Chapter~\ref{CA}, we review some basic characteristics of Clifford algebras, while in Chapter~\ref{SM}, we review some basic characteristics of the standard model.  In Chapter~\ref{CO} we introduce the complex octonions, $\CO$, and show how they can provide a direct route to the unbroken internal symmetries, generated by $su(3)_c$ and $u(1)_{em}$, for one generation of quarks and leptons.  These symmetries appear as  \it unitary MTIS symmetries, \rm which are first introduced in this chapter.  In Chapter~\ref{su2L}, we then combine the $\CH$ and $\CO$ results to show a rudimentary leptonic model.  The unitary MTIS symmetries, found here,  happen to be none other than $su(2)_L$ and $u(1)_Y$ on these states.  Furthermore, we find that this model can offer an explanation as to why $SU(2)$ acts on states of only one chirality (left).   
%In Chapter~\ref{RCHO}, we combine the results for $\CH$ and $\CO$. CHECK Here we make use of the concept of \it unitary MTIS symmetries, \rm which is shown to lead uniquely CHECKKKK to the standard model's complete gauge group, $SU(3)_c \times SU(2)_L \times U(1)_{Y}$, in addition to analogous transformations for spin.  
In Chapter~\ref{GUT}, we review some of the algebraic structure of $SU(5)$ and $Spin(10)$ grand unified theories, as well as the Pati-Salam model.  Finally, in Chapter~\ref{GEN}, we go on to demonstrate how the $SU(3)_c$ and $U(1)_{em}$ representations for exactly three generations of standard model fermions can be found, using (paradoxically) nothing more than the eight-complex dimensional algebra, $\CO$.  

%Finally, in the appendices, we include various underdeveloped ideas from a master's research project on algebraic causal sets, \cite{cm}.  It was proposed in this early essay that these causal sets should represent particle worldlines, as opposed to discrete space-time.

%%%\section{Early-stage ideas,  }

%By this point, readers may have noticed a code of letters, marking the subsections of this thesis, in the Table of Contents.  These letters are meant to inform the reader of the status of the subsection.  

%\medskip

%Subsections marked  are composed of background material, published by other authors.

%\medskip

%Subsections marked  contain published results by this author, found in \cite{UTI}, \cite{Gen}, \cite{qq}.

%\medskip

 %Subsections marked  contain new results by this author.

%\medskip

%%% Subsections marked  in the Table of Contents contain early-stage ideas by this author.  These ideas are at a  speculative stage, requiring  further development and scrutiny.  We are taking the liberty to include these less developed ideas in this thesis, more out of necessity than want, seeing as how some of the basic concepts discussed here have  appeared in recent works,~\cite{QECS} and \cite{Universe}.

%----------------------------------------------------------------------
%----------------------------------------------------------------------

%See equation \ref{eqn_pi} on page \pageref{eqn_pi}.\footnote{A famous equation.}

\chapter{What lies in an algebra?\label{m1}  }

In this first chapter, we introduce an attempt to describe established results in particle physics, while working from the fewest initial assumptions possible.  Our only input will be an algebra, $A$.  

The ideas outlined here originate from a Cambridge Part III research essay, \cite{cm}, \cite{cmnotes} and have provided the underlying intuition for three papers published during these PhD years, \cite{qq}, \cite{Gen}, and \cite{UTI}.  With this being said, the results from those three publications may be considered independently from the ideas sketched here in this chapter, which are still at an early stage of development.
%math second

Much of the material in this chapter was submitted in 2014 to FQXi's annual call for essays in fundamental physics, \cite{fqxi}.

\section{Causality }

\begin{figure}[h!]
\begin{center}
\includegraphics[width=3cm]{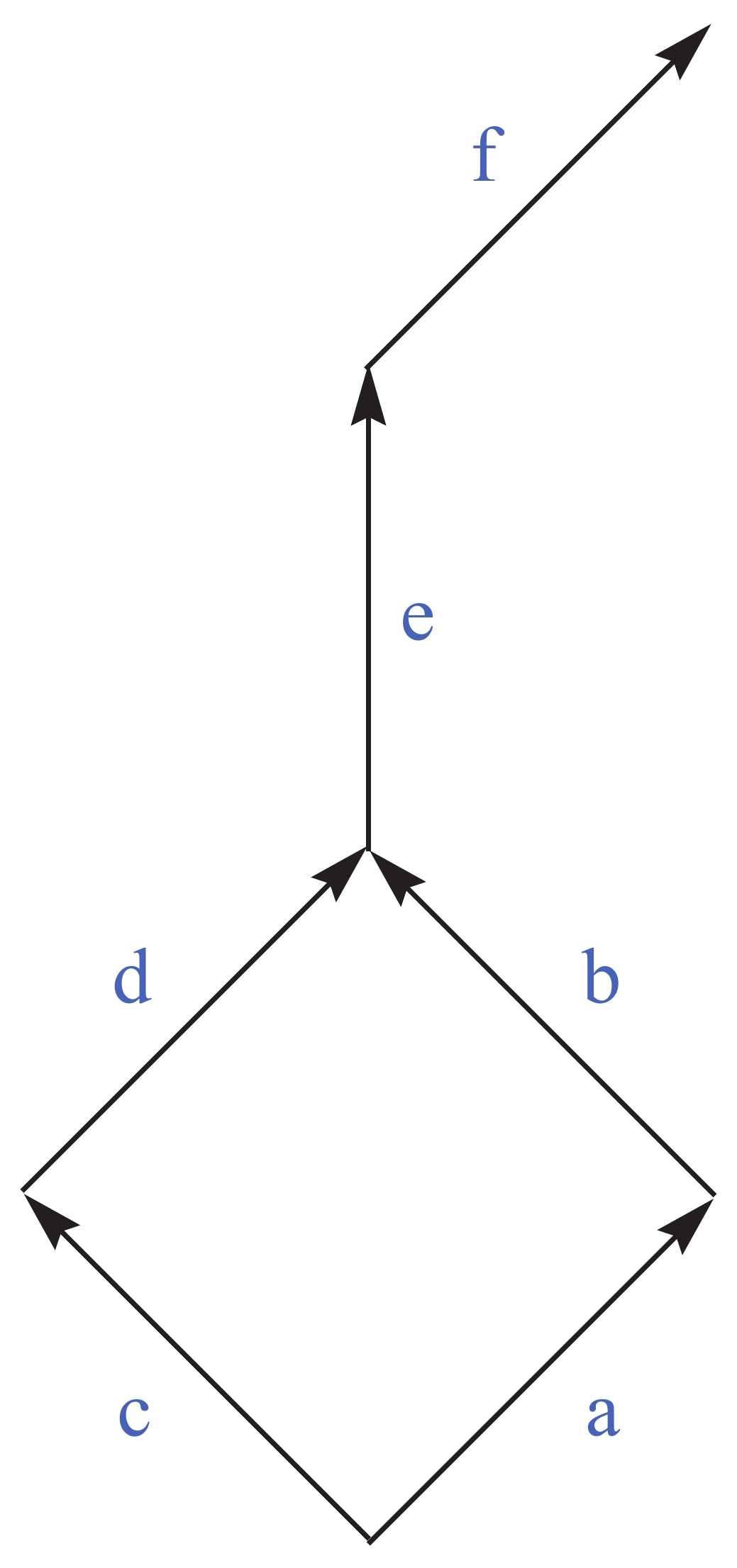}
\caption{\label{cs}
The unevaluated algebraic expression, $f\cdot ( e \cdot( d\cdot c + b \cdot a) )$, gives a causal set.}
\end{center}\end{figure}

Suppose for a moment that nature were represented by an algebra, $A$.  We will start, then, simply with an unevaluated algebraic expression.  Consider for example

\begin{equation}
f \cdot ( e\cdot ( d\cdot c + b\cdot a) ),
\end{equation}

\noindent where $a$, $b$, $c$, $d$, $e$, $f$ are elements of $A$.  Taking multiplication to be the propagation along an edge, and addition to be the joining of two edges at a vertex, it can be seen that this unevaluated algebraic expression gives a causal set.
%\footnote{It should also be important to consider the opposite assignment where propagation along an edge is given by addition.}.  
Please see Figure~(\ref{cs}).  

\vspace{.5cm}

A \it causal set \rm is a set $S$ together with the relation, $\leq$, such that

1.  if $x\leq y$ and $y\leq z$, then $x\leq z$ $\hspace{2mm}\forall x,y,z \in S\hspace{1cm}$ (transitive),

2.  if $x\leq y$ and $y\leq x$, then $x=y$ $\hspace{2mm}\forall x,y \in S\hspace{1cm}$ (non-circular),

3.  for any given $x,z \in S,$ the set of elements $\{y \hspace{1mm}|\hspace{1mm} x\leq y\leq z\}$ is finite $\hspace{.5cm}$ (locally finite). 

In Figure~(\ref{cs}), $S$ is given by the set of vertices, and the relations $\leq$ are indicated by the arrows between those vertices.

It is at this early stage only a conjecture that any (associative) algebraic expression gives a causal set.    For an introduction to causal sets, please see~\cite{dowker}, \cite{rafael} and \cite{fotini}.

\it So it seems to be possible that causality, of all things,
could already appear at the level of an algebra. \rm
%\it So it seems to be possible that causality, of all things, need not be a fundamental input.\rm

\medskip

Now, we would like to propose an unorthodox interpretation for these causal sets.  That is, a causal set is not meant to represent discrete space-time, but is instead meant to represent particle worldlines.  In other words, matter exhibits its own causal structure, and there is no such thing as a space-time point.  This picture has particles with no underlying space-time whatsoever, neither continuous, nor discrete.  

The question we are then asking here is whether or not space-time can be seen to be as surprisingly unnecessary as was the luminiferous aether from one hundred years ago.  Can particles exist \it independently, \rm without the crutch of a fundamental space-time to support them?  

%This is what is meant here by \it space-time independence, \rm which is not to be confused with the weaker condition of background independence.  Clearly, background independent theories are permitted to have space-time points, as is the case with general relativity, as an example.

%This interpretation is closely related to that of Energetic Causal Sets by Cort\^{e}s and Smolin,  \cite{QECS}, \cite{Universe}, and in fact predates that research,~\cite{cm}.  

Similar in spirit is an earlier spaceless  graph model, proposed by Kribs and Markopoulou in \cite{km}, where particles emerge at low energies as noiseless subsystems of quantum information processing structures.  
%More recently, Cort\^{e}s and Smolin have also proposed a spacetime-independent causal set, with an explanation for the emergence of space-time, using Lagrange multipliers, \cite{QECS}, \cite{Universe}.  
These models differ from the popular causal set models of Dowker, Sorkin, and Surya, whose vertices represent space-time points.  In Dowker,  Sorkin, and Surya's models, the vertices specify a position and a time.  In the model presented in this thesis, however, the vertices instead specify the internal degrees of freedom of particles, such as spin, colour and electric charge, etc.  

%Often, in discrete space-time models, Lorentz invariance can be lost when some form of lattice is implemented.  DSS models overcome this by introducing a "sprinkling", which distributes discrete space-time points in a way which is qualitatively Lorentz invariant.  

Over the past ten years, a number of authors have come forward with a variety models, which each call into question the necessity of a fundamental space-time.  In 2005, Piazza  did so by proposing to replace localized regions of space with quantum subsystems in~\cite{fedoglim}, \cite{fedo}, an idea which was later developed in 2007 by Piazza and Costa,~\cite{costapiazza}.  In 2010, Van Raamsdonk, \cite{raam},  proposed building up space-time with quantum entanglement using gauge theory/gravity duality.
%\footnote{It would be of interest to see if our algebraic model (Figure~\ref{cs}) could be connected to \cite{raam}.  For example, we might consider two algebraic composite pure states $| \psi_1 \rangle\otimes\hspace{1mm}|\phi_1\rangle$ and $| \psi_2 \rangle\otimes\hspace{1mm}|\phi_2\rangle$, each propagating along an edge.   Here, addition would not only correspond to the joining of edges at a vertex, but also, the introduction of entanglement.}.  
More recently in 2014, Wieland proposed a model of simplicial gravity, constructed from spinors,~\cite{Wolfgang}.  In the same year, W. Edwards described non-embeddable relational configurations,~\cite{bill}.  Also related to these concepts are works by Kempf in 2013, \cite{achim},   and  Saravani, Aslanbeigi, and Kempf in 2015,  \cite{sak}, which discuss how space-time curvature can be encoded in the vacuum entanglement structure of fields.

%In 2013, Cort\^{e}s and Smolin published work on Energetic Causal Sets,~\cite{QECS}, \cite{Universe}, which describes a spaceless causal set, constructed out of particles at the fundamental level.  This basic concept overlaps significantly with \cite{cmnotes}:  ideas of L. Smolin's PhD student, disclosed to him two years prior.  The notes, \cite{cmnotes}, were written up for L. Smolin in 2011 upon his request, and subsequently emailed to him.   (He later innocently forgot about the notes when~\cite{QECS} and \cite{Universe} were published.)  Readers should note that \cite{km} 

%and also about relevant work, \cite{km}, 

In 2013, Cort\^{e}s and Smolin published work on Energetic Causal Sets,~\cite{QECS}, \cite{Universe}, which describes a spaceless causal set, constructed out of particles at the fundamental level.  This basic concept overlaps significantly with \cite{cmnotes}:  ideas of L. Smolin's PhD student at the time (the present author).  The notes, \cite{cmnotes}, were written up for L. Smolin in 2011 upon his request, and subsequently emailed to him.   (He later innocently forgot about the notes when~\cite{QECS} and \cite{Universe} were published two years later.)  We also point out the relevant earlier work of \cite{km}, who proposed a spaceless causal set of quantum information processing systems, with particles emerging in the low energy limit.

\section{Irreversible time}

Recently, it has been emphasized in the fundamental physics community, \cite{fqxi}, that theories like general relativity do a poor job of encapsulating our experience of \it events unfolding. \rm For example, there is nothing in the theory to explain why events happen, but do not `unhappen'.

It seems, however, that an algebra might well provide such a  notion of irreversible time.    In the case of an algebra, an event is a calculation.  Taking our algebra $A=\mathbb{R}$, for example, an event is the evaluation of 6+3 to give 9, or the evaluation of 5$\cdot$2 to give 10.  

Addition and multiplication are examples of uninvertible binary operations.  Therefore, an event can be seen to be irreversible, it cannot `unhappen'.  For example, if we are given only the output of 9, it is impossible to tell if 9 came from the inputs of 6+3, or 8+1, or 4+5, etc.

% time

Time, then, is simply a sequence of calculations, and is clearly irreversible.  
%It can be seen to be an entirely local concept within the causal set.
The related notion of `now' can be seen to be an entirely \it local \rm concept within the causal set.

%\medskip
% ideals

\section{Particles}

Some carefully chosen algebras, such as the complex Clifford algebra, $A=\CLtwo$, naturally contain subspaces called \it ideals. \rm  Intuitively speaking, an ideal  is a special subspace of an algebra because it can survive multiplication by any element in $A$.  

\it Ideals \rm persisting under multiplication bear a striking resemblance to \it particles \rm persisting under propagation.  The proposal, then, is that particles could be singled out in the algebra, thanks to a mathematical incarnation of Darwin's natural selection.   

\begin{center} Ideals $\sim$ particles.
\end{center}

Given an algebra, $A$, a \it left ideal, \rm $B$, is a subalgebra of $A$ whereby $a \underline{b}$ is in $B$ for all $\underline{b}$ in $B$, and for any $a$ in $A$.  That is, no matter which $a$ we multiply onto $\underline{b}$, the new product, $\underline{b}'\equiv a\underline{b}$, must be in the subspace $B$ (i.e.  the ideal $B$ survives).  It is easy to see how $\underline{b}'\equiv a\underline{b}$ could easily describe, for example, a particle $\underline{b}$ undergoing propagation along $a$.
%\footnote{Note that it is not clear at this point whether or not the association, \it ideals \rm $\sim$ \it particles, \rm can be made for gauge bosons.}

%In particular, \it spinors \rm can be defined as minimal left ideals of Clifford algebras,~\cite{ablam}.  A  \it minimal left ideal \rm is a left ideal which contains no left ideals other than $\{ 0 \} $ and itself.  In other words, it has no non-trivial left ideals inside it.  

%By considering the left action of the algebra $A=\CLtwo$ on itself, $\CLtwo$ can be seen to break down into two (equivalent) 2-complex dimensional minimal left ideals.  Upon considering both the left action and right action of $\CLtwo$ on itself, these two minimal left ideals can then be identified as left- and right-handed Weyl spinors under the Dirac algebra, $\mathbb{C}\otimes Cl(1,3) = \CLfour = \CLtwo \otimes \CLtwo$.  For details, see section~\ref{ideals}.  On the other hand, a similar construction starting from $A = \CLsix$ led to a set of states behaving like a full generation of quarks and leptons in \cite{qq}.

These concepts have a strong connection to well-known physics.  In this text, we will first introduce the notion of  \it generalized ideals. \rm Taking  $A$ to be the complex quaternions,  $\CH \simeq \CLtwo$, we will then see how generalized ideals lead to left- and right-handed Weyl spinors,~\cite{UTI}.  In an analogous construction, starting from the complex octonions, $A = \CO$, generalized ideals will be seen to lead to a set of states behaving like a full generation of quarks and leptons, \cite{qq}.

Ultimately, we intend to merge $\CH$ and $\CO$ together, via a tensor product over $\mathbb{C}$, resulting in the algebra $A=\RCHO$.  Loosely speaking, we will associate $\CH$ with Lorentzian degrees of freedom:  spin and chirality, while the octonionic part of the algebra will give rise to the other internal degrees of freedom, such as colour, weak isospin, and charge.

\begin{center}  $\CH$ $\sim$ Lorentz \hspace{2cm} $\CO$ $\sim$ Other internal:  colour, charge, etc.
\end{center}

The Dixon algebra, $\RCHO$, is the tensor product of the only four normed division algebras over the real numbers:  the real numbers, $\R$, the complex numbers, $\C$, the quaternions, $\h$, and the octonions, $\mathbb{O}$.  Its connection to particle physics was studied indirectly by Casalbuoni \it et al., \rm \cite{Grass}, \cite{wow}, \cite{wowagain}, and later much more extensively by Dixon, \cite{dixon}.  Our goal of identifying standard model structure from $\RCHO$ aligns with that of these earlier authors, however, our implementation of this algebra differs significantly, particularly with respect to chirality and weak isospin, and in our treatment of antiparticles.

\section{Summary and outlook}

We come from sketching a model,  whose only fundamental input is an algebra, $A$.  
From the  algebraic expressions of $A$, we argue that multiple physical concepts can materialize.  %This runs backwards to the usual style with which physical theories are constructed.  
Such resulting physical concepts may include causality, particles, and irreversible time.  Notably, particles may be seen to arise as the algebra's most stable subspaces.

The algebra, $A$,  will be taken to be $\RCHO$ for reasons that will become clear throughout this thesis.  In the future, however, it would be worth investigating whether or not even $\RCHO$ could be simplified.  Explicitly, $\RCHO$ could be replaced with another mathematical object, which approximates it in some limit, but which lacks any occurrence of uncountable infinities, which are inherent to the real number system.  
%It is suspected that uncountable infinities could lead to logical inconsistencies, and are best to be avoided in any fundamental theory of physics.  CHECK WITH ACHIM

With this rough draft of a model, we would like to now develop the idea.  But where to begin?  Perhaps the most straightforward way to see if the algebra $\RCHO$ could stand a chance of eventually producing standard model scattering amplitudes, is to see if it can first produce standard model group representations.  This task will occupy the rest of the thesis.

\chapter{Complex quaternions\label{CH}}

\section{Preamble}

% causal sets have no explanation for 3+1 dimensions  therefore CxH

%  introduce CxH

A significant challenge within  causal set programs has been to explain the existence of 3+1 dimensions.
That is, any causal set chosen at random is highly unlikely to have this particular dimension of our choosing.  

It is for this reason that we propose here causal sets originating from an algebra, $A$.  The idea then is that a careful choice of algebra will impart on the causal set the desired 3+1 Lorentzian structure.  In this section, we will consider $A = \CH,$ the complex quaternions.  Between the left and right action of $\CH$ on itself, we will now introduce an unusually compact way of describing all of the (3+1) Lorentz representations of the standard model of particle physics.

\section{Introduction to $\CH$\label{introCH}  }

Any element of the complex quaternions can be described as the complex linear combination, 
\begin{equation}
c_0 + c_1 \hspace{1mm}i\epsilon_x + c_2 \hspace{1mm}i\epsilon_y+ c_3\hspace{1mm} i{\epsilon_z},
\end{equation}

\noindent where the $c_n \in \mathbb{C}$.  The element $i$ is the usual complex imaginary unit, with $i^2 = -1$, that commutes with all of the elements in the algebra.  The quaternionic imaginary units, ${\epsilon_x}$, ${\epsilon_y}$, and ${\epsilon_z}$ follow the multiplication rules 

\begin{equation}\label{CHmult}
%\begin{displaymath} 
{\epsilon_x}^2={\epsilon_y}^2={\epsilon_z}^2={\epsilon_x}\hspace{1mm}{\epsilon_y}\hspace{1mm}{\epsilon_z}=-1,
%\end{displaymath}
\end{equation}

\noindent which lead to the identities ${\epsilon_x} {\epsilon_y} = -{\epsilon_y}{\epsilon_x} =  {\epsilon_z}, \hspace{0.1cm}  {\epsilon_y} {\epsilon_z} = -{\epsilon_z}{\epsilon_y} =  {\epsilon_x}, \hspace{0.1cm}  {\epsilon_z}{\epsilon_x} = - {\epsilon_x}{\epsilon_z} = {\epsilon_y}.$  The complex quaternions form an associative algebra, meaning that $\left(ab\right)c=a\left(bc\right)$ $\forall$ $a$, $b,$ $c$ $\in\CH$.  (Note that all tensor products will be assumed to be over $\mathbb{R}$ in this text, unless otherwise stated.)

From their behaviour under multiplication, one may associate $i{\epsilon_x}$ with the more familiar Pauli matrix $\sigma_x = \left( \begin{smallmatrix}0&1\\1&0 \end{smallmatrix}  \right)$, $i{\epsilon_y}$ with $\sigma_y = \left( \begin{smallmatrix}0&-i\\i&0 \end{smallmatrix}  \right)$, and $i{\epsilon_z}$ with $\sigma_z = \left( \begin{smallmatrix}1&0\\0&-1 \end{smallmatrix}  \right)$.  Note, however, that the set $\{ i{\epsilon_x}, i{\epsilon_y}, i{\epsilon_z}\}$ transforms more symmetrically under complex conjugation than does $\{\sigma_x, \sigma_y, \sigma_z\}$,

\begin{equation}\begin{array}{lll} \sigma_x^* = \sigma_x, \hspace{13mm}&   \sigma_y^* = -\sigma_y,  \hspace{13mm}&\sigma_z^* = \sigma_z,  \vspace{5mm}  \\
 & \textup{versus} & \vspace{5mm}  \\
  \left(i{\epsilon_x}\right)^* = -i{\epsilon_x}, &   \left(i{\epsilon_y}\right)^* = -i{\epsilon_y},  & \left(i{\epsilon_z}\right)^* = -i{\epsilon_z}. \vspace{2mm}
 
\end{array}\end{equation}

Elements of the form $s\equiv r_1 {\epsilon_x} + r_2 i{\epsilon_x} + r_3 {\epsilon_y} + r_4  i{\epsilon_y} + r_5 {\epsilon_z} + r_6 i{\epsilon_z} $  for $r_n\in \mathbb{R}$ can easily be seen to give a representation of the Lie algebra $sl(2,\mathbb{C})$, using the usual commutator, $\left[ s, s' \right] \equiv ss' - s's$.  

Later in Section~\ref{mc}, we will find that $L\equiv e^{is}\in \CH$ represents an element of $SL(2,\mathbb{C})$, which acts on left-handed Weyl spinors.  These Weyl spinors also reside in $\CH$, and in fact, all of the results obtained throughout this thesis will follow simply from having an algebra $A$ act on itself.

\section{Conjugation  }

In this text, the \it complex conjugate \rm of an element $a$ will be denoted $a^*$.   The conjugate $*$ maps the complex $i\mapsto -i$, in the usual way.  

The \it quaternion conjugate \rm of $a$ will be denoted $\widetilde{a}$, and $\sim$ maps the quaternionic ${\epsilon_x}\mapsto -{\epsilon_x},$ ${\epsilon_y}\mapsto -{\epsilon_y},$ and ${\epsilon_z}\mapsto -{\epsilon_z}$.  

That which we call the \it hermitian conjugate \rm of $a$ will be denoted $a^{\dagger}$, and $\dagger$ is the result of performing both the complex and quaternion conjugates simultaneously:  $i\mapsto -i$, ${\epsilon_x}\mapsto -{\epsilon_x},$ ${\epsilon_y}\mapsto -{\epsilon_y},$ and ${\epsilon_z}\mapsto -{\epsilon_z}$.  

It is important to note that both the quaternion and hermitian conjugates reverse the order of multiplication, as is familiar from matrix multiplication.  For example, $\left(ab\right)^{\dagger} = b^{\dagger}a^{\dagger}$.

% generalized ideals
\section{Generalized ideals }

We define a subalgebra $B$ of an algebra $A$ to be a \it generalized ideal \rm if $m \left( a, \underline{b} \right) \in B,  \hspace{1mm} \forall \underline{b} \in B$ and for any $a \in A$, where $m$ is (generalized) multiplication.  The notion of a generalized ideal was introduced by this author in \cite{UTI}, and differs from the definition of left ideals by generalizing what is meant by `multiplication'.   Directly from the definition, it can be seen that ideals make up the algebra's most robust subspaces, which persist no matter what $a$ is multiplied onto them. 

Starting from the algebra $A=\mathbb{C}\otimes\mathbb{H}$, we will find generalized ideals under three separate notions of generalized multiplication:

\begin{equation}
\begin{array}{lcl}
\textup{\it the complex invariant-action, \rm}  &\hspace{1cm}& m_c ( a, \underline{b} ) \equiv a\hspace{0.5mm}\underline{b}\hspace{0.5mm}P + a^*\hspace{0.5mm}\underline{b}\hspace{0.5mm}P^*,  \vspace{.5cm}\\
\textup{\it the hermitian invariant-action, \rm} & &m_h ( a,\underline{b} ) \equiv a\hspace{0.5mm}\underline{b}\hspace{0.5mm}a^{\dagger}, \vspace{.5cm} \\
\textup{\it and quaternionic invariant-action, \rm } & & m_q ( a,\underline{b} ) \equiv a\hspace{0.5mm}\underline{b}\hspace{0.5mm}\widetilde{a}.
\end{array}
\end{equation}

\noindent  Here, $P$ is a projector in $\CH$, to be defined shortly.  It should be noted that each of these multiplication rules is constructed so as to preserve conjugation-invariant objects.  For example, elements, $\underline{b}$ in $\CH$ with the property $\underline{b}^* = \underline{b}$ will maintain this property under $m_c$, no matter which $a\in\CH$ is multiplied onto them.

Taking $A$ to be $\CH$, we will first show how the complex invariant-action leads to left- and right-handed Weyl spinors, Majorana spinors, and Dirac spinors.  Then we will show how the hermitian invariant-action leads to four-vectors.  Finally, we will show how the quaternionic invariant  action leads to scalars and the field strength tensor.

$$\begin{array}{lcll}
m_c  &\hspace{1cm}\Rightarrow \hspace{1cm}& \psi &\hspace{9mm}(\textup{spinors})\vspace{.3cm}\\
m_h & \hspace{1cm}\Rightarrow \hspace{1cm}& p_{\mu}&\hspace{9mm}(\textup{four-vectors}) \vspace{.3cm} \\
m_q &\hspace{1cm}\Rightarrow \hspace{1cm} & \phi \hspace{2mm}\textup{and}\hspace{2mm} F_{\mu\nu}  &\hspace{9mm} (\textup{scalars and the field strength tensor}).\vspace{.3cm}
\end{array}$$

%\noindent The sense in which the field strength tensor can be thought of as \it effective \rm will be explained later in this text.

\noindent Readers should note that scalars, spinors, four-vectors, and the field strength tensor, mentioned above, account for all of the Lorentz representations of the standard model.

% show CxH gives all of the Lorentz reps

\section{Complex invariant-action, $m_c$\label{mc}  }

\subsection{Preliminaries  }

%  Summarize what it does

We will now find that left- and right-handed Weyl spinors, Majorana spinors, and Dirac spinors are all generalized ideals under the same complex invariant-action, 
\begin{equation}\label{cma}
\underline{b}' =  a\hspace{1mm}\underline{b}\hspace{1mm} P + a^*\hspace{1mm}\underline{b}\hspace{1mm} P^*.  
\end{equation}
\noindent For concreteness, we define $P$ to be the projector $\frac{1}{2}\left(1+i{\epsilon_z}\right)$, although it is  clear that a continuum of other possibilities exist.  $P$ and its complex conjugate, $P^*$, exhibit the properties 
\begin{equation}\begin{array}{lllllll}
PP=P,  & \hspace{1cm} & P^*P^* = P^*, & \hspace{1cm} & PP^* = P^*P = 0,  & \hspace{1cm} & P + P^* = 1.
\end{array}\end{equation}

%  New basis based on sigma z eigenvalues
Before we begin, though, it will be useful to carry out a change of basis from $\{1, \hspace{1mm}i{\epsilon_x}, \hspace{1mm}i{\epsilon_y}, \hspace{1mm}i{\epsilon_z}\}$ to a new, suggestively named basis, $\{\epsilon_{\uparrow \uparrow},  \hspace{1mm} \epsilon_{\downarrow \uparrow},\hspace{1mm} \epsilon_{\uparrow \downarrow},\hspace{1mm} \epsilon_{\downarrow \downarrow} \}$.  This new basis will be linked to the operator, $i{\epsilon_z}\sim \sigma_z$, in that these basis elements will be defined so as to have the properties:
\begin{equation}
\begin{array}{lll}
i{\epsilon_z}\hspace{1mm}\epsilon_{\uparrow \uparrow} = +\hspace{1mm} \epsilon_{\uparrow \uparrow} & \hspace{1cm} &\epsilon_{\uparrow \uparrow}\hspace{1mm}i{\epsilon_z} = +\hspace{1mm} \epsilon_{\uparrow \uparrow}\vspace{3mm} \\

i{\epsilon_z}\hspace{1mm}\epsilon_{\downarrow \uparrow} = -\hspace{1mm}\epsilon_{\downarrow \uparrow} & \hspace{1cm} &\epsilon_{\downarrow \uparrow}\hspace{1mm}i{\epsilon_z} = +\hspace{1mm}\epsilon_{\downarrow \uparrow}\vspace{3mm} \\

i{\epsilon_z}\hspace{1mm}\epsilon_{\uparrow \downarrow} = +\hspace{1mm}\epsilon_{\uparrow \downarrow} & \hspace{1cm} &\epsilon_{\uparrow \downarrow}\hspace{1mm}i{\epsilon_z} = -\hspace{1mm}\epsilon_{\uparrow \downarrow}\vspace{3mm} \\

i{\epsilon_z}\hspace{1mm}\epsilon_{\downarrow \downarrow} = -\hspace{1mm}\epsilon_{\downarrow \downarrow} & \hspace{1cm} &\epsilon_{\downarrow \downarrow}\hspace{1mm}i{\epsilon_z} = -\hspace{1mm}\epsilon_{\downarrow \downarrow}.\vspace{3mm} \\
\end{array}
\end{equation}
%\noindent In this formalism, spin \{$\uparrow$, $\downarrow$\} and chirality \{L, R\} can be seen to be analogues of each other.  

\noindent In terms of the old basis, we define these new basis vectors to be
\begin{equation}
\begin{array}{lll} 
\epsilon_{\uparrow \uparrow} \equiv \frac{1}{2}\left(1+i{\epsilon_z}\right) & \hspace{1cm} & \epsilon_{\downarrow \uparrow} \equiv \frac{1}{2}\left({\epsilon_y}+i{\epsilon_x}\right) \vspace{3mm} \\

\epsilon_{\uparrow \downarrow} \equiv \frac{1}{2}\left(-{\epsilon_y}+i{\epsilon_x}\right)  & \hspace{1cm} & \epsilon_{\downarrow \downarrow} \equiv \frac{1}{2}\left(1-i{\epsilon_z}\right).

\end{array}
\end{equation}

\noindent For convenience, we also include here the old basis in terms of the new one,

\begin{equation}
\begin{array}{lll} 
1 = \epsilon_{\uparrow \uparrow}+\epsilon_{\downarrow \downarrow} & \hspace{1cm} & \epsilon_x = -i\left(\epsilon_{\downarrow\uparrow}+\epsilon_{\uparrow\downarrow}\right) \vspace{3mm} \\

\epsilon_y = \epsilon_{\downarrow\uparrow}-\epsilon_{\uparrow\downarrow}  & \hspace{1cm} & \epsilon_z = -i\left(\epsilon_{\uparrow \uparrow}-\epsilon_{\downarrow \downarrow}   \right).

\end{array}
\end{equation}

% psi _L = ... psi_R = ...  suggestively named
% props under P

\subsection{Weyl and Dirac spinors as generalized ideals\label{WD}  }

Let us now identify two subspaces, which partition the algebra $\CH$.  These subspaces will be given the suggestive nomenclature $\Psi_{\textup{L}}$ and $\Psi_{\textup{R}}$, and be defined as
\begin{equation}\begin{array}{lll}
\Psi_{\textup{L}} \equiv \psi^{\uparrow}_{\textup{L}} \hspace{1mm} \epsilon_{\uparrow \uparrow} +  \psi^{\downarrow}_{\textup{L}} \hspace{1mm} \epsilon_{\downarrow \uparrow}  & \hspace{1cm} &  \Psi_{\textup{R}} \equiv \psi^{\uparrow}_{\textup{R}} \hspace{1mm} \epsilon_{\uparrow \downarrow} +  \psi^{\downarrow}_{\textup{R}} \hspace{1mm} \epsilon_{\downarrow \downarrow},
\end{array}\end{equation}
\noindent where $\psi^{\uparrow}_{\textup{L}}$, $\psi^{\downarrow}_{\textup{L}}$, $\psi^{\uparrow}_{\textup{R}}$, $\psi^{\downarrow}_{\textup{R}}$ $\in \mathbb{C}$.  Readers may notice that spin and chirality are analogues of each other in this formalism, as transitions between spin states occur via left multiplication, and transitions between $\textup{L}$ and $\textup{R}$ occur via right multiplication.

Straightforward calculation shows that  
\begin{equation} \begin{array}{lll}
\Psi_{\textup{L}}\hspace{1mm} P = \Psi_{\textup{L}} & \hspace{1cm} & \Psi_{\textup{L}}\hspace{1mm} P^* = 0,\vspace{3mm}\\
\Psi_{\textup{R}}\hspace{1mm} P = 0 & \hspace{1cm} &   \Psi_{\textup{R}}\hspace{1mm} P^* = \Psi_{\textup{R}}.
\end{array}\end{equation}

%  both of these are stable under m_c

Now, the reader may confirm that $\Psi_{\textup{L}}$ and $\Psi_{\textup{R}}$ are left ideals, as defined in Chapter~\ref{m1}.  That is, no matter which $a_1$ and $a_2$  are left multiplied onto them, there exists some ${\Psi_{\textup{L}}}'  \equiv {\psi^{\uparrow}_{\textup{L}}}' \hspace{1mm} \epsilon_{\uparrow \uparrow} +  {\psi^{\downarrow}_{\textup{L}}}'\hspace{1mm} \epsilon_{\downarrow \uparrow} $ and ${\Psi_{\textup{R}}}'  \equiv {\psi^{\uparrow}_{\textup{R}}}' \hspace{1mm} \epsilon_{\uparrow \downarrow} +  {\psi^{\downarrow}_{\textup{R}}}' \hspace{1mm} \epsilon_{\downarrow \downarrow}$ such that
\begin{equation}\begin{array}{lll}
a_1\Psi_{\textup{L}} = {\Psi_{\textup{L}}}'  &  \hspace{1cm}  &   a_2\Psi_{\textup{R}} = {\Psi_{\textup{R}}}' .
\end{array}\end{equation}
\noindent In other words, the L and R subspaces are each stable under left multiplication.  

With this knowledge in hand, it is now almost trivial to see that $\Psi_{\textup{L}}$ and $\Psi_{\textup{R}}$ are each generalized ideals under the complex invariant-action, $m_c$.  That is, for any $a\in\CH$, there exists a $\Psi_{\textup{L}}'$ such that
\begin{equation}
\Psi_{\textup{L}}' =  a\Psi_{\textup{L}} =  a\Psi_{\textup{L}}P  = a\Psi_{\textup{L}}P + a^*\Psi_{\textup{L}}P^* = m_c(a,\Psi_{\textup{L}} ).
\end{equation}
\noindent Likewise, for any $a^*\in\CH$, there exists a $\Psi_{\textup{R}}'$ such that
\begin{equation}
\Psi_{\textup{R}}' =  a^*\Psi_{\textup{R}} =  a^*\Psi_{\textup{R}}P^*  = a\Psi_{\textup{R}}P + a^*\Psi_{\textup{R}}P^* = m_c(a,\Psi_{\textup{R}} ).
\end{equation}

\noindent It is furthermore easily seen that the set of elements of the form $\Psi_{\textup{D}} \equiv \Psi_{\textup{L}} + \Psi_{\textup{R}}$ spans all of $\CH$, and hence trivially qualifies as a generalized ideal under $m_c$.  

%  input L as multiplyer  - psi_L psi_R trf as Weyl spinors in QFT

Taking now $a$ to be $L = e^{is}$ from Section~\ref{introCH} gives the transformation
\begin{equation}
{\Psi_{\textup{D}}}' = m_c(L,\Psi_{\textup{D}} ) = L\Psi_{\textup{L}} + L^* \Psi_{\textup{R}}.
\end{equation}
\noindent The reader is encouraged to confirm that indeed $\Psi_{\textup{D}}$ transforms as a Dirac spinor, $\Psi_{\textup{L}}$ transforms as a left-handed Weyl spinor, and $\Psi_{\textup{R}}$ transforms as a right-handed Weyl spinor under $SL(2,\mathbb{C})$.  In other words, the complex coefficients within $\Psi_{\textup{D}}$ transform exactly as would the complex components of a four-dimensional column vector, representing a Dirac spinor, from standard quantum field theory.    It is for this reason that the names $\Psi_{\textup{D}}$, $\Psi_{\textup{L}}$, and $\Psi_{\textup{R}}$ were given to these generalized ideals early on.

% no epsilon matrix
\subsection{A seamless new way to conjugate Weyl spinors\label{lr}   }
%\medskip
%\noindent \it A seamless new way to conjugate Weyl spinors. \rm
%\medskip

%How it's usually done

In quantum field theory textbooks, it is typically explained that a left handed Weyl spinor, $\Psi_{\textup{L}}$, can be conjugated so as to give a right-handed Weyl spinor.  The procedure necessary to do so, in the usual matrix-and-column-vector formalism, entails (1) complex conjugating the left-handed spinor's two components, and (2) multiplying this column vector by the matrix $\epsilon = -i\sigma_y=\left( \begin{smallmatrix}0&-1\\1&0 \end{smallmatrix}  \right)$, 

\begin{equation}\label{old} \begin{array}{lllll}\left[\hspace{1mm}\psi^{\uparrow}_{\textup{L}} , \hspace{1mm} \psi^{\downarrow}_{\textup{L}}\hspace{1mm}\right]^{\top} &  \rightarrow (1) \rightarrow  & \left[\hspace{1mm}{\psi^{\uparrow}_{\textup{L}}}^* , \hspace{1mm} {\psi^{\downarrow}_{\textup{L}}}^*\hspace{1mm}\right]^{\top} & \rightarrow (2) \rightarrow & \left[\hspace{1mm}-{\psi^{\downarrow}_{\textup{L}}}^* , \hspace{1mm} {\psi^{\uparrow}_{\textup{L}}}^*\hspace{1mm}\right]^{\top}.
\end{array}\end{equation}

Notice what happens, though, if we write these Weyl spinors instead in terms of the $\CH$ algebra:

\begin{equation}\label{new}\Psi_{\textup{L}}^* = {\psi^{\uparrow}_{\textup{L}}}^* \hspace{1mm} {\epsilon_{\uparrow \uparrow}}^* +  {\psi^{\downarrow}_{\textup{L}}}^* \hspace{1mm} {\epsilon_{\downarrow \uparrow}}^* = -{\psi^{\downarrow}_{\textup{L}}}^* \hspace{1mm} {\epsilon_{\uparrow \downarrow}} + {\psi^{\uparrow}_{\textup{L}}}^* \hspace{1mm}{\epsilon_{\downarrow \downarrow}}.
\end{equation}

\noindent That is, we arrive at the exact same result using nothing more than the complex conjugate:  $i\mapsto-i$.

\vspace{1cm}

Now, in order to return back to the original left-handed Weyl spinor, in the usual matrix-and-column-vector formalism, we must again $(1)$ complex conjugate the spinor's components, but this time, $(2)'$ multiply by the new matrix $-\epsilon$,

\begin{equation} \begin{array}{lllll}
 \left[\hspace{1mm}-{\psi^{\downarrow}_{\textup{L}}}^* , \hspace{1mm} {\psi^{\uparrow}_{\textup{L}}}^*\hspace{1mm}\right]^{\top} & \rightarrow (1) \rightarrow & \left[\hspace{1mm}-{\psi^{\downarrow}_{\textup{L}}} , \hspace{1mm} {\psi^{\uparrow}_{\textup{L}}}\hspace{1mm}\right]^{\top} & \rightarrow (2)' \rightarrow& 
\left[\hspace{1mm}\psi^{\uparrow}_{\textup{L}} , \hspace{1mm} \psi^{\downarrow}_{\textup{L}}\hspace{1mm}\right]^{\top}. 
\end{array}\end{equation}

\noindent In comparison, the same result is achieved in $\CH$ more simply by taking the complex conjugate twice,
\begin{equation}\label{psicc}{\Psi_{\textup{L}}}^{**} = \Psi_{\textup{L}}.
\end{equation}

Readers should note that $*: \Psi_{\textup{L}}\mapsto \Psi_{\textup{R}}$ is a basis-independent statement in our formalism.  It holds, regardless of whether $\Psi_\textup{L}$ is written in the Weyl, Dirac, or Majorana basis of the Dirac algebra.   To the best of this author's knowledge, \cite{UTI} was the first instance where this more streamlined method of conjugating spinors has been proposed.   

% there is no epsilon
% what was its purpose?
% it took care of complex conjugating basis vectors
% sheds light on a deficiency of qft:  and in fact any theory written in terms of matrix and column vectors - have lost the information by not writing explicittly what the basis vectors are.  in this case, basis vectors transform non-trivially under C*

\vspace{.3cm}

In considering these two formalisms, one comes to notice that no $\epsilon$-type object was needed in the $\CH$ case.  Somehow, the complex conjugate $i\mapsto -i$ automatically encoded the information given by $\epsilon$.  It is natural to ask, then, why $\epsilon$ is necessary in the matrix-and-column-vector formalism.

After comparing equations (\ref{old}) and (\ref{new}), one sees that $\epsilon$ was what was needed so as to account for the complex conjugation of basis vectors.  In the $\CH$ formalism, we naturally take the complex conjugate of \it both \rm the coefficients, and the basis vectors.  On the other hand, there is no notion of complex conjugating basis vectors in the matrix-and-column-vector formalism.  The matrix $\epsilon$ needs to be introduced so as to account for this.  The conjugation of Weyl spinors, as explained here, then exposes a naturalness to the purely algebraic formalism.

%  Majorana

\subsection{Majorana spinors as generalized ideals  }

%  Majorana invariant under complex conjugate

Given our description above of Dirac spinors, it is now easy to build Majorana spinors.  Any Dirac spinor splits into two Majorana spinors, given by 
\begin{equation} \begin{array}{lll}\Psi_{\textup{M}_1} \equiv  \frac{1}{2}\left( \Psi_{\textup{D}} + \Psi_{\textup{D}}^* \right) &\hspace{1cm} & \Psi_{\textup{M}_2} \equiv  \frac{1}{2}\left( \Psi_{\textup{D}} - \Psi_{\textup{D}}^* \right). 
\end{array}\end{equation}
\noindent Clearly, $\Psi_{\textup{M}_1}$ is invariant under complex conjugation, $\Psi_{\textup{M}_1}^* = \Psi_{\textup{M}_1}$, while $\Psi_{\textup{M}_2}$ acquires a minus sign, $\Psi_{\textup{M}_2}^* = -\Psi_{\textup{M}_2}$.  The reader is encouraged to confirm that $\Psi_{\textup{M}_1}$ and $\Psi_{\textup{M}_2}$ both constitute generalized ideals, and that $m_c(L, \Psi_{\textup{M}})$ gives their transformation under $SL(2, \mathbb{C})$.

%  Summarize spinors

\vspace{1cm}

To summarize, we have just found that left- and right-handed Weyl spinors, Dirac spinors, and Majorana spinors are all simply  generalized ideals of the same complex invariant-action, $m_c$.  Furthermore, in every case, their transformation under $SL(2, \mathbb{C})$ can be described succinctly as $\Psi ' =m_c(L, \Psi)$.

%  4 vectors

\section{Hermitian invariant-action, $m_h$ \label{mh}  }

We will now find that the hermitian invariant-action, $m_h$, leads to new generalized ideals, which behave like four-vectors, 

\begin{equation}
\underline{b} ' = a\hspace{1mm} \underline{b}\hspace{1mm} a ^{\dagger}.
\end{equation}

Any element of $\mathbb{C}\otimes\mathbb{H}$ can be written as a sum of hermitian $p\equiv p_0 + p_1i{\epsilon_x} +p_2i{\epsilon_y} + p_3i{\epsilon_z}$ and anti-hermitian $\hat{p}\equiv i\hat{p}_0 + \hat{p}_1{\epsilon_x} +\hat{p}_2{\epsilon_y} + \hat{p}_3{\epsilon_z}$ parts, where the $p_n$ and $\hat{p}_n$ $\in \mathbb{R}$.  %The element $\widetilde{a}$ is the parity conjugate of $a$, obtained by sending ${\epsilon_x}$, ${\epsilon_y}$, ${\epsilon_z}$ $\mapsto$ $-{\epsilon_x}$, $-{\epsilon_y}$, $-{\epsilon_z}$ and reversing the order of multiplication.
 As $a p a ^{\dagger}$ is hermitian and $a \hat{p} a ^{\dagger}$ is antihermitian for any $a \in \mathbb{C}\otimes\mathbb{H}$, it is clear that these two subspaces form generalized ideals under the multiplication $m_h( a,\underline{b} ) = a \underline{b} a ^{\dagger}$.

Just as was done in the case for spinors, we may now set $a=L$, where $L$ represents elements of $SL(2, \mathbb{C})$.  The hermitian element, $p$, transforms as $p'= LpL^{\dagger}$.  Matching components, one finds that under this transformation law, $p$ transforms as a contravariant four-vector under the Lorentz group, \cite{gs}.  Taking the complex conjugate of $p$ describes the transformation a covariant four-vector,  ${p^*}'= L^*p^*\widetilde{L}$.  It can be seen that the antihermitian case for $\hat{p}$ follows analogously.

As an example, let us consider the momentum $p = p_0 + p_1i{\epsilon_x} +p_2i{\epsilon_y} + p_3i{\epsilon_z}$ under a rotation about the $z$ axis by an angle $\theta$.  This rotated momentum is given by

\begin{equation} \begin{array}{c}
p' =  \exp{\left(-\frac{\theta{\epsilon_z}}{2}\right)}\hspace{1mm}p\hspace{1mm} \exp{\left(\frac{\theta{\epsilon_z}}{2}\right)} = \left(  \cos{\frac{\theta}{2}} - {\epsilon_z}\sin{\frac{\theta}{2}}    \right) p \left(    \cos{ \frac{ \theta }{2} } + {\epsilon_z}\sin{\frac{\theta}{2}}   \right)=
\\ \\
p_0 + \left( p_1 \cos{\theta} + p_2 \sin{\theta}  \right) i{\epsilon_x} + \left(  p_2 \cos{\theta} - p_1 \sin{\theta} \right)i{\epsilon_y} + p_3i{\epsilon_z},
\end{array}\end{equation}

\noindent as expected. 

From this example, readers may note that even though the object $p = p_0 + p_1i{\epsilon_x} +p_2i{\epsilon_y} + p_3i{\epsilon_z}$ looks to have indices which are all tied off, it does not represent a Lorentz scalar.  It transforms as a four-vector.  Furthermore, even though the coefficients $p_{\mu}$ have space-time indices, they do not transform directly on their own.   This is in contrast to objects in the formalism of standard QFT, which would have real numbers, $p_{\mu}$, transforming under the Lorentz group.  Here, we have only the complete  $p = p_0 + p_1i{\epsilon_x} +p_2i{\epsilon_y} + p_3i{\epsilon_z}$ transforming under the Lorentz group.

 As shown in~\cite{gs}, scalars can be constructed  between a covariant vector $p$ and contravariant vector $q$, as $\frac{1}{2}\left( pq + \widetilde{pq} \right)$, which is simply the real part of $pq$.  Indeed, when $q=p^*$, this gives $p_0^2 -p_1^2-p_2^2-p_3^2$.

\vspace{1cm}

In brief, we have just seen that four-vectors can be represented simply by generalized ideals under the hermitian multiplicative action, $m_h$.  For the hermitian case, contravariant four vectors transform as $p'= LpL^{\dagger}$, and covariant four-vectors transform as ${p^*}'= L^*p^*\widetilde{L}$, where $L$ represents and element of $SL(2, \mathbb{C})$.

% scalars and F

\section{Quaternionic invariant-action, $m_q$\label{mq}  }

Finally, we will now consider the quaternionic invariant-action, $m_q$.  This new action leads to generalized ideals, which are more commonly known as Lorentz scalars, $\phi$, and the field strength tensor, $F$,

\begin{equation} \underline{b} ' = a\hspace{1mm}\underline{b} \hspace{1mm}\widetilde{a}.
\end{equation}

$\mathbb{C}\otimes\mathbb{H}$ can be partitioned once again, this time into subspaces of the form $\phi \in \mathbb{C} $ and $F = \left( F^{32} +iF^{01}\right){\epsilon_x} + \left( F^{13}+iF^{02} \right){\epsilon_y}  + \left( F^{21}+iF^{03}  \right){\epsilon_z}  $, with $F^{mn} \in \mathbb{R}$.  As $\widetilde{\phi} = \phi$, and $\widetilde{F} = -F$, it is clear that each of these two subspaces is closed under the multiplication $\underline{b}' = a \underline{b} \widetilde{a}$ from any element $a$ of the algebra.  Hence, they each constitute generalized ideals under $m_q$.

Lorentz transformations on these two generalized ideals can be found, again, by replacing $a$ with $L$.  For our scalar, $\phi$, we have $\phi' = L \phi \widetilde{L}$.  Since $\phi \in \mathbb{C}$, it commutes with every element in $\CH$, and so $\phi' = L \phi \widetilde{L} =  L\widetilde{L} \phi$.  It is then easily confirmed that $\widetilde{L} = L^{-1}$ $\hspace{1mm}\forall L$, so that $\phi$ is indeed a Lorentz scalar, $\phi' = \phi$.

In~\cite{gs} it is shown that the usual field strength tensor is represented by $F$, which transforms as $F' = L F \widetilde{L} $ under the Lorentz group.  $F ^* = \left( B_1 +iE_1\right){\epsilon_x} + \left( B_2+iE_2 \right){\epsilon_y}  + \left( B_3+iE_3  \right){\epsilon_z} $ gives the  field strength $F_{\mu\nu}$, while $F $ gives $F^{\mu\nu}$.  
%and the Hodge dual $*F^{\mu\nu}$ is the unexpectedly simple $-iF$.  NOOO!!!!  see lounesto p38
%%CHECK!!!%%  checked  p 966 oct 30 2015 willow binder

%Readers may notice that we have been referring to $F = \left( B^1 -iE^1\right){\epsilon_x} + \left( B^2-iE^2 \right){\epsilon_y}  + \left( B^3-iE^3  \right){\epsilon_z}$ as the \it effective \rm field strength tensor, and not simply the field strength tensor.  This detail will be explained in Section~(\ref{Fmunu}), once the Dirac algebra has been introduced.

Readers are referred to the upcoming Section \ref{Fmunu}, where the description of $F_{\mu\nu}$ will be extended so as to satisfy parity transformations, defined there.

\vspace{1cm}

In summary, we have just seen that the Lorentz scalar, $\phi$, and the field strength tensor, $F$, are simply generalized ideals under the quaternionic invariant-action, $m_q$.  These generalized ideals transform as $\phi' = L\phi \widetilde{L} = \phi$ and $F' = L F \widetilde{L}$ under the Lorentz group, where $L$ represents an element of $SL(2, \mathbb{C})$.

%\section{Complex conjugate as a local metric   }

%We point out here a curious effect of the complex conjugate throughout this work.  Namely, the complex conjugate relates left-handed Weyl spinors to right-handed Weyl spinors;  the complex conjugate relates covariant four-vectors to contravariant four-vectors;  the complex conjugate relates the field strength tensor with indices up to the field strength tensor with indices down:
%\begin{equation}\begin{array}{lllll} \Psi_{\textup{L}}\hspace{3mm} \leftarrow * \rightarrow \hspace{3mm}\Psi_{\textup{R}} &\hspace{1cm}&  p_{\mu} \hspace{3mm} \leftarrow * \rightarrow \hspace{3mm}
%p^{\hspace{.5mm}\mu}      &\hspace{1cm}&      F_{\mu\nu} \hspace{3mm} \leftarrow * \rightarrow \hspace{3mm}F^{\mu\nu}.     \end{array}\end{equation}
%In other words, the complex conjugate seems to encode the action of  flat, or local, space-time metrics.  

%A project worth investigating, then, would be to see how concepts relating to general relativity could be rewritten in this framework.   This is a pursuit which would likely bear some connection to existing descriptions of gravity in terms of geometric algebra, known as gauge theory gravity,~\cite{dorlas}, \cite{ggg}.

\section{Bilinears   }

Bilinears and other scalars can now be built by combining the various ideal representations, whose $SL(2, \mathbb{C})$ factors, $L$, fit together like lock and key.  Noting that $\widetilde{L} = L^{-1}$, let us consider for example the real part of $\Psi_{\textup{L}}^{\dagger} i \partial \hspace{0.05cm} \Psi_{\textup{L}} +  \Psi_{\textup{R}}^{\dagger} i {\partial}^* \hspace{0.05cm} \Psi_{\textup{R}} $, where  $\partial \equiv \partial_t -i{\epsilon_x}\hspace{1mm} \partial_x- i{\epsilon_y}\hspace{1mm} \partial_y- i{\epsilon_z} \hspace{1mm}\partial_z$.  Under a Lorentz transformation, 
%%CHECK THIS - ITS PROBABLY WRONG - - -no it's ok.   checked
\begin{equation}\begin{array}{r}\left<\Psi_{\textup{L}}^{\dagger} i \partial \hspace{0.05cm} \Psi_{\textup{L}} +  \Psi_{\textup{R}}^{\dagger} i {\partial}^* \hspace{0.05cm} \Psi_{\textup{R}} \right>'
= \left<\Psi_{\textup{L}}^{\dagger} L^{\dagger} \hspace{0.15cm} L^* i \partial \widetilde{L} \hspace{0.15cm} {L}  \Psi_{\textup{L}} + \Psi_{\textup{R}}^{\dagger} \widetilde{L} \hspace{0.15cm} L i {\partial}^* L^{\dagger} \hspace{0.15cm} {L}^*  \Psi_{\textup{R}}\right>  \vspace{3mm}\\
=\left< \Psi_{\textup{L}}^{\dagger} i \partial \hspace{0.05cm} \Psi_{\textup{L}} +  \Psi_{\textup{R}}^{\dagger} i {\partial}^* \hspace{0.05cm} \Psi_{\textup{R}}\right>,
\end{array}\end{equation}
\noindent where $\left<\hspace{1mm}\cdots\right>$ means to take the real part.  

This scalar is the same as the scalar between Dirac spinors of quantum field theory, at a fixed space-time point, ${\overline{\Psi}}_D i \hat{\partial} {\Psi}_D = {{\Psi}}^{\dagger}_D  \beta i \gamma^{\alpha} \partial_{\alpha} {\Psi}_D$, in the usual matrix-and-column-vector formalism.  
%Note, however, that in the $\CH$ formalism, we had no need to explicitly introduce an object corresponding to the $4\times4$ matrix $\beta = \left( \begin{smallmatrix}0&I\\ I&0 \end{smallmatrix}  \right)$.  In fact, the reader may notice that we did not need to introduce any objects corresponding to the usual gamma-matrices, either.  
%%THINK ABOUT THIS ARGUMENT.  WHY ARE THERE NO LR CROSSTERMS?

\section{Summary\label{adv}  }

We have just found a set of generalized ideals, originating from three generalized notions of multiplication, $m_c$, $m_h$, and $m_q$.  These generalized ideals led directly to left- and right-handed Weyl spinors, Dirac spinors, and Majorana spinors, four-vectors, scalars, and the field strength tensor.  This accounts for \it all \rm of the Lorentz representations of the standard model.  

\noindent Furthermore, we found that Lorentz transformations can be described concisely by 

\begin{equation}\underline{b}' = m(L,\underline{b}),  
\end{equation}

\noindent (or by  $\underline{b}' = m(L^*,\underline{b})$).  These results were obtained using nothing but the algebra $\CH$ acting on itself.   

\vspace{1cm}

\noindent In contrast to the usual matrix-and-column-vector formalism of  QFT, we point out a number of advantages offered by the $\CH$ formalism:

\vspace{0.5cm}

\noindent * A compact way:  $\underline{b}' = m(L,\underline{b})$, to describe all of the Lorentz representations of the standard model

%\noindent* A limited number of degrees of freedom available, suggesting a cut off for the total  spin of fundamental particles
%\noindent* No need to write in gamma matrices explicitly

%\noindent* A less redundant way to write down the field strength tensor, $F_{\mu\nu}$

\noindent* A seamless new way of conjugating Weyl spinors, using only the complex conjugate, $i\mapsto-i$

%\noindent * More generally speaking, a new description of the flat space, or local metric, via $i\mapsto-i$

%\noindent* A simpler way to arrive at the Hodge dual for $F_{\mu\nu}$

%\noindent * A more unified view of these representations, in terms of generalized ideals

\noindent* A minimalistic formalism, making use of just a single algebra acting on itself

\noindent *  A conceptually new description of Lorentz representations as stable subspaces (generalized ideals) of an algebra.

\noindent *  A possible explanation as to why higher spin states are not seen experimentally in fundamental particle physics.  Higher spin representations of the Lorentz group should be naturally excluded here, if the representation exceeds the number of dimensions describable by $\CH$.

% jordan algebras

% Conne's NCG

% why conjugates?   other automs

%  could a graviton be modelled?

\section{Outlook}

It is plain to see that $m_c$, $m_h$, and $m_q$ each correspond to an involution of the $\CH$ algebra.  That is, $m_c$ corresponds to  $*$, $m_h$ corresponds to $\dagger$, and $m_q$ corresponds to $\sim$.  The complex quaternions, though, have more discrete symmetries available than is shown here, and it would be interesting to see what could come from the corresponding constructions of generalized ideals. 
% CHECK ARE THESE ALL AUTOMORPHISMS??

Another proposal is to take the concept of generalized multiplication further by considering \it generalized automorphisms. \rm  Just as automorphisms preserve the multiplicative structure of an algebra, generalized automorphisms are defined to preserve the \it generalized \rm multiplicative structure of an algebra.  These generalized automorphisms might then qualify as valid candidates for gauge symmetries, in the context of $\RCHO$. 

Another interesting lead to consider is whether or not this work has a connection to Connes' non-commutative geometry, \cite{conneslott}, \cite{boylefar}, \cite{barrett}.  That is, the complex and/or hermitian invariant-actions, $m_c$ and $m_h$, introduced here seem to bear some resemblance to the Dirac operators found there.  Could the results presented here fit into that formalism?  If not, could this work suggest admissible alterations to Connes' axioms?
%%CHECK THIS  yep.  ok check p 986 willow binder nov 2 2015
%bear a resemblance is right

Yet another line of questioning, currently under investigation, is to see how these $\CH$ representations (and later, $\RCHO$ representations) relate to Jordan algebras.

%Finally, it is natural to ask whether or not it could be possible to model a spin-2 graviton within this framework. 

\chapter{Clifford algebras\label{CA}  }

\section{Motivation  }
% Clifford algebras are important

It is difficult to ignore the presence of Clifford algebras within elementary particle physics.  Clifford algebras inevitably appear whenever spinors do, and ultimately underlie the algebra of some well-known grand unified theories, \cite{agut}.  It was in fact argued in \cite{Grass}, \cite{wow}, and \cite{wowagain} that Clifford algebras alone are the source of internal structure for quarks and leptons.  
%Much of the standard model work by Dixon also focussed on the Clifford algebra structure of $\RCHO$.

%  Dirac algebra

%  Grand unified theories

%  Nature doesn't choose every Clifford algebra, it just chooses some of them.  Which ones?

But one may ask, where do these Clifford algebras come from?   Certainly, there is an infinite number of Clifford algebras available, and nature appears to choose only some of them.  How does she make that choice?

%  Look to Clifford algebras arising from generalized ideals of RCHO

%  Clifford algebras were introduced so as to extend quaternions

The connection between the division algebras, $\mathbb{R}$, $\mathbb{C}$, $\mathbb{H}$, $\mathbb{O}$, and Clifford algebras is unmistakable.  In fact, Clifford algebras were introduced in 1878 by William Kingdon Clifford as an extension of the quaternions, \cite{cliff}.

Later, in Section~\ref{chain}, we will show how the action of $\CO$ on itself leads to the Clifford algebra $\CLsix$.  From this Clifford algebra we will ultimately extract a one-generation description of quarks and leptons, followed by some further indications of a three-generation model.

% Intro to Clifford algebras

\section{Definition  }

In \cite{paddy}, \cite{west}, a Clifford algebra over $\mathbb{R}$ is defined to be an associative algebra, which is generated by $n$ elements, $e_i$.  These $n$ generators exhibit the properties

\begin{equation}\label{CAdefn}
\{e_i, e_j \}\equiv e_ie_j + e_je_i = 2\eta_{ij} \mathbb{I}, 
\end{equation}

\noindent where the entries $\eta_{ij}=0$ for $i\neq j$, and $\eta_{ii} = \pm 1$ $\forall i$.  The symbol $\mathbb{I}$ represents the identity.   A Clifford algebra over $\mathbb{R}$ with $p$ generators having the property $e_i^2 = +\mathbb{I}$, and $q$ generators having the property $e_i^2 = -\mathbb{I}$, is referred to as $Cl(p,q)$.  

In this text, we will be interested mostly in Clifford algebras over $\mathbb{C}$, referred to as $\mathbb{C}l(n)$.  Clearly, taking the algebra's field to be $\mathbb{C}$, instead of $\mathbb{R}$, erases the Clifford algebra's signature, $(p,q)\mapsto(n)$.

%None-the-less, any complex Clifford algebra can be rewritten in terms of Clifford algebras over $\mathbb{R}$.

% e.g. CxH as Cl2  and Cl3

Complex Clifford algebras, $\CLn$, with even $n$, each have only one irreducible representation.  This irreducible representation has $2^{n/2}$ complex dimensions.  For $\CLn$ with $n$ odd, there are two inequivalent irreducible representations, each with $2^{(n-1)/2}$ complex dimensions.  In this thesis, we will be concerned mostly with $\mathbb{C}l(n)$ for $n$ even.

Below, we include a couple of helpful tables from~\cite{loun}, detailing how Clifford algebras may be faithfully represented by matrices over the rings $\mathbb{R}$, $\C$, $\h$, $^2\R\equiv\R\oplus\R$, or $^2\h\equiv\h\oplus\h$.  Here, the notation $\mathbb{A}(d)$ refers to $d\times d$ matrices over the ring $\mathbb{A}$.

\vspace{6mm}

\noindent \bf Real Clifford algebras $Cl(p,q)$ for $p+q<8$ \rm \vspace{2mm}

%\begin{sideways}
\scalebox{0.65}{
$ \begin{array}{lcccccccccccccccccccccccccccccc} \label{rca}
\overrightarrow{\mathbf{p-q}} & & \mathbf{-7} & & \mathbf{-6}& & \mathbf{-5}& & \mathbf{-4}& & \mathbf{-3}& & \mathbf{-2}& & \mathbf{-1}& & \mathbf{0}& & \mathbf{1}& & \mathbf{2}& & \mathbf{3}& & \mathbf{4}& & \mathbf{5}& & \mathbf{6}& & \mathbf{7}   \vspace{5mm}\\

\downarrow \mathbf{p+q} & & & & & & & & & & & & & & & & & & & & & & & & & & & & & &  \vspace{3mm}\\
                         \mathbf{0} & & & & & & & & & & & & & & & & \R& & & & & & & & & & & & & &  \vspace{3mm}\\
                         \mathbf{ 1}& & & & & & & & & & & & & & \C& & & &^2\R & & & & & & & & & & & &  \vspace{3mm}\\
                         \mathbf{2 }& & & & & & & & & & & &\h & & & &\R(2) & & & & \R(2)& & & & & & & & & &  \vspace{3mm}\\
                         \mathbf{3} & & & & & & & & & & ^2\h& & & & \C(2)& & & & ^2\R(2)& & & & \C(2)& & & & & & & &  \vspace{3mm}\\
                         \mathbf{ 4}& & & & & & & &\h(2) & & & & \h(2) & & & &\R(4) & & & &\R(4) & & & &\h(2) & & & & & &  \vspace{3mm}\\
                         \mathbf{5 }& & & & & &\C(4) & & & & ^2\h(2)& &  & & \C(4)& & & &^2\R(4) & & & &\C(4) & & & & ^2\h(2)& & & &  \vspace{3mm}\\
                         \mathbf{6 }& & & &\R(8) & & & &\h(4) & & & & \h(4) & & & &\R(8) & & & & \R(8)& & & & \h(4)& & & & \h(4) & &  \vspace{3mm}\\
                         \mathbf{ 7}& & ^2\R(8)& & & & \C(8)& & & &^2\h(4) & &  & &\C(8) & & & & ^2\R(8)& & & &\C(8) & & & &^2\h(4) & &  & &\C(8) 
\end{array}$
%\end{sideways}
}
%$$ \resizebox{1.05\hsize}{!}{ 

%\begin{array}{lcccccccccccccccccccccccccccccc} \label{rca}
%\overrightarrow{\mathbf{p-q}} & & \mathbf{-7} & & \mathbf{-6}& & \mathbf{-5}& & \mathbf{-4}& & \mathbf{-3}& & \mathbf{-2}& & \mathbf{-1}& & \mathbf{0}& & \mathbf{1}& & \mathbf{2}& & \mathbf{3}& & \mathbf{4}& & \mathbf{5}& & \mathbf{6}& & \mathbf{7}   \vspace{5mm}\\

%\downarrow \mathbf{p+q} & & & & & & & & & & & & & & & & & & & & & & & & & & & & & &  \vspace{3mm}\\
                   %      \mathbf{0} & & & & & & & & & & & & & & & & \R& & & & & & & & & & & & & &  \vspace{3mm}\\
                  %       \mathbf{ 1}& & & & & & & & & & & & & & \C& & & &^2\R & & & & & & & & & & & &  \vspace{3mm}\\
                  %       \mathbf{2 }& & & & & & & & & & & &\h & & & &\R(2) & & & & \R(2)& & & & & & & & & &  \vspace{3mm}\\
                   %      \mathbf{3} & & & & & & & & & & ^2\h& & & & \C(2)& & & & ^2\R(2)& & & & \C(2)& & & & & & & &  \vspace{3mm}\\
                   %      \mathbf{ 4}& & & & & & & &\h(2) & & & & \h(2) & & & &\R(4) & & & &\R(4) & & & &\h(2) & & & & & &  \vspace{3mm}\\
                   %      \mathbf{5 }& & & & & &\C(4) & & & & ^2\h(2)& &  & & \C(4)& & & &^2\R(4) & & & &\C(4) & & & & ^2\h(2)& & & &  \vspace{3mm}\\
                    %     \mathbf{6 }& & & &\R(8) & & & &\h(4) & & & & \h(4) & & & &\R(8) & & & & \R(8)& & & & \h(4)& & & & \h(4) & &  \vspace{3mm}\\
                    %     \mathbf{ 7}& & ^2\R(8)& & & & \C(8)& & & &^2\h(4) & &  & &\C(8) & & & & ^2\R(8)& & & &\C(8) & & & &^2\h(4) & &  & &\C(8) 
%\end{array}
%}$$

\vspace{6mm}

\noindent Furthermore, for  Clifford algebras of larger $p+q$, we have the isomorphisms
\begin{equation} Cl(p,q+8)\simeq Cl(p,q)\otimes\R(16), 
\end{equation}
\noindent and similarly, 
\begin{equation}Cl(p+8,q)\simeq Cl(p,q)\otimes\R(16).
\end{equation}

\vspace{10mm}

\noindent On the other hand, for the case of complex Clifford algebras, we have $\CLn\simeq\C(2^{n/2})$ for $n$ even, and $\CLn\simeq \hspace{1mm}^2\C(2^{(n-1)/2})$ for $n$ odd.

\vspace{5mm}

\noindent \bf Complex Clifford algebras $\C l(n)$ for $n<8$ \rm \vspace{2mm}

\begin{tabular}{cccccccccccccccccc}\label{cca}

$\mathbf{n}\rightarrow$ & & $\mathbf{0 }$ & & $\mathbf{1}$  & &$\mathbf{ 2} $ & &$ \mathbf{3}$  & &$ \mathbf{4}$  & &$ \mathbf{5}$  & & $\mathbf{6} $ & & $\mathbf{7} $ \vspace{3mm}\\

$\mathbb{A}(d) \rightarrow$ & &$\C$& &$ ^2\C $& &$ \C(2) $& & $^2\C(2) $& & $ \C(4) $& & $^2\C(4) $& & $\C(8) $& &$^2\C(8)$

\end{tabular}

\vspace{1cm}

\section{Examples of Clifford algebras from $\CH$  }

\begin{figure}[h!]
\begin{center}
\includegraphics[width=7cm]{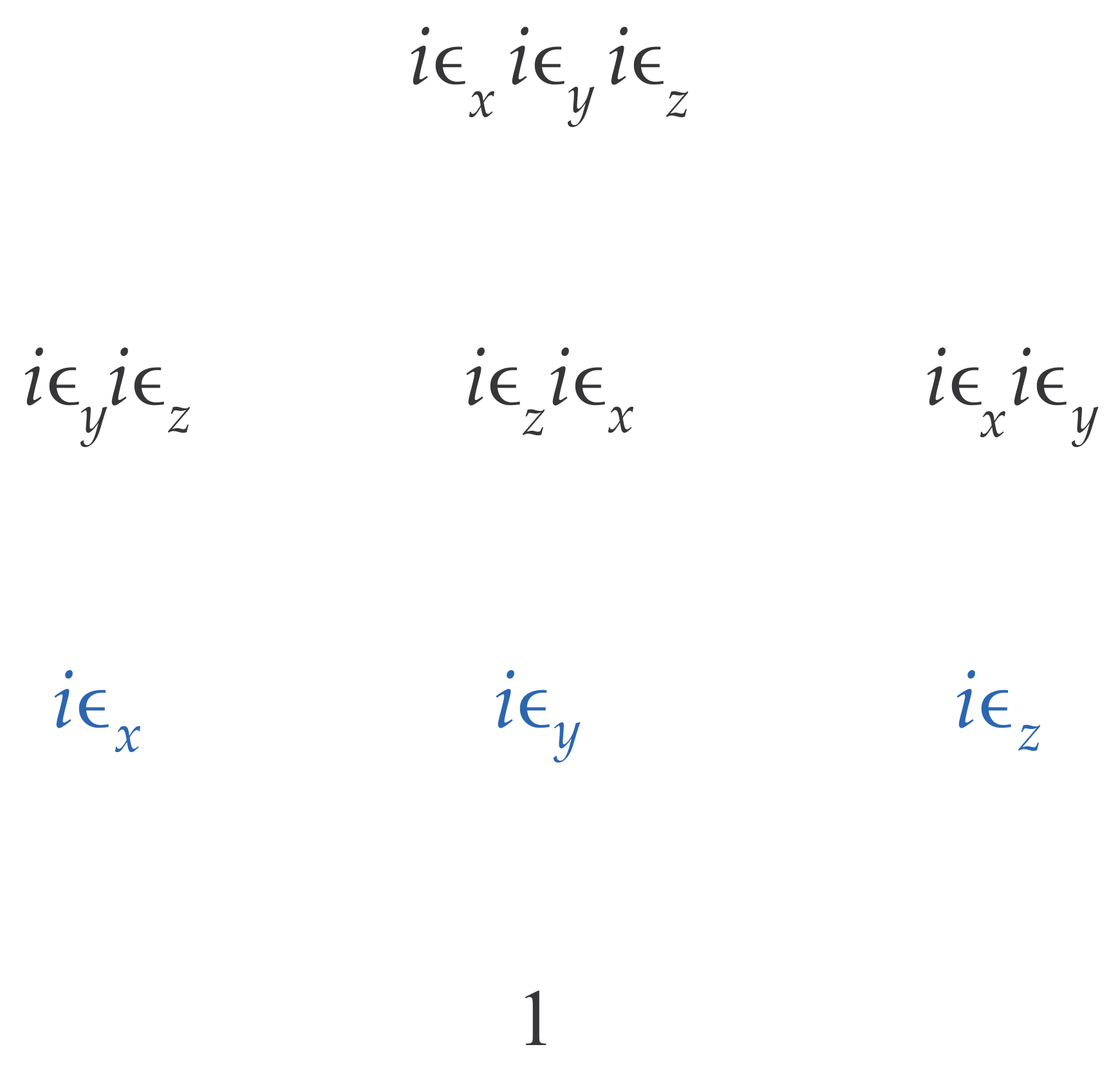}
\caption{\label{cl3}
The algebra $\CH$ written so as to show its $Cl(3,0)$ structure.  Here, the Clifford algebra's zero-vector is 1, its generating vectors are $i{\epsilon_x}$, $i{\epsilon_y}$, $i{\epsilon_z}$, its bivectors are ${i\epsilon_y}{i\epsilon_z} = -{\epsilon_x}$, \hspace{1mm} ${i\epsilon_z}{i\epsilon_x} = -{\epsilon_y}$, \hspace{1mm}${i\epsilon_x}{i\epsilon_y} = -{\epsilon_z}$, and its 3-vector is ${i\epsilon_x}{i\epsilon_y}{i\epsilon_z} = i$.  These multi-vectors are understood to be taken over $\R$.}
\end{center}\end{figure}

Given the multiplication properties from equation~(\ref{CHmult}), it is straightforward to see that $\CH$ gives a representation of the real Clifford algebra, $Cl(3,0)$.  Please see Figure~(\ref{cl3}).

\noindent Alternately, $\CH$ also gives a representation of $\CLtwo$, as shown in Figure~(\ref{cl2}).
\begin{figure}[h!]\label{cl2}\begin{center}
\includegraphics[width=5cm]{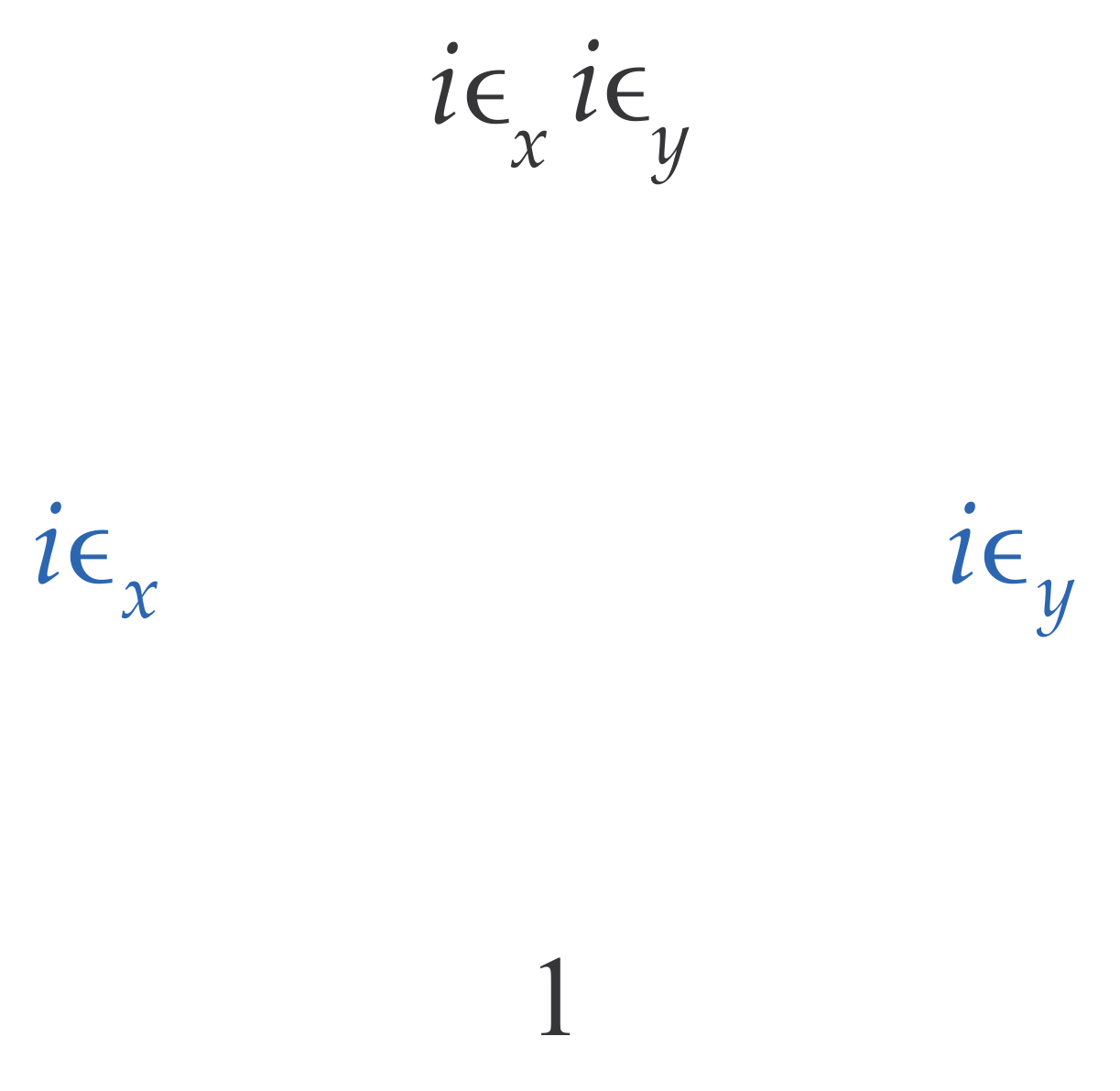}
\caption{\label{cl2}
The algebra $\CH$ written so as to show its $\CLtwo$ structure.  Here, the Clifford algebra's zero-vector is 1, its generating vectors are $i{\epsilon_x}$ and $i{\epsilon_y}$, its bivector is ${i\epsilon_x}{i\epsilon_y} = -{\epsilon_z}$.  These multi-vectors are understood to be taken over $\C$.}
\end{center}\end{figure}

% complex cliff algebras have these irreps

\newpage

It was mentioned above that complex Clifford algebras, $\mathbb{C}l(n)$, each have just a single irreducible representation when $n$ is even, and have two inequivalent irreducible representations when $n$ is odd.  In the even case, that irreducible representation is $2^{n/2}$-complex-dimensional, whereas, in the odd case, each of the two irreducible representations are $2^{(n-1)/2}$-complex-dimensional.  These irreducible representations are commonly known as  spinors. 
%CHECK THIS

Consider the example of the complex quaternions acting on themselves from the left.  This action gives a representation of $\CLtwo$, so that $n=2$ in this case.  We would then expect a single $2^{n/2}$-complex-dimensional irreducible representation for that Clifford algebra.  This irreducible representation is none other than the familiar 2-complex-dimensional Weyl spinor.  
% - CHECK WHOLE THING 2 paragraphs

\section{Minimal left ideals  }

% but we want simply an algebra acting on itself.  therefore min left ideals
For the remainder of this text, we will be interested only in Clifford algebras over $\mathbb{C}$, where $n$ is even.  The most common way to model these algebras is to represent them as $2^{n/2}\times2^{n/2}$ complex matrices, as mentioned above.  These complex matrices then act on spinors in the form of $2^{n/2}$-complex-dimensional column vectors.  
%%%ARE THESE CALLED SPINORS AT THIS POINT?

%%%%%ARE IRREPS OF CAs DEFINED AS SPINORS OR DO YOU NEED THE SPIN GROUP?  COULD ALSO DEFINDE MLIS

We would like, however, to avoid resorting to this matrix-and-column-vector formalism, which posits two separate entities:  matrices, and column vectors.   Instead, we will opt for a more streamlined formalism, based simply on a single algebra acting on itself.  The motivation for such a formalism originates from the algebraic model sketched in Chapter~\ref{m1}.

%  minimal left ideals - defn
It should not be surprising that this goal is an obtainable one.  That is, there exist $2^{n/2}$-complex-dimensional subalgebras within these Clifford algebras, which serve as irreducible representations.  These subalgebras are called \it minimal left ideals, \rm and provide one way of defining spinors,~\cite{ablam}.

Given an algebra, $A$, a \it left ideal, \rm $B$, is a subalgebra of $A$ whereby $a \underline{b}$ is in $B$ for all $\underline{b}$ in $B$, and for any $a$ in $A$.  That is, no matter which $a$ we multiply onto $\underline{b}$, the new product, $\underline{b}'\equiv a\underline{b}$, must be in the subspace $B$ (i.e.  the ideal $B$ survives).  

Now, a \it minimal left ideal \rm is a left ideal which contains no other left ideals other than $\{0\}$ and itself.

%  protocol steps - ablam

\section{How to identify minimal left ideals  }

We will now summarize for the reader a procedure for finding minimal left ideals in complex Clifford algebras, where $n$ is even.  This procedure is described in~\cite{ablam}, which also accommodates other types of Clifford algebra.  

In the special case of $\CH$,  this procedure will allow us to recover the Weyl spinors $\Psi_{\textup{L}}$ or $\Psi_{\textup{R}}$, which we found earlier via the complex multiplicative action, $m_c$.    We will then move on to the algebra of $\CO$, which can be seen to generate $\CLsix$.  This leads us to minimal left ideals behaving as a full generation of quarks and leptons.

%  Define Quadratic form / Bilinear form
\subsection{Some definitions  \label{defns}}

Let $V$ be the vector space over a field $\mathbb{F}$.  A \it quadratic form, \rm $Q$, is a map $Q:V\rightarrow \mathbb{F}$ such that $\forall v, w\in V$ and $\lambda \in \mathbb{F}$, 
\begin{equation} Q(\lambda v) = \lambda^2 Q(v)
\end{equation}
\noindent and such that the map $B:V\times V\rightarrow \mathbb{F}$,
\begin{equation}B: (v,w) \mapsto Q(v+w)-Q(v)-Q(w)
\end{equation}
\noindent is linear in both $v$ and $w$.  This $V,$ together with its quadratic form, is called a \it quadratic vector space. \rm  

We will be especially interested in the quadratic vector space given by $V$ and $Q$, where $V$ is the $n$-dimensional vector space spanned by the generating elements $e_j$ of definition (\ref{CAdefn}).  Here, the quadratic form will be given by $Q(v) = \{v,v\}$, and the corresponding bilinear form is then $B(v,w) = 2\{v,w\}.$  Note that $V$ here is only the generating subspace of our Clifford algebra, it does not represent the entire Clifford algebra.

%  Our example is {,}

%  Define 

A subspace $U$ of $V$ is said to be \it totally isotropic \rm if $B(\alpha_i, \alpha_j)=0$ $\forall$ $\alpha_i, \alpha_j \in U.$  In the following construction, we will be interested in \it maximal totally isotropic subspaces \rm of $V$, which are defined to be totally isotropic subspaces of $V$ with maximal dimension.  As explained in \cite{ablam}, for Clifford algebras over $\mathbb{C}$ with $n$ even, the dimension of any maximal totally isotropic subspace will be $n/2$.  
In fact, we find that the generating space splits into exactly two maximal totally isotropic subspaces, each with the dimension $n/2$.
%That is, the maximal totally isotropic subspace makes up exactly half of the Clifford algebra's generating space.

It is interesting to note that our maximal totally isotropic subspace, $U$ of $V,$ generates a Grassmann algebra, seeing as how $B(\alpha_i, \alpha_j)= 2\{\alpha_i,\alpha_j\}=0\hspace{1mm}$ $\forall$ $\alpha_i, \alpha_j \in U.$  This fact merits notice, especially in consideration of recent work by L\'{a}szl\'{o}, \cite{andras}.

% MTIS space
%  Build Vac
% MLI = CA Vac

\subsection{Procedure\label{proc}  }

With these definitions in hand, the procedure for constructing minimal left ideals is straightforward:

$$\begin{array}{ll}
 1. &  \textup{First identify a quadratic vector space} \hspace{1mm} \left(V,Q\right).\hspace{2mm}  \textup{In particular, we will take}\hspace{1mm} V\hspace{1mm} \textup{to be} 
 
 \\ & \textup{the generating space of a given Clifford algebra, spanned by the generators}\hspace{1mm} \{e_j\}
 \\ &\vspace{.5cm} \textup{over}\hspace{1mm} \C, \hspace{1mm}\textup{where}\hspace{1mm} j=1\dots n.  \hspace{2mm}\textup{We will take our quadratic form to be} \hspace{1mm}Q(v) = \{v,v\}.
\\
  2. & \textup{Identify an MTIS (maximal totally isotropic subspace) $U$ of $V$ as the largest possible }\\
& \vspace{.5cm}  \textup{subspace of $V$ such that $\{\alpha_i,\alpha_j\} = 0 \hspace{2mm}\forall \alpha_i, \alpha_j\in U$.}
\\
  3. &  \textup{Define the nilpotent object $\Omega \equiv \alpha_1 \alpha_2 \cdots \alpha_{n/2}$, where the $\alpha_i$ are linearly independent }
\\  \vspace{.5cm} & \textup{basis vectors of $U$.}
\\
 4. &  \textup{Construct the projector, $\Omega\Omega^{\dagger}$, where $\dagger$ takes $i\mapsto-i$, $e_j\mapsto-e_j$, and reverses the order}
 \\ & \vspace{.5cm}\textup{of multiplication. }\\
  5. &  \textup{Left multiply the entire Clifford algebra onto $\Omega\Omega^{\dagger}$ to arrive at a minimal left ideal:  }
 \\ & \mathbb{C}l(n)\Omega\Omega^{\dagger}. 
\end{array}$$
\medskip

It should be noted that the above is not the only way in which minimal left ideals may be constructed; for details, please see~\cite{ablam}. 

The projector, $\Omega\Omega^{\dagger}$, is an example of a \it primitive idempotent. \rm   A primitive idempotent, $P$, is defined in~\cite{loun} to be an idempotent ($P^2=P$), which is not the sum of two annihilating idempotents, $P\neq P_1+P_2$, where $P_1P_2 = P_2P_1 = 0$.

%  weyl spinors as min left ideals

%\section{MTIS symmetries}

%Going beyond the construction of minimal left ideals in~\cite{ablam}, we would now like to point out some symmetries in their construction.  These symmetries will be called \it unitary MTIS symmetries, \rm or simply \it MTIS symmetries. \rm  These MTIS symmetries will be of central importance for us later on, as they will lead to gauge symmetries when the full $\RCHO$ algebra is considered.

%In the case of complex Clifford algebras, $\CLn$, with even $n$, the generating space can always be partitioned into two maximally totally isotropic subspaces, \cite{ablam}, each of dimension $n/2$.

\section{Minimal left ideals in $\CH$  }

As a specific example, let us now build a minimal left ideal in $\CH\simeq\CLtwo$, as per Figure (\ref{cl2}).   Following the procedure from Section~(\ref{proc}),

$$\begin{array}{ll} 
 1. &  \textup{Our vector space, $V,$ is spanned by the elements $i{\epsilon_x}$ and $i{\epsilon_y}$ over $\mathbb{C}$.  The quadratic form is} \\
& \vspace{.5cm} \textup{given by $Q(v) = \{v,v\}$.}
 \\
  2. & \textup{An MTIS of $V$ is just one-complex-dimensional in this case, and can be spanned}\\
 & \vspace{.5cm} \textup{by $\alpha_1 = \epsilon_{\downarrow\uparrow} = \frac{1}{2}\left({\epsilon_y} +i {\epsilon_x}\right)$.  (Another option would be $\epsilon_{\uparrow\downarrow}= \frac{1}{2}\left(-{\epsilon_y} +i {\epsilon_x}\right)$).} \\
 3. &  \textup{Since $U$ is only one-dimensional here, the nilpotent object $\Omega$ is given simply by} \\ 
 & \vspace{.5cm} \Omega =\alpha_1 =  \epsilon_{\downarrow\uparrow}. 
\\
 4. &  \vspace{.5cm} \textup{The primitive idempotent, $\Omega\Omega^{\dagger}$, is then  $\Omega\Omega^{\dagger} = \epsilon_{\downarrow\uparrow}{\epsilon_{\downarrow\uparrow}}^{\dagger} = \epsilon_{\downarrow\downarrow}$.}
\\5. & \textup{Our minimal left ideal is finally given by $\Psi_{\textup{R}}  = \CH \hspace{1mm}\Omega\Omega^{\dagger} = \psi^{\uparrow}_{\textup{R}} \hspace{1mm} \epsilon_{\uparrow \downarrow} +  \psi^{\downarrow}_{\textup{R}} \hspace{1mm} \epsilon_{\downarrow \downarrow} $.}
\end{array}$$
\medskip

\noindent  Readers will notice that this minimal left ideal matches $\Psi_{\textup{R}}$, found as a generalized ideal  in Section~(\ref{mc}).  Furthermore, redoing this procedure by taking the maximal totally isotropic subspace to be instead spanned by $\epsilon_{\uparrow\downarrow}$ yields the familiar $\Psi_{\textup{L}}  =  \psi^{\uparrow}_{\textup{L}} \hspace{1mm} \epsilon_{\uparrow \uparrow} +  \psi^{\downarrow}_{\textup{L}} \hspace{1mm} \epsilon_{\downarrow \uparrow}$ from before.

So, we have just shown that we can use this procedure to build a left- or right-handed Weyl spinor from the left action of $\CH$ on itself.

\subsection{Fock space structure\label{fock}  }

With a little relabelling, it becomes obvious that these minimal left ideals naturally exhibit Fock space structure.  Taking $\alpha_1 = \epsilon_{\downarrow\uparrow}$, it then follows that $\alpha_1^{\dagger} = \epsilon_{\uparrow\downarrow}$, which have the anticommutation relations,

\begin{equation}\begin{array}{lll} \{\alpha_1, \alpha_1\} =  \{\alpha_1^{\dagger}, \alpha_1^{\dagger}\} = 0 & \hspace{1cm} &  \{\alpha_1, \alpha_1^{\dagger}\} = 1.
\end{array}\end{equation}

\noindent Defining the (formal) vacuum states to be $v\equiv \Omega\Omega^{\dagger}$ and $v^{*}\equiv \Omega^{\dagger}\Omega$ for $\Psi_{\textup{R}}$ and $\Psi_{\textup{L}}$, respectively, we have

\begin{equation}  \Psi_{\textup{R}}  =  \psi^{\uparrow}_{\textup{R}} \hspace{1mm} \epsilon_{\uparrow \downarrow} +  \psi^{\downarrow}_{\textup{R}} \hspace{1mm} \epsilon_{\downarrow \downarrow} =   \psi^{\uparrow}_{\textup{R}} \hspace{1.5mm} \alpha_1^{\dagger}v +  \psi^{\downarrow}_{\textup{R}} \hspace{1.5mm} v
\end{equation}

\begin{equation}   \Psi_{\textup{L}} = \psi^{\uparrow}_{\textup{L}} \hspace{1mm} \epsilon_{\uparrow \uparrow} +  \psi^{\downarrow}_{\textup{L}} \hspace{1mm} \epsilon_{\downarrow \uparrow} = \psi^{\uparrow}_{\textup{L}} \hspace{1.5mm} v^* +  \psi^{\downarrow}_{\textup{L}} \hspace{1.5mm} \alpha_1v^*.
\end{equation}

\noindent That is, $\alpha_1^{\dagger}$ acts as a raising operator from the vacuum $v$ within $\Psi_{\textup{R}}$, and $\alpha_1^{*\dagger} = -\alpha_1$ acts as a raising operator from the vacuum $v^*$ within $\Psi_{\textup{L}}$.  It should be clear to the reader that $v$ and $v^*$ represent vacua only in an algebraic sense, and are not meant to represent the zero-particle state.

Such Fock space structure will reappear in other constructions, for example, when we build minimal left ideals from the algebra $\CO$.  Moreover, it comes up again in the work of L\'{a}szl\'{o}, \cite{andras}.

\section{The Dirac algebra \label{da}  }

%  Cl2 from the right as CA ww^T  + CA w ~ a 2C dim spinor
It is not enough for us here to consider minimal left ideals for just $\Psi_{\textup{L}}$ or for just $\Psi_{\textup{R}}$, separately.  We would like also to combine these two objects together into a single four-complex-dimensional Dirac spinor, as shown in Section~\ref{WD}.

It had been mentioned earlier that chirality, L and R, is the analogue of spin, $\uparrow$ and $\downarrow$, in this formalism.  One can confirm that left multiplying $\Psi_{\textup{L}}$ and $\Psi_{\textup{R}}$ by $\CH$ causes rotation between their spin states, $\uparrow$ and $\downarrow$, whereas right multiplying $\Psi_{\textup{L}}$ and $\Psi_{\textup{R}}$ by $\CH$ causes rotation between L and R.

% representation of Cl_4 ~ Cl2xCl2
The left action and right action of $\CH$ on itself each give a representation of $\CLtwo$.  Furthermore, taking $\ell,$ $a$, $r$ $\in\CH$, one can easily show that $\ell a = ar$ $\forall a$ $\Rightarrow$ $\ell, r$ $\in\mathbb{C} \subset \CH$.  That is, the left action of $\CH$ on itself cannot be re-expressed as the right action of $\CH$ on itself, and vice versa.  It is a property of $\CH$ that these two actions are distinct, a feature which does not appear in the case of $\CO$.  Furthermore, the associativity of $\CH$ ensures that the left and right action commute with each other.  Hence, the underlying Clifford algebraic structure when considering both left and right multiplication of $\CH$ on itself is $\CLtwo\otimes_{\mathbb{C}}\CLtwo$, where the tensor product is over $\mathbb{C}$. 

As $\mathbb{C}l(n) \otimes_{\mathbb{C}} \CLtwo \simeq \mathbb{C}l(n+2)$ for $n\in\mathbb{Z}\geq 0,$ we then find that $\CLtwo\otimes_{\mathbb{C}}\CLtwo\simeq \CLfour$.  Furthermore, as the complexification of a real Clifford algebra, $Cl(p,q)$ acts to erase its signature, $\mathbb{C}\otimes Cl(p,q) \simeq \mathbb{C}l(p+q)$, we then find that $\CLfour\simeq \mathbb{C}\otimes Cl(1,3)$.  

In other words, the left and right action of $\CH$ on itself gives a representation of the Dirac algebra.

%  Dirac algebra rep

It is straightforward for readers to confirm this fact.  Explicitly, the gamma matrices $\gamma^0$, $\gamma^1$, $\gamma^2$, and $\gamma^3$, written in the Weyl basis, have the following correspondences in $\CH$:

\begin{equation}\begin{array}{ccc}
\gamma^0 = \left(   \begin{array}{cc} 0 & I \\ I & 0 
\end{array}\right) \hspace{2mm} \mapsto \hspace{2mm} 1\hspace{1mm} | \hspace{1mm} i{\epsilon_x}    &  \hspace{1cm} &
\gamma^1 = \left(   \begin{array}{cc} 0 & \sigma_x \\ -\sigma_x & 0 
\end{array}\right)  \hspace{2mm}\mapsto \hspace{2mm} i{\epsilon_x}\hspace{1mm} | \hspace{1mm} {\epsilon_y}   \vspace{5mm}\\

\gamma^2 = \left(   \begin{array}{cc} 0 & \sigma_y \\ -\sigma_y & 0 
\end{array}\right) \hspace{2mm} \mapsto \hspace{2mm} i{\epsilon_y}\hspace{1mm} | \hspace{1mm} {\epsilon_y}    &  \hspace{1cm} &
\gamma^3 = \left(   \begin{array}{cc} 0 & \sigma_z \\ -\sigma_z & 0 
\end{array}\right)  \hspace{2mm}\mapsto \hspace{2mm} i{\epsilon_z}\hspace{1mm} | \hspace{1mm} {\epsilon_y}. 
%% THESE HAVE BEEN DOUBLE CHECKED AGAINST ZEE.  MATCH
\end{array}\end{equation}

\noindent These findings are likely to appear in future joint work with G. Fiore from INFN.  

Here, we make use of the bar notation from earlier authors, \cite{deleoGUT}.  By definition, the operator $a\hspace{1mm} | \hspace{1mm} b$ acting on some element $c$, for $a,b,c\in\CH$, is given simply by $acb$.  As an example, consider the object corresponding to $\gamma^0$  acting on $\Psi_{\textup{D}}\in\CH$,

\begin{equation} \begin{array}{l}    1\hspace{1mm}  \Psi_{\textup{D}} \hspace{1mm} i{\epsilon_x}  =  1\hspace{1mm} \left( \Psi_{\textup{L}}+ \Psi_{\textup{R}}\right) \hspace{1mm} i{\epsilon_x}\vspace{3mm}\\ =  \left( \psi^{\uparrow}_{\textup{L}} \hspace{1mm} \epsilon_{\uparrow \uparrow} +  \psi^{\downarrow}_{\textup{L}} \hspace{1mm} \epsilon_{\downarrow \uparrow}  + \psi^{\uparrow}_{\textup{R}} \hspace{1mm} \epsilon_{\uparrow \downarrow} +  \psi^{\downarrow}_{\textup{R}} \hspace{1mm} \epsilon_{\downarrow \downarrow}  \right)  \hspace{1mm} i{\epsilon_x}\vspace{3mm}\\   = \left( \psi^{\uparrow}_{\textup{R}} \hspace{1mm} \epsilon_{\uparrow \uparrow} +  \psi^{\downarrow}_{\textup{R}} \hspace{1mm} \epsilon_{\downarrow \uparrow}  + \psi^{\uparrow}_{\textup{L}} \hspace{1mm} \epsilon_{\uparrow \downarrow} +  \psi^{\downarrow}_{\textup{L}} \hspace{1mm} \epsilon_{\downarrow \downarrow}  \right).
\end{array}\end{equation}  

\noindent This flips chirality, as one would expect.  

From these operators, an object corresponding to $\gamma^5 \equiv i\gamma^0\gamma^1\gamma^2\gamma^3$ can be found to be 

\begin{equation} \gamma^5 = \left(   \begin{array}{cc} -I & 0 \\ 0 & I 
\end{array}\right)  \hspace{2mm} \mapsto \hspace{2mm} -1\hspace{1mm} | \hspace{1mm} i{\epsilon_z}.
\end{equation}

\noindent This thereby confirms the role of $P$ and $P^*$ as $\frac{1}{2}\left(1\mp\gamma^5\right)$, introduced early on in equation~(\ref{cma}).
%CHECKED projector signs against Zee.  works

More generally, it can now be seen that $  \Psi_{\textup{D}} =   \Psi_{\textup{L}}+  \Psi_{\textup{R}}$ introduced previously, does in fact behave as a Dirac spinor, acted upon by a complex quaternionic representation of the Dirac algebra, $\mathbb{C}\otimes Cl(1,3)$.

%\section{Minimal left ideals as operators}

\section{Right action required\label{rar}  }

Readers may have noticed that the left action of $\CH$ on itself was not enough to describe $\Psi_{\textup{L}}+\Psi_{\textup{R}}$ together as a single irreducible representation.  That is, the left action alone has only $\Psi_{\textup{L}}$ (or $\Psi_{\textup{R}}$) as an irrep.  However, the combined Dirac spinor, $\Psi_{\textup{L}}+\Psi_{\textup{R}}$, does become an irreducible representation of a Clifford algebra when the right action of $\CH$ is also included.  Readers will find this theme coming up again in future chapters.

\section{Parity and the field strength tensor\label{Fmunu}  }

The work in this section is to appear in a joint paper, together with G. Fiore from INFN.
%p was described as

In Section~\ref{mh}, we introduced four-vectors, $p$,  as generalized ideals under the hermitian invariant-action, $m_h$.   These can be described as 

\begin{equation}p =  p_0 + p_1i{\epsilon_x} +p_2i{\epsilon_y} + p_3i{\epsilon_z}, 
\end{equation}

\noindent which transform under the Lorentz group as $p' = m_h(L,p) = LpL^{\dagger}$, for $L\in SL(2,\C)$.

%F was described as

Furthermore, in Section~\ref{mq}, we introduced the  field strength tensor, $F$, as a generalized ideal under the quaternionic invariant-action, $m_q$.  Explicitly, we had

\begin{equation}F = \left( B_1 -iE_1\right){\epsilon_x} + \left( B_2-iE_2 \right){\epsilon_y}  + \left( B_3-iE_3  \right){\epsilon_z},
\end{equation}

\noindent which transforms under the Lorentz group as $F' =m_q(L, F)=LF\widetilde{L}$.  Of course, we also studied spinors and scalars in previous sections, but these will not be relevant for us at the moment.

%parity appears to be effected by compl conj

Given these two Lorentz representations, we might wonder how parity transformations are to be carried out.  Under a parity transformation, we would expect

\begin{equation}\begin{array}{lll}p& \hspace{2mm}\mapsto& \hspace{2mm}p' =  p_0 - p_1i{\epsilon_x} -p_2i{\epsilon_y} - p_3i{\epsilon_z}, \vspace{6mm}\\
F&  \hspace{2mm}\mapsto\hspace{2mm}& F' = \left( B_1 +iE_1\right){\epsilon_x} + \left( B_2+iE_2 \right){\epsilon_y}  + \left( B_3+iE_3  \right){\epsilon_z}.
\end{array}
\end{equation}

%however, would like to reserve this for charge conjugation

Now, it might be tempting to consider the quaternionic conjugate as the parity conjugate, since $\widetilde{\epsilon_x} = -\epsilon_x$, $\widetilde{\epsilon_y} = -\epsilon_y$, and $\widetilde{\epsilon_z} = -\epsilon_z$.  However, the quaternionic conjugate sends $F\mapsto -F$, which does not give the desired result:  $E_i\mapsto-E_i$, and $B_j\mapsto B_j$.  So how might we define a parity transformation?

%Another suggestion might be to use the complex conjugate, $i\mapsto-i$.  This conjugate does map $p$ and $F$ properly, and therefore merits further investigation.  However, we point out that the complex conjugate is expected to align more closely with the operation of charge conjugation, $C$, familiar in quantum field theory.  A full comparison between the algebraic conjugates, and quantum field theory's $C$, $P$, and $T$ discrete transformations is the subject of future work.

%Readers may notice, however, that the complex conjugate, $i\mapsto-i$, does transform $p$ and $F$ as would a parity transformation.  Given that complex conne might wonder, \it why? \rm

%now that we have dirac algebra, we can generalize our defn of 4-vs and field strength tensors

In Section~\ref{da}, we demonstrated a representation of the Dirac algebra, $\C\otimes Cl(1,3)$, using both left and right multiplication of $\CH$ on itself.  From this Clifford algebra, it is then possible to define a parity transformation:

\begin{equation}\label{parity} \gamma^0\hspace{2mm}\mapsto\hspace{2mm}\gamma^0 = 1\hspace{1mm}|\hspace{1mm}i\epsilon_x,\hspace{1cm} \gamma^j\hspace{2mm}\mapsto\hspace{2mm}-\gamma^j = -i\epsilon_j \hspace{1mm}| \hspace{1mm}\epsilon_y,
\end{equation}

\noindent for $j = 1,2,3.$  
%can come up with a better defn for parity

%
So, the question is now:  how do we use this parity transformation on the $\gamma^{\mu}$ to induce a parity transformation on $p$ and $F$?  
%The reader is encouraged to take a moment now, so as to answer this question for herself.

The resolution to this question is not obvious, because $p$ and $F$ are objects which are derived from $\CH\simeq\CLtwo$ structure, while the $\gamma^{\mu}$ belong in $\left(\CH\right)\otimes_{\C}\left(\CH\right)\simeq\C\otimes Cl(1,3) \simeq \CLfour$.  One solution, which we will now show, is to generalize the $\CLtwo$ objects, $p$ and $F$, to two new $\CLfour$ operators, $\hat{p}$ and $\hat{F}$.  We might hope, then, that under the right conditions, $\hat{p}$ and $\hat{F}$ would reduce to $p$ and $F$.   In summary, we would like to see the parity transformations on $\hat{p}$ and $\hat{F}$, given by equation~\ref{parity}, automatically induce parity transformations on $p$ and $F$.

%define p
Let us then define $\hat{p}$ to be

\begin{equation}\hat{p} \equiv p_{\mu} \gamma^{\mu} = p_0 \hspace{1mm}1\hspace{1mm}|\hspace{1mm}i\epsilon_x + p_j\hspace{1mm} i\epsilon_j \hspace{1mm}| \hspace{1mm}\epsilon_y, 
\end{equation}

\noindent for $p_{\mu}\in\R$, for $\mu = 0,1, 2, 3$, and $j=1,2,3.$ This is none other than the usual $p_{\mu} \gamma^{\mu}$ of quantum field theory.

%define F

Making use of  the Clifford algebraic description of~\cite{loun}, let us define $\hat{F}$ to be 

\begin{equation} \hat{F} \equiv F_{\mu\nu}\gamma^{\mu\nu}  = F_{0i}\gamma^{0i} + F_{jk}\gamma^{jk} = E_i\hspace{1mm}\epsilon_{i} \hspace{1mm}|\hspace{1mm} \epsilon_z - B_j\hspace{1mm} \epsilon_j\hspace{1mm}|\hspace{1mm}1,
\end{equation}

\noindent for $\mu, \nu = 0,1, 2, 3$, $F_{\mu\nu}\in \R$,  $\gamma^{\mu\nu}\equiv \frac{1}{2}\left[\gamma^{\mu}, \gamma^{\nu}\right]$, and $i, j = 1,2,3$.

%show p , F, spinor transform properly

It is then trivial to see that under a parity transformation given by 

\begin{equation} \gamma^0\hspace{2mm}\mapsto\hspace{2mm}\gamma^0 = 1\hspace{1mm}|\hspace{1mm}i\epsilon_x,\hspace{1cm} \gamma^j\hspace{2mm}\mapsto\hspace{2mm}-\gamma^j = -i\epsilon_j \hspace{1mm}| \hspace{1mm}\epsilon_y,
\end{equation}

\noindent for $j = 1,2,3,$ the operator, $\hat{p}$ transforms as 

\begin{equation}\hat{p}\hspace{2mm}\mapsto\hspace{2mm} \hat{p}' = p_0 \hspace{1mm}1\hspace{1mm}|\hspace{1mm}i\epsilon_x - p_j\hspace{1mm} i\epsilon_j \hspace{1mm}| \hspace{1mm}\epsilon_y,
\end{equation}

\noindent and the operator, $\hat{F}$ transforms as

\begin{equation} \hat{F}\hspace{2mm}\mapsto\hspace{2mm} \hat{F}' = -E_i\hspace{1mm}\epsilon_{i} \hspace{1mm}|\hspace{1mm} \epsilon_z - B_j\hspace{1mm} \epsilon_j\hspace{1mm}|\hspace{1mm}1,
\end{equation}

\noindent as we would expect.

So, we  have a new description for four-momenta, $\hat{p}$, and the field strength tensor, $\hat{F}$, which transform as they should under parity.  We would now like to know how these relate to the generalized ideals, $p$ and $F$ that we found earlier in Sections~\ref{mh} and \ref{mq}.  

%new description reduces to generalized ideals when considered as operators acting on spinors.

We find that the operators $\hat{p}$ and $\hat{F}$ reduce to the generalized ideals $p$ and $F$, when $\hat{p}$ and $\hat{F}$ are taken to be operators, \it acting on spinors \rm in $\CH$.  Incorporating the object $\gamma^0$, we find that $\hat{p}$ reduces to $p$, or $p^*$, on Weyl spinors,

\begin{equation} \gamma^0 \hat{p} \hspace{1mm}\Psi_{\textup{L}} = p^*\hspace{1mm}\Psi_{\textup{L}}, \hspace{1cm}  \gamma^0 \hat{p} \hspace{1mm}\Psi_{\textup{R}} = p\hspace{1mm}\Psi_{\textup{R}}.
\end{equation}

%\noindent or in other words,

%\begin{equation} \gamma^0 \hat{p}\Psi_{\textup{D}} = m_c(p^*, \Psi_{\textup{D}}).
%\end{equation}

\noindent Similarly, we find that $\hat{F}$ reduces to $F$, or $F^*$, on Weyl spinors as

\begin{equation}\hat{F}\hspace{1mm}\Psi_{\textup{L}} =-F^*\hspace{1mm}\Psi_{\textup{L}}, \hspace{1cm} \hat{F}\hspace{1mm}\Psi_{\textup{R}} =-F\hspace{1mm}\Psi_{\textup{R}}.
\end{equation}

\noindent Readers may note that a parity transformation on $\hat{p}$ and $\hat{F}$ now automatically induces the correct parity transformation on $p$ and $F$, when these objects are taken to be operators on spinors.

Finally, we point out that this definition of parity, 

\begin{equation} \gamma^0\hspace{2mm}\mapsto\hspace{2mm}\gamma^0 = 1\hspace{1mm}|\hspace{1mm}i\epsilon_x,\hspace{1cm} \gamma^j\hspace{2mm}\mapsto\hspace{2mm}-\gamma^j = -i\epsilon_j \hspace{1mm}| \hspace{1mm}\epsilon_y,
\end{equation}

\noindent sends $\gamma^5 \mapsto -\gamma^5$, so that left- and right-handed Weyl spinors are swapped under this transformation, as we would expect.

\medskip

In summary, we have shown that the parity conjugate of the  $\C\otimes Cl(1,3) \simeq \left(\CH\right)\otimes_{\C}\left(\CH\right)$ algebra may be used to induce a parity conjugate on our  generalized ideals in $\CH$.   

Note that we have also demonstrated a way to write down four-vectors, and the field strength tensor, in the Dirac algebra formalism, which makes use of the generalized ideals found in Sections~\ref{mh} and \ref{mq}.  Namely, 

\begin{equation}\hat{p}\hspace{1mm} \Psi_{\textup{D}} =  \gamma^0\hspace{1mm} m_c(p^*, \Psi_{\textup{D}}), \hspace{1cm} \hat{F}\hspace{1mm}\Psi_{\textup{D}} = m_c(-F^*, \Psi_{\textup{D}}).
\end{equation}

\chapter{Standard model of particle physics\label{SM}  }

\section{What it is\label{is}  }

% mosaic - developed over several decades (ram)

The standard model of particle physics is the result of decades of collaboration, which began roughly in the 1930s, and converged finally on its current state in 1979, \cite{ramjbsm}.  It is a mosaic of our best efforts in particle physics over that half century.  In the decades since 1979, the standard model has seen little in the way of alterations, and yet has survived rigorous experimental testing, nearly completely unscathed.

%JJ Thomson electron discovered >100 years ago
%1930s QED

A brief history:  In 1928, P.A.M. Dirac set up the foundations for quantum electrodynamics (QED), which was later generalized to incorporate the neutrino by E. Fermi, \cite{ramjbsm}.  Feynman, Schwinger, and Tomonaga subsequently developed renormalization theory for QED, \cite{kibble}.  In the mid-1950s, M. Gell-Mann  proposed \it strangeness \rm as a new quantum number, while 
%in 1956, 
T.D. Lee and C. N. Yang suggested methods for detecting parity violation of the weak force.  In that same decade, the W boson was suggested as a mediator of the weak force, following the work of  C.N. Yang and R.L. Mills,~\cite{ramjbsm}.

In the 1960s, G. Zweig, and independently, M. Gell-Mann, proposed that subatomic particles (now known as quarks) were the constituents of baryonic matter.  Electroweak theory was developed by S. Glashow, A. Salam, J.C. Ward, and S. Weinberg, in that same decade, \cite{ramjbsm}.  A mechanism to impart mass on gauge bosons was then proposed by three groups independently:  first of all, R. Brout and F. Englert, then, P.W. Higgs, and finally G. Guralnik, C. R. Hagen, and T. Kibble, all  in 1964, \cite{kibble}.  In 1971, G. 't Hooft showed that the Glashow-Salam-Ward-Weinberg electroweak model was renormalizable, \cite{thooft}. M. Gell-Mann, H. Fritzsch, and H. Leutwyler's work in the 1970s resulted in what we know as quantum chromodynamics (QCD).  For a more detailed history of the standard model, readers are encouraged to consult~\cite{ramjbsm}.

% representations

The particle content of the standard model can be characterized by labelling particles according to how they transform under the standard model's gauge group, $SU(3)_c\times SU(2)_L\times U(1)_{Y}$.  (More accurately, the standard model's gauge group is $SU(3)_c\times SU(2)_L\times U(1)_{Y}/{\mathbb{Z}_6}$, which will be discussed in Chapter~\ref{GUT}.)  Here, $Y$ stands for weak hypercharge, and we will be using the weak hypercharge conventions found in~\cite{BM}.  Readers may note the rather arbitrary-looking collection of hypercharges displayed below.  %Values for the particle masses were taken from the Particle Data Group website, http://pdg.lbl.gov

%\vspace{.3cm}

\noindent \bf Fermionic (matter) content of the standard model \rm

$$\begin{array}{lccr}\label{smferm}

\underline{\textup{Symbol} } \hspace{2mm} &  \hspace{2mm} \underline{SU(3)_c}\hspace{2mm}   &   \hspace{2mm}{\underline{SU(2)_L}}\hspace{2mm} & \hspace{2mm} \underline{U(1)_{Y}}   \vspace{6mm}\\

\left(\nu_e, \hspace{1mm} e^- \right)_L   & 1 & 2 & -1/2   \vspace{3mm}\\

\left(\nu_{\mu}, \hspace{1mm} {\mu}^- \right)_L   & 1 & 2 & -1/2   \vspace{3mm}\\

\left(\nu_{\tau}, \hspace{1mm} {\tau}^- \right)_L   & 1 & 2 & -1/2   \vspace{3mm}\\

\left(u, \hspace{1mm} d\right)_L   & 3 & 2 & 1/6   \vspace{3mm}\\

\left(c, \hspace{1mm} s\right)_L   & 3 & 2 & 1/6   \vspace{3mm}\\

\left(t, \hspace{1mm} b\right)_L   & 3 & 2 & 1/6   \vspace{9mm}\\

e^-_R   & 1 & 1 & -1   \vspace{3mm}\\

{\mu}^-_R   & 1 & 1 & -1   \vspace{3mm}\\

{\tau}^-_R   & 1 & 1 & -1   \vspace{3mm}\\

u_R   & 3 & 1 & 2/3   \vspace{3mm}\\

c_R   & 3 & 1 & 2/3   \vspace{3mm}\\

t_R   & 3 & 1 & 2/3   \vspace{3mm}\\

d_R   & 3 & 1 & -1/3   \vspace{3mm}\\

s_R   & 3 & 1 & -1/3   \vspace{3mm}\\

b_R   & 3 & 1 & -1/3   \vspace{3mm}\\

\end{array}$$

Each of the particles above has an anti-particle partner, which transforms as the representation conjugate to that of the original particle.  For example,  $\left(\bar{t}, \hspace{1mm} \bar{b}\right)_R$ transforms as  $\bar{3}$ under $SU(3)_c$, as a $\bar{2}\simeq 2$ under $SU(2)_L$, and has weak hypercharge $-1/6$.

 The standard model also has 12 gauge bosons, whose charges  can be summarized by the following table.

\vspace{1cm}

\noindent \bf Gauge bosonic content of the standard model (prior to electroweak symmetry breaking)\rm

$$\begin{array}{lccc}\label{smbos}

\underline{\textup{Symbol} } \hspace{2mm} &  \hspace{2mm} \underline{SU(3)_c}\hspace{2mm}   &   \hspace{2mm}{\underline{SU(2)_L}}\hspace{2mm} & \hspace{2mm} \underline{U(1)_{Y}}   \vspace{6mm}\\

G^{a}_{\mu} & 8 & 1 & 0 \vspace{3mm}\\

W^{b}_{\mu} & 1 & 3 & 0 \vspace{3mm}\\

B_{\mu} & 1 & 1 & 0 \vspace{3mm}\\
\end{array}$$

Finally, we also have the Higgs boson, which is a scalar particle under the Lorentz group, and transforms under the gauge groups according to the following table.

\vspace{1cm}

\noindent \bf Higgs field \rm

$$\begin{array}{lccc}\label{smbos}

\underline{\textup{Symbol} } \hspace{2mm} &  \hspace{2mm} \underline{SU(3)_c}\hspace{2mm}   &   \hspace{2mm}{\underline{SU(2)_L}}\hspace{2mm} & \hspace{2mm} \underline{U(1)_{Y}}   \vspace{6mm}\\

\phi & 1 & 2 & 1/2 \vspace{3mm}\\

\end{array}$$

\vspace{1cm}

The standard model is believed to be largely valid over a vast range of scales, from the Hubble radius, $10^{30}cm$, down to subatomic scales of $10^{-16}cm$, \cite{ramjbsm}.  It has successfully predicted the existence of a number of particles, including for example, the  $W$ and $Z$ bosons, \cite{Lang},  the top quark, tau-neutrino, \cite{ramjbsm}, and most recently, the Higgs boson, \cite{CMSjhep}, \cite{CMSnat1}, \cite{CMSnat2}, \cite{ATLAS}.

%ranges from 10^(-16)cm to 10^30cm Hubble radius
% predicted all these particles   W, Z, charm quark (Lang)
% predicted top quark mass
%predicted Higgs

Given the striking complexity of our known universe, it is hard to believe that so much of it can be described by such a short list of elementary particles.  Beyond its particle content, the standard model also has a relatively concise list of free parameters, 19 of them, whereby 13 come from the Yukawa sector, 2 from the Higgs sector, 3 from gauge couplings, and 1 from an apparently absent QCD term,~\cite{Mann}, \cite{Robinson}.  
%rather condensed particle list - chess
%good:  only 19 parameters - 13 yukawa 2 Higgs 3 gauge coupling const 1 thetaQCD (Robinson)

\section{What it is not  }

If the standard model is to be faulted for anything, that fault would lie almost exclusively in its incompleteness.  The first  legitimate fracture in the standard model came with the discovery of neutrino oscillations, which imply that neutrinos do indeed have mass, \cite{Mann}.  Having said that, the standard model is easily amended to accommodate massive neutrinos, albeit, at the expense of having to accept a number of additional free parameters in the theory.  

%neutrinos have mass - sun expos

In the absence of any major conflict with experiment, much criticism of the standard model is based in the theory's apparent inability to go beyond.  The standard model does not explain  dark matter, nor dark energy.  Nor, does it describe baryon asymmetry.  Furthermore, the standard model has consistently defied unification with gravity.  

%dark matter
%dark energy
%gravity
%baryon asymmetry

\section{What it should be  }

%worse: accounts for these things, but does not explain them

One might not criticize the standard model too harshly, for simply remaining silent about many of the open problems in physics.  However, what is less benign is that the standard model does at times account for various aspects of nature, and yet fails to explain them.

For example, the standard model singles out a particular gauge group, $SU(3)_c\times SU(2)_L\times U(1)_{Y}/{\mathbb{Z}_6}$, yet does not explain where this group came from.  It further specifies a list of particle representations for this group, Section~\ref{is}, without explaining why these representations were chosen over any other possible set.  The standard model does not explain why those representations are organized into three generations.  It does not explain why $SU(2)_L$ acts on left-handed spinors, but not on right-handed spinors.   It does not explain the apparently ad hoc arrangement of hypercharges.  Furthermore, the standard model does not explain the values of its 19 parameters.

%why these groups?
%why these representations?  
%why three generations?
%why charge quantized?  - important because allows for atoms to be neutral
%why SU(2) acts on L vs R?
%why these hyper charges?  ugly

%bad:  19 parameters, more if you include neutrino mass
%why mass ranges over 5 orders of magnitude? (Langacker)
%strong CP problem - why coeff so small?  (Langacker)
%in comparison to Planck mass - differences are huge (Ram)

% perhaps all of this is due to the SM's patchwork history
% lacks a spine, or unifying structure

Perhaps the gaps in our understanding could be attributed to the patchwork way in which the standard model came to be.  One might argue that what is needed is some form of unifying principles to smooth over the seams.  These unifying principles might come in the form of grand unified theories, supersymmetry, M-theory, or perhaps something else.

%  intro to O
\chapter{One generation of quarks and leptons from $\CO$\label{CO}  }

In Chapter~\ref{CH}, we saw how a four-complex-dimensional algebra, $\CH$, supplies an unusually compact description of all the Lorentz representations of the standard model.     However, there is more to the standard model than just the Lorentz group.   Beyond spin and chirality, standard model fermions also exhibit colour, weak isospin, and charge.  Could these physical features also be the result of some algebra acting on itself?  If so, which one?
%CHECK THAT CHIRALITY MEANS L OR R HANDED

In our work with $\CH$, the reader may have noticed the trivial identity, 

\begin{equation} \CH = \mathbb{R}\otimes\CH. \end{equation} 

\noindent That is, the algebra we have dealt with up until now is the tensor product of three of the four division algebras.  It is then only natural to ask,  what about $\RCHO$?  The octonionic algebra, $\mathbb{O},$ possesses eight degrees of freedom, and it is difficult not to notice: so does a generation of quarks and leptons.  (Neutrino, charged lepton, three up-type quarks and three down-type quarks).

Despite its counter-culture status, the octonions have long drawn the curiosity of generations of physicists.  The algebra is known to appear without warning in apparently disparate areas of mathematics, within algebra, geometry, and topology.  However, despite its ubiquity, its practical uses in physics have remained elusive, due to the algebra's non-associativity, which must be handled with care.  In the following chapters, we aim to demonstrate to the reader that the octonions' non-associativity is not a impediment, but instead a gift, and that this misunderstood algebra is really at the heart of the standard model of particle physics.

The findings which we will now describe make a case in support of those who have been long advocating for the existence of a connection between non-associative algebras and particle theory, \cite{GGquarks}, \cite{GGstats}, \cite{adreams}, \cite{okubo}, \cite{dixon}, \cite{matter}, \cite{UTI}, \cite{Gen}, \cite{qq},  \cite{boylefar}, \cite{mth}, \cite{magic} \cite{jjI}, \cite{jjII}, \cite{jjIII}, \cite{jjIV}, \cite{e6md}, \cite{Grass}, \cite{wow}.  

One of the earliest breakthroughs along these lines belong to G\"{u}naydin and G\"{u}rsey, \cite{GGquarks}, who showed $SU(3)_c$ quark structure in the split octonions.  Later, in \cite{GGstats}, they  showed anti-commuting ladder operators within that model.  Our new results stem from the octonionic chromodynamic quark model of \cite{GGstats}, and are meant to replace the provisional charges of \cite{UTI}.   

In the following pages, we will extend G\"{u}naydin and G\"{u}rsey's findings of quark structure under $SU(3)_c$ by further demonstrating 

1.  lepton structure and 

2.  a natural $U(1)_{em}$ symmetry.  

\noindent In other words, using only the complex octonions, we will complete the particle content of the model to include a full generation of quarks and leptons, under not only $SU(3)_c$, but under both of the standard model's two unbroken gauge symmetries.

\section{A summary of the results to come  }

Using only the algebra of the complex octonions, which we will introduce, we expose an intrinsic structure to a generation of quarks and leptons.  This algebraic structure mimics familiar quantum systems, which have a vacuum state acted upon by raising and lowering operators.     In this case, the neutrino poses as the vacuum state, and electrons and quarks pose as the excited states.  These results are simply the analogue of the Fock space we found earlier in Section~\ref{fock}.

With these raising and lowering operators in hand, we are then able to construct a number operator in the usual way,

\begin{equation}N= \sum_i \alpha_i^{\dagger}\alpha_i.
\end{equation}

\noindent It will be seen that $N$ has eigenvalues given by $\{0,1,1,1,2,2,2,3\}$.  At first sight, these eigenvalues might not look familiar, that is, until they are divided by 3.  $N/3$ has eigenvalues $\{0,\hspace{.5mm}\frac{1}{3},\hspace{.5mm}\frac{1}{3},\hspace{.5mm}\frac{1}{3}, \hspace{.5mm}\frac{2}{3},\hspace{.5mm}\frac{2}{3},\hspace{.5mm}\frac{2}{3},\hspace{.5mm}1\}$, which can now be recognized as the electric charges of a neutrino (or anti-neutrino), a triplet of anti-down quarks, a triplet of up quarks, and a positron.   We will then define our electric charge, $Q$, as

\begin{equation}\label{Q}
Q\equiv\frac{N}{3}.
\end{equation}

\noindent  As $N$ must take on integer values, $Q$ must be quantized.  Hence, the model offers at this early stage a straightforward explanation for the quantization of electric charge.

As we will show, the remaining states within a generation are related to these particles by complex conjugation, and hence are acted upon by $-Q^*$ in the usual way. 

The anti-commuting ladder operators, mentioned above, can be seen to possess a certain symmetry generated uniquely by $su(3)_c$ and $u(1)_{em}$, the two unbroken symmetries of the standard model.  These generators of $SU(3)_c$ do indeed match those of~\cite{GGquarks}, and fit in perfectly with $Q$, the $U(1)_{em}$ generator mentioned above.  Under these symmetries, we find that our Fock space, and its complex conjugate, together transform as would a full generation of quarks and leptons.

%CHECK CHECK CHECK

Ours is certainly not the first instance where G\"{u}naydin and G\"{u}rsey's model has been adapted.  As an extension of their model, \cite{dixon}, \cite{matter}, Dixon describes electric charge  as a mix of quaternionic and octonionic objects.    It would be interesting to see if a ladder system could be found, which alternately gives Dixon's $Q$ as a number operator.   Readers are encouraged to see \cite{dixon}, \cite{matter}, or other examples of his extensive work.

Since the time of first writing, more octonionic chromo-electrodynamic models have been found.  Most noteworthy of all were three papers written in the late 1970s, \cite{Grass}, \cite{wow}, and \cite{wowagain}, which could also be considered as extensions of G\"{u}naydin and G\"{u}rsey's model,~\cite{GGstats}.  In these papers, the authors use two separate ladder systems:  system (a) fits with the octonionic ladder operators of \cite{GGstats}, and system (b) is introduced as quaternionic.  By combining the two systems, they describe the electric charge generator not as a number operator, but as the difference  between the number operators of the two systems.  References \cite{Grass}, \cite{wow}, and \cite{wowagain} are important papers, worth careful reconsideration by the community. 

%Our results differ from earlier versions in that we will be constructing a generation of quarks and leptons explicitly as \it minimal left ideals \rm of a Clifford algebra, generated only by the complex octonions.  In doing so, we will use just a single octonionic ladder system, with its complex conjugate.  This in turn allows us to (1) define electric charge more simply as $Q=N/3$, and (2) expose a more direct route to the two unbroken gauge symmetries of the standard model.  Furthermore, our formalism naturally relates particles and anti-particles using only the complex conjugate, $i\mapsto -i$, which is not a feature of these earlier models.  Finally, as our generation of quarks and leptons will be constructed from Clifford algebra elements, not column vectors, we will then be free to model  weak isospin, using \it right multiplication \rm of the original Clifford algebra onto these minimal left ideals.

Our results differ from earlier versions in that we will
be constructing a generation of quarks and leptons explicitly
as \it minimal left ideals \rm of a Clifford algebra, generated
only by the complex octonions. In doing so, we will use just
a single octonionic ladder system, with its complex conjugate.
This in turn allows us to define electric charge
more simply as Q = N/3, thereby exposing a more direct
route to the two unbroken gauge symmetries of the
standard model. Furthermore, our formalism naturally
relates particles and anti-particles using only the complex
conjugate, $i \mapsto -i$, which is not a feature of these earlier
models. Finally, as our generation of quarks and leptons
will be constructed from Clifford algebra elements,
not column vectors, we will then be free to model weak
isospin, using \it right multiplication \rm of this same Clifford
algebra onto these minimal left ideals.

\section{Introduction to $\CO$  }
%  intro to O

The complex octonions are not a tool commonly used in physics, so we introduce them here.

\begin{figure}[h!] \begin{center}
\includegraphics[width=7cm]{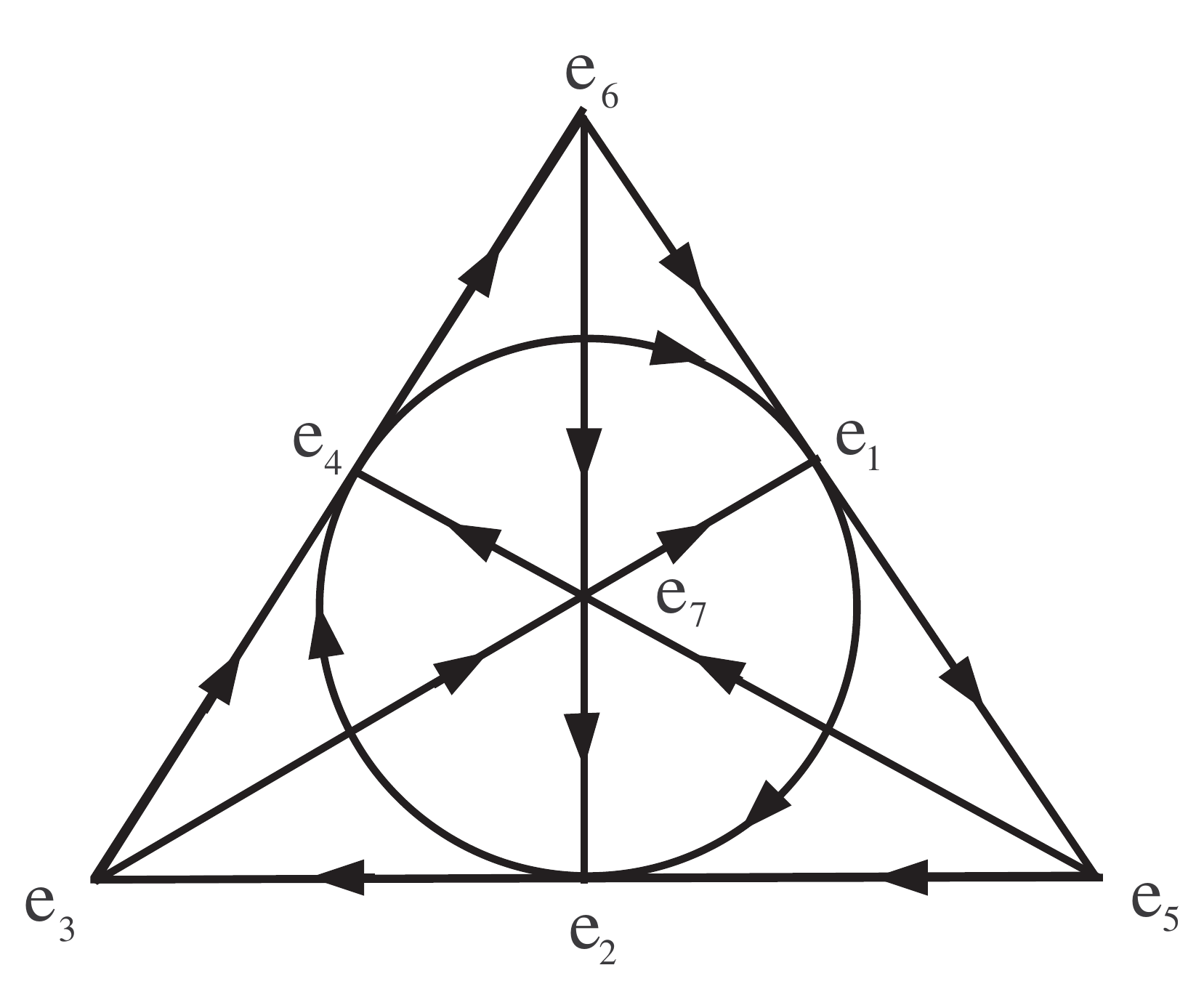}
\caption{\label{fano}
Multiplication of octonionic imaginary units}
\end{center}\end{figure}

The generic element of $\mathbb{C} \otimes \mathbb{O}$ is written $ \sum_{n=0}^7 A_n e_n $, where the $A_n$ are complex coefficients.  The $e_n$ are octonionic imaginary units $\left(e_n^2=-1\right)$, apart from $e_0=1$, which multiply according to Figure~\ref{fano}.  The complex imaginary unit $i$ commutes with the octonionic $e_n$.

Any three imaginary units on a directed line segment in Figure~\ref{fano} act as if they were a triplet of Pauli matrices, $\sigma_m$.  (More precisely, they behave as $-i\sigma_m$.)  For example, $e_6e_1 =-e_1e_6= e_5,$ $e_1e_5=-e_5e_1=e_6,$ $e_5e_6=-e_6e_5=e_1,$ $e_4e_1=-e_1e_4=e_2$, etc.  

The multiplication rules for these imaginary units can be defined by setting $e_1e_2 = e_4$, and then applying the following rules,  as shown in \cite{baez}.  

\begin{equation}\begin{array}{l} 
e_ie_j = -e_je_i \hspace{1cm} i\neq j, \vspace{2mm}\\
e_ie_j = e_k \hspace{2mm} \Rightarrow  \hspace{2mm}e_{i+1}e_{j+1} = e_{k+1},\vspace{2mm}\\
e_ie_j = e_k \hspace{2mm}\Rightarrow \hspace{2mm} e_{2i}e_{2j} = e_{2k}. 
\end{array}\end{equation}

It is indeed true that the octonions form a non-associative algebra, meaning that the relation $(ab)c =a(bc)$ does not always hold.   The reader can check this by finding three imaginary units, which are not all on the same line segment, and substituting them as in $a$, $b$, and $c$.  

The octonionic automorphism group is $G_2$, which is a 14-dimensional exceptional Lie group.  Its Lie algebra may be represented by the generators 

\begin{equation}\begin{array}{lll}\label{g2}

\Lambda_1 = \frac{i}{2}\big(e_1\left(e_5\hspace{2mm} \cdot\hspace{1mm}\right)-e_3\left(e_4 \hspace{2mm}\cdot\hspace{1mm}\right) \big),  &\hspace{1mm}&    \Lambda_8 = \frac{i}{2\sqrt{3}}\big(e_1\left(e_3\hspace{2mm} \cdot\hspace{1mm}\right)+e_4\left(e_5 \hspace{2mm}\cdot\hspace{1mm}\right)   -    2e_2\left(e_6 \hspace{2mm}\cdot\hspace{1mm}\right) \big) , \vspace{3mm} \\

\Lambda_2 = -\frac{i}{2}\big(e_1\left(e_4\hspace{2mm} \cdot\hspace{1mm}\right)+e_3\left(e_5 \hspace{2mm}\cdot\hspace{1mm}\right) \big),  &\hspace{1mm}& g_9 =   -\frac{i}{2\sqrt{3}}\big(e_1\left(e_5\hspace{2mm} \cdot\hspace{1mm}\right)+e_3\left(e_4 \hspace{2mm}\cdot\hspace{1mm}\right)   +    2e_2\left(e_7 \hspace{2mm}\cdot\hspace{1mm}\right) \big),\vspace{3mm} \\

\Lambda_3 = \frac{i}{2}\big(e_4\left(e_5\hspace{2mm} \cdot\hspace{1mm}\right)-e_1\left(e_3 \hspace{2mm}\cdot\hspace{1mm}\right) \big) , &\hspace{1mm}& g_{10} = \frac{i}{2\sqrt{3}}\big(e_1\left(e_4\hspace{2mm} \cdot\hspace{1mm}\right)-e_3\left(e_5 \hspace{2mm}\cdot\hspace{1mm}\right)   +    2e_6\left(e_7 \hspace{2mm}\cdot\hspace{1mm}\right) \big), \vspace{3mm} \\

 \Lambda_4 = \frac{i}{2}\big(e_2\left(e_5\hspace{2mm} \cdot\hspace{1mm}\right)+e_4\left(e_6 \hspace{2mm}\cdot\hspace{1mm}\right) \big),  
 &\hspace{1mm}&  g_{11} =   -\frac{i}{2\sqrt{3}}\big(e_4\left(e_6\hspace{2mm} \cdot\hspace{1mm}\right)-e_2\left(e_5 \hspace{2mm}\cdot\hspace{1mm}\right)   +    2e_1\left(e_7 \hspace{2mm}\cdot\hspace{1mm}\right) \big),\vspace{3mm} \\

\Lambda_5 = \frac{i}{2}\big(e_5\left(e_6\hspace{2mm} \cdot\hspace{1mm}\right)-e_2\left(e_4 \hspace{2mm}\cdot\hspace{1mm}\right) \big),  &\hspace{1mm}&  g_{12} = -\frac{i}{2\sqrt{3}}\big(e_2\left(e_4\hspace{2mm} \cdot\hspace{1mm}\right)+e_5\left(e_6 \hspace{2mm}\cdot\hspace{1mm}\right)   -    2e_3\left(e_7 \hspace{2mm}\cdot\hspace{1mm}\right) \big), \vspace{3mm} \\

\Lambda_6 = \frac{i}{2}\big(e_1\left(e_6\hspace{2mm} \cdot\hspace{1mm}\right)+e_2\left(e_3 \hspace{2mm}\cdot\hspace{1mm}\right) \big),  &\hspace{1mm}& g_{13} =   -\frac{i}{2\sqrt{3}}\big(-e_1\left(e_6\hspace{2mm} \cdot\hspace{1mm}\right)+e_2\left(e_3 \hspace{2mm}\cdot\hspace{1mm}\right)   +   2e_4\left(e_7 \hspace{2mm}\cdot\hspace{1mm}\right) \big),\vspace{3mm} \\

\Lambda_7 = \frac{i}{2}\big(e_1\left(e_2\hspace{2mm} \cdot\hspace{1mm}\right)+e_3\left(e_6 \hspace{2mm}\cdot\hspace{1mm}\right) \big),  
&\hspace{1mm}&  g_{14} = \frac{i}{2\sqrt{3}}\big(-e_1\left(e_2\hspace{2mm} \cdot\hspace{1mm}\right)+e_3\left(e_6 \hspace{2mm}\cdot\hspace{1mm}\right)   +   2e_5\left(e_7 \hspace{2mm}\cdot\hspace{1mm}\right) \big) ,

\end{array}\end{equation}

\noindent acting on the octonions.  Here, the nested brackets indicate that the generators are constructed from chains of octonions, multiplying from right to left.  

The eight objects, $\Lambda_i$, generate $SU(3)\subset G_2$.    This $SU(3)$ may be defined as the subgroup of $G_2$ which leaves the octonionic unit $e_7$ invariant.  Of course,  alternate $SU(3)$ subgroups of $G_2$ may be found, which correspond to other imaginary units. For a more thorough introduction of $\mathbb{O}$ see~\cite{baez}, \cite{conway}, \cite{okubo}.

Finally, we define three notions of conjugation on an element $a$ in $\mathbb{C}\otimes\mathbb{O}$.  The \it complex conjugate \rm of $a$, denoted $a^*$, maps the complex $i\mapsto -i$, as would be expected.  The \it octonionic conjugate \rm of $a$, denoted $\tilde{a}$, takes each of the octonionic imaginary units $e_n\mapsto -e_n$ for $n=1,\dots 7$.  That which we will call the \it hermitian conjugate \rm of $a$, denoted $a^{\dagger}$, performs both of these maps simultaneously, $i \mapsto -i$ and $e_n \mapsto -e_n$ for $n=1,\dots 7$.  The conjugate and the hermitian conjugate each reverse the order of multiplication, as is familiar from the hermitian conjugate of a product of matrices.

%  O to Cl6

\section{Octonionic chain algebra\label{chain}  }

As a non-associative algebra, the octonions can at times seem temperamental.  Equations involving this algebra can quickly become unwieldy, due to the need to repeatedly specify the order of multiplication, by use of brackets.  The assumptions we are accustomed to making in associative algebras now do not always apply, and one might be led to wonder how (associative) groups can be described with the (non-associative) octonions.  

In this section, however,  we explain how this is not the conundrum that it might seem to be.  Every multiplication between two octonions can be considered as a linear map of one octonion on the other.  As maps are associative by definition, this gives a way of re-describing the action of octonions as an associative algebra.

\vspace{.5cm}

It is plain to see that left-multiplying one complex octonion, $m$, onto another, $f$, provides a map from $f\in\mathbb{C}\otimes\mathbb{O}$ to $f'\equiv mf \in\mathbb{C}\otimes\mathbb{O}$.  Subsequently left-multiplying by another complex octonion, $n$, provides another map: $f\mapsto  f'' \equiv n(mf)$.  We will call this map $\overleftarrow{nm}$, where the arrow is in place so as to indicate the order in which multiplication occurs.  We may extend the chain further by left-multiplying by $p\in\mathbb{C}\otimes\mathbb{O}$, giving  $\hspace{1mm}\overleftarrow{pnm}:f\mapsto p(n(mf))$, and so on.

In an associative algebra, $A$, the exercise of building up chains in order to make new maps would be futile.  That is, for $m_1, m_2, f$ in an associative algebra, $m_2(m_1f)$ can always be summarized as $(m_2m_1)f=m'f$, where $m'\equiv m_2m_1 \in A$.  However, as the complex octonions form a non-associative algebra, building chains does in fact lead to new maps.  For example, consider the map $\overleftarrow{e_3e_4}$ acting on $f=\left(e_6+ie_2 \right)$. 

\begin{equation}\overleftarrow{e_{3}e_{4}}\left(e_6+ie_2 \right) =  e_3 \left( e_4 \left( e_6+ie_2  \right) \right) =  -1 +ie_7.
\end{equation}

\noindent This is not the same as 

\begin{equation}
\left( e_3  e_4 \right) \left( e_6+ie_2  \right)= \left( e_6 \right) \left(e_6+ie_2  \right)=  -1 -ie_7 , 
\end{equation}

\noindent and in fact there exists no $a \in \mathbb{C}\otimes\mathbb{O}$ such that $\overleftarrow{e_{3}e_{4}}\left( e_6+ie_2 \right) = a \left( e_6+ie_2 \right)$.

Addition and multiplication are easy to define on this set of chains; we will refer to the resulting algebra as the \it complex octonionic chain algebra, \rm $\mathbb{C}\otimes\overleftarrow{\mathbb{O}}$, or simply the \it chain algebra \rm for short.  Addition of two maps $N =\overleftarrow{\cdots n_3 n_2 n_1}$ and $P=\overleftarrow{\cdots p_3 p_2 p_1}$ $\in \mathbb{C}\otimes\overleftarrow{\mathbb{O}}$ on $f$ is given by $\left[N+P\right]f = Nf+Pf,$ where the $n_i$ and $p_j$ $\in \mathbb{C}\otimes\mathbb{O}$.  Multiplication, $\circ$, is given simply by the composition of maps,

\begin{equation}\left[P\circ N\right]f =P(N(f))= \overleftarrow{\cdots p_3 p_2 p_1\cdots n_3 n_2 n_1}f.
\end{equation}

As the composition of maps is always associative, $\mathbb{C}\otimes\overleftarrow{\mathbb{O}}$ is an associative algebra. Unconvinced readers are encouraged to check explicitly that $ \left[\left[A \circ B\right] \circ C \right]f = \left[A \circ \left[B \circ C \right]\right]f$ $\forall$ $A,B,C\in \mathbb{C}\otimes\overleftarrow{\mathbb{O}}$ and $\forall$ $f\in\mathbb{C}\otimes\mathbb{O}$.

In analogy with the $\CO$ case, three notions of conjugation can be defined on $\COOO$.  Here, the complex conjugate is the same as before: $i\mapsto -i$.  For the octonionic and hermitian conjugates, the definition is also the same as for the $\CO$ case, bearing in mind that these conjugates now reverse the order of the chain algebra's multiplication, $\circ$.

Looking more closely at the chains, we notice quickly that

\begin{equation}\label{flip}
\overleftarrow{\cdots e_ae_b\cdots} f = -\hspace{1.5mm}\overleftarrow{\cdots e_be_a\cdots} f \hspace{0.5cm} \forall f \in \mathbb{C}\otimes\mathbb{O},
\end{equation}

\noindent for $a,b = 1,2,\dots 7$, when $a\neq b$.  Furthermore,

\begin{equation}\label{snip}
\overleftarrow{\cdots e_ie_je_je_k\cdots} f = -\hspace{1.5mm}\overleftarrow{\cdots e_ie_k\cdots} f \hspace{0.5cm} \forall f \in \mathbb{C}\otimes\mathbb{O},
\end{equation}

\noindent when $i, k = 0, 1, 2, \dots 7$ and $j=1,2, \dots 7$.  With these properties, the chains acting on $\CO$ provide a representation of the Clifford algebra $\CLseven$, where $\{i\overleftarrow{e_1}, i\overleftarrow{e_2}, \dots i\overleftarrow{e_7}\}$, acting on $f$, forms the generating set of vectors, \cite{baez}.  

It turns out, though, that $\COOO$ does not give a faithful representation of $\CLseven$.  There exists an additional symmetry, which identifies  two monomial chains with the same map.  For example, $\overleftarrow{e_1e_2e_3}f =-\overleftarrow{e_4e_5e_6e_7}f,$ $\overleftarrow{e_5e_7}f =-\overleftarrow{e_1e_2e_3e_4e_6}f,$ $\overleftarrow{e_7}f =\overleftarrow{e_1e_2e_3e_4e_5e_6}f$, etc.  These 64 equations (duality relations) are readily found by making use of equations (\ref{flip}) and (\ref{snip}), and also the following form of the identity: $\overleftarrow{e_0}f=-\overleftarrow{e_1e_2e_3e_4e_5e_6e_7}f$.  We then see that any element of $\mathbb{C}\otimes\overleftarrow{\mathbb{O}}$ may be represented as a complex linear combination of chains, of no more than three $e_j$s in length.

\begin{figure}[h!]
\begin{center}
\includegraphics[width=10cm]{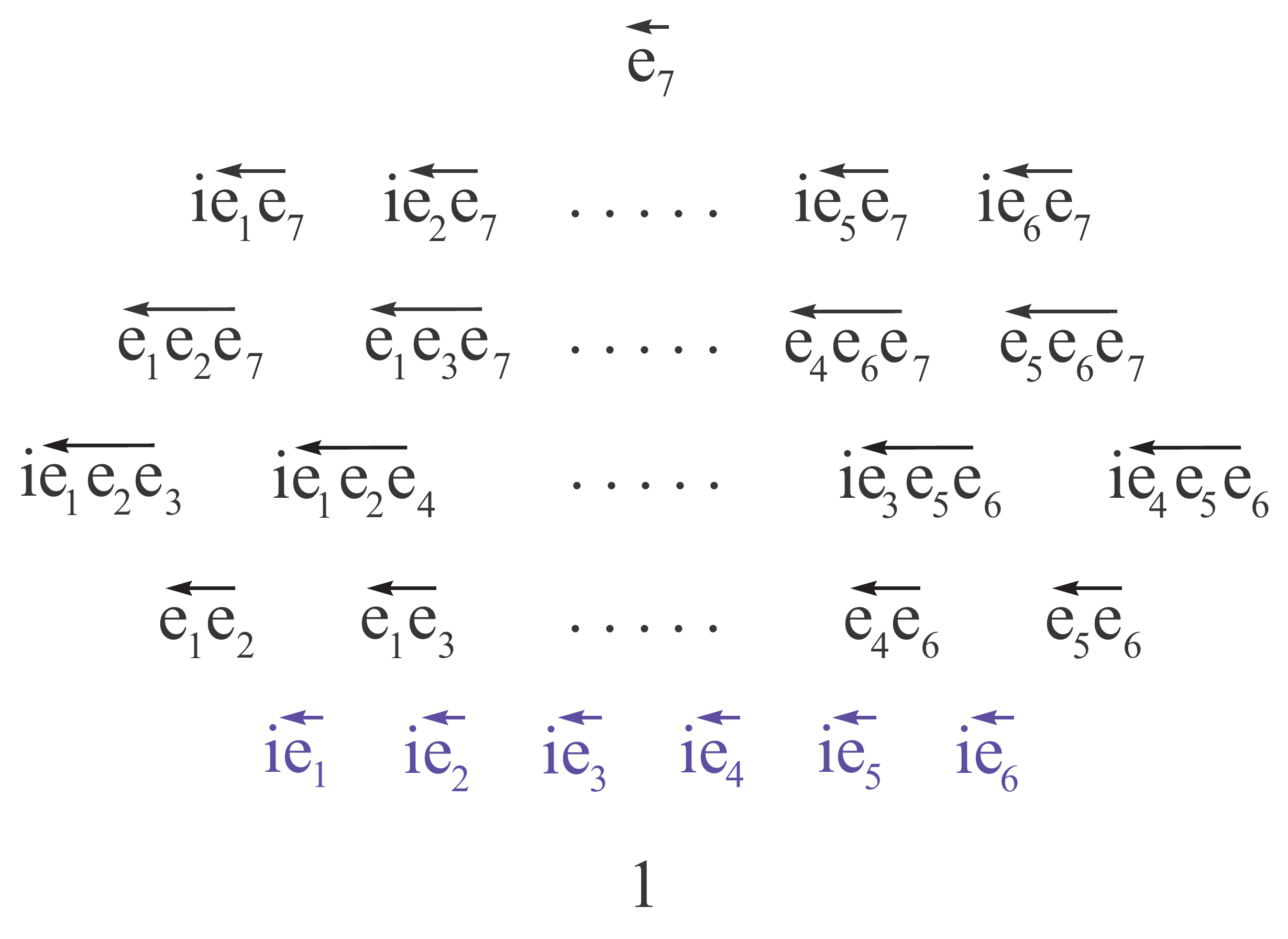}
\caption{\label{cliff6}
The 64-complex-dimensional octonionic chain algebra gives a representation of $\CLsix$.  This octonionic chain algebra is a space of maps acting through left multiplication onto any element $f\in\mathbb{C}\otimes\mathbb{O}$.}
\end{center}
\end{figure}

The reader is encouraged to check that $\mathbb{C}\otimes\overleftarrow{\mathbb{O}}$ faithfully represents the 64-complex-dimensional Clifford algebra $\CLsix$, generated by the set $\{i\overleftarrow{e_1}, i\overleftarrow{e_2}, \dots i\overleftarrow{e_6}\}$, acting on $f$.  Figure~(\ref{cliff6}) depicts the complex octonionic chain algebra, organized so as to demonstrate its $\CLsix$ structure.  Starting from the bottom, we have the zero-vector, $1$ acting on $f$, the generating vectors, $\{i\overleftarrow{e_1}, i\overleftarrow{e_2}, \dots i\overleftarrow{e_6}\}$ acting on $f$, the bivectors, $\{\overleftarrow{e_1e_2},\dots  \overleftarrow{e_5e_6}\}$ acting on $f$, and so on.  Note that we make regular use the identity $\overleftarrow{e_7}f=\overleftarrow{e_1e_2e_3e_4e_5e_6}f$ so as to avoid writing long chains of multivectors involving only the generators $i\overleftarrow{e_1}, \hspace{.7mm}i\overleftarrow{e_2}, \dots i\overleftarrow{e_6}$.  

For readers more comfortable with matrices and column vectors, one may loosely think of $\COOO$ as a space of $8\times8$ complex matrices, whereas the elements, $f$, would be represented by 8-complex-dimensional column vectors.

\newpage

Now that we have described octonionic left multiplication, by making use of an associative algebra, one might wonder if we have lost something by neglecting right multiplication.  It turns out, though, that unlike in the $\CH$ case, every complex octonion multiplied from the right may be re-expressed as a sum of chains of octonions, multiplying from the left,

\begin{equation}\label{COOOgen}
%\begin{array}{l}
fm = c_0f  + \sum_{i=1}^7 c_i \overleftarrow{e_i}f + \sum_{j=2}^7\sum_{i=1}^{j-1}c_{ij} \overleftarrow{e_{i}e_{j}}f + \sum_{k=3}^7\sum_{j=2}^{k-1}\sum_{i=1}^{j-1}c_{ijk} \overleftarrow{e_{i}e_{j}e_{k}}f,
%\end{array}
\end{equation}

\noindent for $f$, $m$ $\in\CO$, and  for some $c_0, c_i,  c_{ij}, c_{ijk} \in \mathbb{C}$.  In other words,  these right-multiplication maps are already accounted for by $\COOO$.

%\newpage

%Readers may also be concerned that in focussing on the associative algebra, $\COOO$, we are losing information from the complex octonions, $\CO$, which was tied up in their non-associativity.  This does not appear to be the case, however, since, as demonstrated above, it was the non-associativity of the octonions which caused a larger space of maps to be created.  It seems, then, that the non-associativity of the octonions re-emerges in $\COOO$ as a property necessary to produce an algebra of 64 complex dimensions.

Finally, one might ask, if we are moving to a Clifford algebraic description of octonionic multiplication via $\CLsix$ anyway, why not just start with $\CLsix$ in the first place?  The answer to this question is two-fold.  First of all, in starting only from Clifford algebras, one would be hard-pressed to know which Clifford algebras to choose.  That is, an infinite number of Clifford algebras exist, and there appears to be no reason to choose one over any other.  Secondly, with an octonionic description of $\CLsix$, we will be able to map particles into ant-particles, and vice versa, using only the complex conjugate, $i\mapsto -i$.  This is typically not the case when Clifford algebras are expressed as matrices with complex components, as was already shown in the $\CH \simeq \CLtwo$ case of Section~\ref{lr} for left- and right- handed Weyl spinors.

%  one-gen model  - MLI S^u

\section{A system of ladder operators\label{syst}  }

With the algebra $\COOO$ in hand, we can now set out to find a system of ladder operators within the octonionic chain algebra.  Consider $\alpha_1 \equiv \frac{1}{2}\left(-e_5+ie_4\right)$,  $\alpha_2 \equiv \frac{1}{2}\left(-e_3+ie_1\right),$ and $\alpha_3 \equiv \frac{1}{2}\left(-e_6+ie_2\right)$, similar to that defined in~\cite{GGquarks}.  For all $f$ in $\mathbb{C}\otimes\mathbb{O}$, and assuming right-to-left multiplication, these three lowering operators obey the anticommutation relations

\begin{equation}\label{alpha}
\{\alpha_i, \alpha_j\}f =\alpha_i(\alpha_jf)+\alpha_j (\alpha_if) = 0 
\end{equation}

\noindent for all $i,j = 1,2,3.$  The above can be seen as a generalization of the system in~\cite{GGstats}.  That is, \cite{GGstats} is recovered by restricting the general $f$ in $\mathbb{C}\otimes\mathbb{O}$ to $f=1$.  

In another slight deviation from \cite{GGstats}, we define raising operators as $\alpha_1^{\dagger} = \frac{1}{2}\left(e_5+ie_4\right)$,  $\alpha_2^{\dagger} = \frac{1}{2}\left(e_3+ie_1\right),$ and $\alpha_3^{\dagger} = \frac{1}{2}\left(e_6+ie_2\right)$, which obey

\begin{equation}\label{alphad}
\{\alpha_i^{\dagger}, \alpha_j^{\dagger}\}f = 0 \hspace{0.7cm}\textup{for all} \hspace{0.2cm} i,j = 1,2,3.
\end{equation}

\noindent We finally also have 

\begin{equation}\label{alphaalphad}
\{\alpha_i,\alpha_j^{\dagger}\}f=\delta_{ij}f \hspace{0.7cm} \textup{for all} \hspace{0.2cm}i,j = 1,2,3.
\end{equation}

With the purpose simplifying notation, we will now implicitly assume all multiplication to be carried out starting at the right, and moving to the left, as was shown in equation~(\ref{alpha}).  That is, we will now not write these brackets in explicitly, nor will we include an arrow going from right to left, specifying the direction of multiplication.   

Furthermore,  we will now be concerned only with operators, such as the $\alpha_i$, as opposed to the object $f$.  This being the case, it will now be understood that all equations will hold over all $f$ in $\CO$, even though $f$ will not be mentioned explicitly.    For example, we will now write equation~(\ref{alpha}) simply as

\begin{equation}
\{\alpha_i, \alpha_j\} =\alpha_i\alpha_j+\alpha_j \alpha_i = 0 \hspace{1cm}\textup{for all}\hspace{0.2cm} i,j = 1,2,3.
\end{equation}

As mentioned earlier, these operators acting on $f$ may be viewed as $8\times8$ complex matrices acting on $f,$ an eight-complex-dimensional column vector.  Taking into account the above paragraph, our equations from here on in can be considered as relations only between the matrices.

Another way of restating the above is to say that the operators are simply elements of $\COOO$, and we will be considering only the elements of $\COOO$ from here on in.

\section{Complex conjugation's analogue\label{conj}  }

Under complex conjugation, we find an analogous ladder system.  Consider $\alpha_1^* = \frac{1}{2}\left(-e_5-ie_4\right)$,  $\alpha_2^* = \frac{1}{2}\left(-e_3-ie_1\right),$ and $\alpha_3^* = \frac{1}{2}\left(-e_6-ie_2\right)$.  These three lowering operators obey the anticommutation relations

\begin{equation}
\{\alpha_i^*, \alpha_j^*\} = 0 \hspace{0.7cm}\textup{for all}\hspace{0.2cm} i,j = 1,2,3.
\end{equation}

We define raising operators as $\tilde{\alpha}_1 = \frac{1}{2}\left(e_5-ie_4\right)$,  $\tilde{\alpha}_2 = \frac{1}{2}\left(e_3-ie_1\right),$ and $\tilde{\alpha}_3 = \frac{1}{2}\left(e_6-ie_2\right)$, which obey

\begin{equation}
\{\tilde{\alpha}_i, \tilde{\alpha}_j\} = 0 \hspace{0.7cm}\textup{for all} \hspace{0.2cm} i,j = 1,2,3.
\end{equation}

\noindent Finally, we have also

\begin{equation}\label{alphastar}
\{\alpha_i^*,\tilde{\alpha}_j\}=\delta_{ij}\hspace{0.7cm} \textup{for all} \hspace{0.2cm}i,j = 1,2,3.
\end{equation}

Using these ladder operators, we will now build \it minimal left ideals, \rm which can be seen to mimic the set of quarks and leptons of the standard model.

\section{Minimal left ideals\label{susd}  }

%  mc =>  S^u + S^d

We are now proposing to represent quarks and leptons using minimal left ideals within our space of octonionic operators, $\COOO\simeq\CLsix$.  That is, within the space of the $\alpha_i$, $\alpha^{\dagger}_j$, and their products.  A pair of these ideals, $S^u$ and $S^d$, will be constructed using our procedure from Section~\ref{proc}:

$$\begin{array}{ll} 
 1. &  \textup{Our vector space, $V,$ is spanned by the elements $ie_1, ie_2, \dots ie_6$ over $\mathbb{C}$, in keeping } \\
& \vspace{.5cm} \textup{with Figure~\ref{cliff6}.  The quadratic form is given by $Q(v) = \{v,v\}$.}
 \\
  2. & \textup{An MTIS of $V$ is three-complex-dimensional in this case, and can be spanned}\\
 & \textup{by $\alpha_1 = \frac{1}{2}\left(-e_5+ie_4\right)$, $\alpha_2 = \frac{1}{2}\left(-e_3+ie_1\right),$ and $\alpha_3 = \frac{1}{2}\left(-e_6+ie_2\right)$.} \vspace{.5cm}\\

 3. &  \vspace{.5cm}\textup{Our nilpotent object is then given by $\omega  \hspace{1mm}\equiv \hspace{1mm} \alpha_1\alpha_2\alpha_3.$} \\ 

 4. &  \vspace{.5cm} \textup{This leads to the primitive idempotent,   $\omega\omega^{\dagger} = \alpha_1\alpha_2\alpha_3\alpha_3^{\dagger}\alpha_2^{\dagger}\alpha_1^{\dagger}$.}

\\5. & \textup{Our minimal left ideal is finally given by $S^u\equiv\COOO\hspace{1mm}\oot$, below.}
\end{array}$$
\medskip

%\noindent which lead to the identies $\omega ^{\dagger}\oot = \omega ^{\dagger}$ and $\oot\oot = \oot$.

The eight-complex-dimensional minimal left ideal for the first ladder system is given by

\begin{equation}\begin{array}{lrlcll}  \label{su}
 &S^{u}  \equiv&&&& \vspace{\spacer}\\
&& &\mathcal{V}  \hspace{1mm} \omega \omega ^{\dagger}   & & \vspace{\spacer}\\

   & +  \hspace{1mm}\bar{\mathcal{D}}^{\textup{r}} \hspace{1mm}  \alpha_{1}^{\dagger}\oot    & + &   \hspace{1mm}\bar{\mathcal{D}}^{\textup{g}}  \hspace{1mm} \alpha_{2}^{\dagger}\oot    & + & \hspace{1mm}\bar{\mathcal{D}}^{\textup{b}}  \hspace{1mm} \alpha_{3}^{\dagger}\oot   \vspace{\spacer}\\

    & +\hspace{1mm}\mathcal{U}^{\textup{r}} \hspace{1mm}  \alpha_{3}^{\dagger}\alpha_{2}^{\dagger} \oot   & + & 
\hspace{1mm}\mathcal{U}^{\textup{g}} \hspace{1mm}  \alpha_{1}^{\dagger}\alpha_{3}^{\dagger} \oot  & + &  \hspace{1mm}\mathcal{U}^{\textup{b}} \hspace{1mm}  \alpha_{2}^{\dagger}\alpha_{1}^{\dagger} \oot  \vspace{\spacer} \\

  & & +&\hspace{1mm}\mathcal{E}^{+}  \hspace{.7mm}  \alpha_{3}^{\dagger}\alpha_{2}^{\dagger}\alpha_{1}^{\dagger} \oot,  & &
\end{array}\end{equation}

\noindent where $\mathcal{V},$  $\bar{\mathcal{D}}^{\textup{r}}, \dots \mathcal{E}^{+} $ are 8 suggestively named complex coefficients.  

As 
\begin{equation} \alpha_i\hspace{1mm}\oot = 0 \hspace{1cm} \forall i,
\end{equation}

\noindent $\oot$ plays the role of the vacuum state, where the term \it vacuum \rm is used loosely.  Again, it is not to be interpreted as a zero-particle state.  However, readers may recognize the similarity between $S^u$ and a Fock space.

The conjugate system of Section~\ref{conj} analogously leads to

\begin{equation}\begin{array}{lrlcll}  \label{sd}
&S^{d} \equiv&&&& \vspace{\spacer}\\

 & & &\bar{\mathcal{V}}  \hspace{1mm} \oto   & & \vspace{\spacer}\\

   &   \hspace{1mm}\mathcal{D}^{\textup{r}} \hspace{1mm}  \alpha_{1}\oto    & + &   \hspace{1mm}\mathcal{D}^{\textup{g}}  \hspace{1mm} \alpha_{2}\oto    & + & \hspace{1mm}\mathcal{D}^{\textup{b}}  \hspace{1mm} \alpha_{3}\oto   \vspace{\spacer}\\

    & +\hspace{1mm}\bar{\mathcal{U}}^{\textup{r}} \hspace{1mm}  \alpha_{3}\alpha_{2} \oto  & + & 
\hspace{1mm}\bar{\mathcal{U}}^{\textup{g}} \hspace{1mm}  \alpha_{1}\alpha_{3} \oto  & + &  \hspace{1mm}\bar{\mathcal{U}}^{\textup{b}} \hspace{1mm}  \alpha_{2}\alpha_{1} \oto \vspace{\spacer} \\

  & & +&\hspace{1mm}\mathcal{E}^{-}  \hspace{.7mm} \alpha_3\alpha_{2}\alpha_{1} \oto,  & &
\end{array}\end{equation}

\noindent where $\bar{\mathcal{V}},$  $\mathcal{D}^{\textup{r}}, \dots \mathcal{E}^{-} $ are eight complex coefficients.   

This new ideal, (\ref{sd}), is linearly independent from the first, (\ref{su}), in the space of octonionic operators.   Clearly, the two are related via the complex conjugate, $i \mapsto -i$.  In fact, the complex conjugate is \it all \rm that is needed in order to map particles into anti-particles, and vice versa.  This was a feature in the models of~\cite{GGstats}, \cite{Gen}, and also in the context of left- and right-handed Weyl spinors in~\cite{UTI} and Section~\ref{lr} of this thesis.  It was not a feature in \cite{Grass}, \cite{wow}, \cite{wowagain}, \cite{dixon}, or \cite{matter},  where an additional quaternionic algebra was implemented in order to obtain $S^d$.

The Clifford algebra, $\CLsix$, is known to have just a single 8-complex-dimensional irreducible representation, as in $S^u$, above.  In this text, we will none-the-less be including the conjugate ideal, $S^d$, in analogy to our inclusion of both left- and right-handed Weyl spinors in Section~\ref{rar}.  Just as in the case with Weyl spinors, rotations between $S^u$ and $S^d$ are enacted via \it right \rm multiplication onto these ideals.  $S^u$ and $S^d$ can then be combined into a single irreducible representation under $\CLsix \otimes_{\C} \CLtwo$, where the factor of $\CLtwo$ accounts for the right action that mixes these two spinors with each other. Unlike in the earlier literature, this additional factor of $\CLtwo$ will originate from right multiplication of our original octonionic operators on these ideals, instead of having to introduce an entirely new Clifford algebra, $\CLtwo$.  This topic of doubling the spinor space will come up again in Chapter~\ref{GEN}.

As a final note, we point out that another interesting way to obtain anti-particles could be to use the conjugate $\dagger$, instead of $*$.  In that case, the two minimal left ideals would not be entirely linearly independent from each other.  That is, we would find a special Majorana-like property unique to the neutrino:  $\left(\oot\right)^{\dagger} = \oot$.

% this is Cl6 and Cl2

% alpha^t alpha object

\section{MTIS symmetries:  $su(3)_c$ and $u(1)_{em}$\label{MTIS}  }

Having obtained these minimal left ideals, we would now like to know how they transform, so as to justify the labels we gave to their coefficients in equations (\ref{su}) and (\ref{sd}).  Up until now, however, we have not specified under which groups these spinors transform.

A popular choice in the literature for this symmetry group is take the Clifford algebra's spin group.  For example, the well-known ``SO(10)" grand unified theory is built from the $Spin(10)$ group acting on 16-dimensional spinors.  (Here, spinors are defined as irreducible representations of $Spin(10)$.)  We point out, though, that $Spin(10)$ is a 45-dimensional group, meaning that 33 generators will need to be explained away in order to arrive finally at $SU(3)_c\times SU(2)_{\textup{L}} \times U(1)_{\textup{Y}}$.  Could there (alternately, or additionally) be another type of symmetry, which leads directly to $SU(3)_c\times SU(2)_{\textup{L}} \times U(1)_{\textup{Y}}$?  Or to the surviving  unbroken symmetries, $SU(3)_c\times U(1)_{em}$?

We would now like to point out some symmetries in the construction of our minimal left ideals.  These symmetries will be called \it unitary MTIS symmetries, \rm or simply \it MTIS symmetries. \rm  (Readers may wish to refer back to Section~\ref{defns}, where maximal totally isotropic subspaces were first defined.)  We propose here to consider MTIS symmetries  when building  covariant derivatives in gauge theories. 
%especially in the low-energy limit.

In the case of complex Clifford algebras, $\CLn$, with even $n$, the generating space can always be partitioned into two maximal totally isotropic subspaces, \cite{ablam}, each of dimension $n/2$.  For $\CLsix$ which we have here, the generating space spanned by $\{ie_1, ie_2, \dots ie_6\}$ is partitioned into an MTIS spanned by $\{\alpha_1, \alpha_2, \alpha_3 \}$, and another MTIS spanned by $\{\alpha_1^{\dagger}, \alpha_2^{\dagger}, \alpha_3 ^{\dagger}\}$.  Loosely speaking, MTIS symmetries will preserve this structure.

\bf Unitary MTIS symmetries. \rm  We are interested in operator transformations of the form $e^{i\phi_k g_k}
\hspace{1mm}ie_j \hspace{1mm}e^{-i\phi_k g_k}$, where $\phi_k\in\mathbb{R}$ and $g_k\in\COOO$.  Already, with this constraint, we find that the anti-commutation relations of equations (\ref{alpha}), (\ref{alphad}), and (\ref{alphaalphad}) are preserved.  Furthermore, as the name indicates, we will restrict our attention to those transformations under which each MTIS is closed.   That is, to first order, the $\alpha_i$ rotate only into themselves, and the $\alpha_i^{\dagger}$ rotate only into themselves, 

\begin{equation}  \large[ \hspace{2mm} g_k, \hspace{2mm}\sum_i b_i \alpha_i  \hspace{2mm} \large] = \sum_j c_j \alpha_j   \hspace{1.5cm}  \large[  \hspace{2mm}g_k, \hspace{2mm}\sum_i b_i' {\alpha_i}^{\dagger}  \hspace{2mm} \large] = \sum_j c_j' {\alpha_j}^{\dagger} ,
\end{equation}

\noindent for some complex coefficients, $c_j, c_j'$.  Here, $b_i, b_i'$ are some given complex coefficients, and  $g_k$ is a generator of the MTIS symmetry.  

Finally, we demand that the group transformation on $\alpha_j$ commute with hermitian conjugation, $\dagger$,

\begin{equation}  e^{i\phi_k g_k}\hspace{1mm} \alpha^{\dagger}_j\hspace{1mm} e^{-i\phi_k g_k} = \left(e^{-i\phi_k g_k}\right)^{\dagger}\hspace{1mm} \alpha^{\dagger}_j \hspace{1mm}\left(e^{i\phi_k g_k}\right)^{\dagger}.
\end{equation}

Under these conditions, in the case of $\CLsix$, our unitary MTIS symmetries are then found to be generated uniquely by $su(3)$ and $u(1)$.  Explicitly, the $SU(3)$ generators are given by the $\Lambda_i$ of equation~(\ref{g2}), and the $U(1)$ generator is found to be given by $Q$ of equation~(\ref{Q}).

\section{Ladders to $su(3)_c$ and $u(1)_{em}$ \label{ladders}  }

 We will now find a compact way of describing the generators of the MTIS symmetries, $su(3)$ and $u(1)$, and apply them to our minimal left ideals.

Consider $\alpha \equiv c_1\alpha_1 + c_2\alpha_2 +c_3\alpha_3$ and $\alpha' \equiv c'_1\alpha_1 + c'_2\alpha_2 +c'_3\alpha_3$, where the $c_i$ and $c'_j$ are complex coefficients.  We can then build hermitian operators, $\mathcal{H}$, of the form 

\begin{equation}
\mathcal{H}\equiv \alpha'^{\dagger}\alpha  +   \alpha^{\dagger}\alpha'.
\end{equation}

\noindent Taking the most general sum of these objects results in nine hermitian operators:

\begin{equation}\label{unique}
\sum_{\mathcal{H}} \mathcal{H}\hspace{1.2mm} = \hspace{1.2mm}r_0\hspace{0.7mm}Q+\hspace{0.7mm}\sum_{i=1}^8 r_i\Lambda_i,
\end{equation}

\noindent where $r_0$ and $r_i$ are real coefficients.  $Q$ is our electromagnetic generator from equation~(\ref{Q}), and the eight $\Lambda_i$ can be seen to generate $SU(3)$.   Indeed, these $\Lambda_i$ coincide with those described in equation~(\ref{g2}), which generate a subgroup of the octonionic automorphism group, $G_2$.  It should be noted that earlier authors, \cite{Grass}, \cite{wow}, \cite{wowagain}, were quite close to these results, but required a second ladder system in order to build $Q$ (and incidentally also all of the isospin down type states). 

%The result of equation~(\ref{unique}) is worth emphasizing.  That is, the simple form, $\sum_{\mathcal{H}} \mathcal{H}$ (the generators of U(3)) lead \it uniquely \rm to the generators of the two unbroken gauge symmetries of the standard model. 

%\begin{equation}\begin{array}{lll} \alpha &\equiv &c_1\alpha_1 + c_2\alpha_2 +c_3\alpha_3, \vspace{\spacer}
%\\  \alpha' &\equiv &c'_1\alpha_1 + c'_2\alpha_2 +c'_3\alpha_3,
%\end{array}
%\end{equation}
%\noindent where $c_i, c'_j\in \mathbb{C}$. 

%The automorphism group of the octonions is the exceptional Lie group $G_2$, and $SU(3)\subset G_2$.  In~\cite{GGquarks}, the authors introduce $SU(3)$ as a subgroup of $G_2$ which rotates the ladder operators only amongst themselves.  

In terms of ladder operators, the $SU(3)$ generators take the form

\begin{equation}\begin{array}{ll}\label{liealg}
\Lambda_1= -\alpha_2^{\dagger}\alpha_1    -\alpha_1^{\dagger}\alpha_2           &                \Lambda_2=i\alpha_2^{\dagger}\alpha_1 -i\alpha_1^{\dagger}\alpha_2       \vspace{\spacer}\\
\Lambda_3=\alpha_2^{\dagger}\alpha_2  -\alpha_1^{\dagger}\alpha_1
& \Lambda_4=-\alpha_1^{\dagger}\alpha_3-\alpha_3^{\dagger}\alpha_1        \vspace{\spacer}
\\ \Lambda_5=-i\alpha_1^{\dagger}\alpha_3 +i\alpha_3^{\dagger}\alpha_1& \Lambda_6=-\alpha_3^{\dagger}\alpha_2-\alpha_2^{\dagger}\alpha_3
\vspace{\spacer}
\\ \Lambda_7=i\alpha_3^{\dagger}\alpha_2-i\alpha_2^{\dagger}\alpha_3 & \Lambda_8=  -\frac{1}{\sqrt{3}}\left[\alpha_1^{\dagger}\alpha_1  + \alpha_2^{\dagger}\alpha_2  -2\alpha_3^{\dagger}\alpha_3\right],
\end{array}\end{equation}

\noindent all eight of which can be seen to commute with $Q$, and its conjugate.  

%It is straightforward to see that the set of $\Lambda_i$ and $Q$ together generate the group U$(3) = \textup{SU}_c$(3)$\times$U$_{em}$(1).  This group arises naturally in our construction of minimal left ideals, in that describes the internal symmetry of our MTIS.  Unitary MTIS symmetries will be important in the following sections, where they will be described in more detail.
 
We take the operators $\alpha$ and $\alpha^{\dagger}$ to transform according to

\begin{equation}\label{transf}
e^{i\sum\mathcal{H}} \hspace{1mm}\alpha \hspace{1mm}e^{-i\sum\mathcal{H}}\hspace{3mm} \textup{and} \hspace{3mm} e^{i\sum\mathcal{H}} \hspace{1mm}\alpha^{\dagger} \hspace{1mm}e^{-i\sum\mathcal{H}},
\end{equation}

\noindent respectively.  The reader may confirm that $\alpha$ transforms as a $3$ and $\alpha^{\dagger}$ transforms as a $\bar{3}$ under $SU(3)$,  consistent with the results of~\cite{GGstats}.

Now, the minimal left ideal, $S^u$, transforms as 

\begin{equation}\label{transfspin}
e^{i\sum\mathcal{H}} \hspace{1mm}S^u \hspace{1mm}e^{-i\sum\mathcal{H}} =e^{i\sum\mathcal{H}} \hspace{1mm}S^u  ,
\end{equation}

\noindent  where the equality holds because $\omega^{\dagger}\alpha_i^{\dagger} = 0$ for all $i$.   

We now identify the subspaces of $S^{u}$ by specifying their electric charges with respect to $U(1)_{em}$, and also which irreducible representation they belong to under $SU(3)_c$.  Clearly, $i$, $j$ and $k$ are meant to be distinct from each other in any given row.

\begin{equation}\begin{array}{cccc}\label{upcharge}
\hspace{.75cm}\underline{Q}\hspace{.75cm}   &    \hspace{.75cm}\boldmath{\underline{\Lambda}} \hspace{.75cm}&\hspace{.75cm} \underline{S^u} \hspace{.75cm}& \hspace{.75cm}\underline{\textup{ID}} \hspace{.75cm}\vspace{6mm}\\

0 & 1 & \oot & \nu \hspace{1mm} \left(\textup{or}\hspace{1mm} \bar{\nu}\right)   \vspace{\spacer}\\

1/3 & \bar{3} & \alpha_i^{\dagger}\oot & \bar{d}_i  \vspace{\spacer}\\

2/3 & 3 &  \alpha_i^{\dagger}\alpha_j ^{\dagger}\oot & u_k \vspace{\spacer}\\

1 & 1 & \alpha_i^{\dagger}\alpha_j ^{\dagger}\alpha_k^{\dagger}\oot & e^+    \vspace{\spacer}
\end{array}\end{equation}

\noindent So, here we identify a neutrino, $\nu$, (or antineutrino, $\bar{\nu}$), three anti-down type quarks, $\bar{d}_i$, three up-type quarks, $u_k$, and a positron, $e^+$.

%Imposing the condition that the complex conjugate commute with the transformations of~(\ref{transf}) leads to $\alpha^*$ and $\tilde{\alpha}$ transforming as
%\begin{equation}
%e^{-i\sum\mathcal{H}^*} \hspace{1mm}\alpha^* \hspace{1mm}e^{i\sum\mathcal{H}^*}\hspace{3mm} \textup{and} \hspace{3mm} e^{-i\sum\mathcal{H}^*} \hspace{1mm}\tilde{\alpha} \hspace{1mm}e^{i\sum\mathcal{H}^*},
%\end{equation}

%\noindent so that $\alpha^*$ transforms as a $\bar{3}$ and $\tilde{\alpha}$ transforms as a $3$ under $SU(3)_c$.  This is to be expected, of course, as $\alpha^* = -\alpha^{\dagger}$ and $\tilde{\alpha} = -\alpha$.

As the minimal left ideal, $S^d$, is related to $S^u$ by complex conjugation, we then see that it transforms as 

\begin{equation}\label{transfspin}
e^{-i\sum\mathcal{H}^*} \hspace{1mm}S^d \hspace{1mm}e^{i\sum\mathcal{H}^*} =e^{-i\sum\mathcal{H}^*} \hspace{1mm}S^d  ,
\end{equation}

\noindent where the equality holds because $\omega \alpha_i =0$ for all $i$.  This leads to the table below.  

\begin{equation}\begin{array}{cccc}\label{downcharge}
\hspace{.75cm}-\underline{Q}^*\hspace{.75cm}   &    \hspace{.75cm}\boldmath{-\underline{\Lambda}^*} \hspace{.75cm}&\hspace{.75cm} \underline{S^{d}} \hspace{.75cm}& \hspace{.75cm}\underline{\textup{ID}} \hspace{.75cm}\vspace{6mm}\\

0 & 1 & \oto&\bar{\nu} \hspace{1mm} \left(\textup{or}\hspace{1mm} \nu\right)   \vspace{\spacer}\\

-1/3 & 3 & \alpha_i\oto & d_i\vspace{\spacer}\\

-2/3 & \bar{3} &  \alpha_i\alpha_j \oto & \bar{u}_k \vspace{\spacer}\\

-1 & 1 & \alpha_i\alpha_j \alpha_k\oto   & e^-   \vspace{\spacer}
\end{array}\end{equation}

\noindent Here, we identify an antineutrino, $\bar{\nu}$, (or a neutrino, $\nu$), three down-type quarks, $d_i$,  three anti-up type quarks, $\bar{u}_k$, and the electron, $e^-$.  

%By this point, readers may have noticed a rather curious fact.  That is, given a space which breaks down into the correct $SU(3)_c$ representations and the correct $U(1)_{em}$ representations for a generation of particles, there is no reason to have expected the two sets of representations to have fit together.  However, it appears that spontaneously, they have.

\begin{figure}[h!]\begin{center}
\includegraphics[width=10cm]{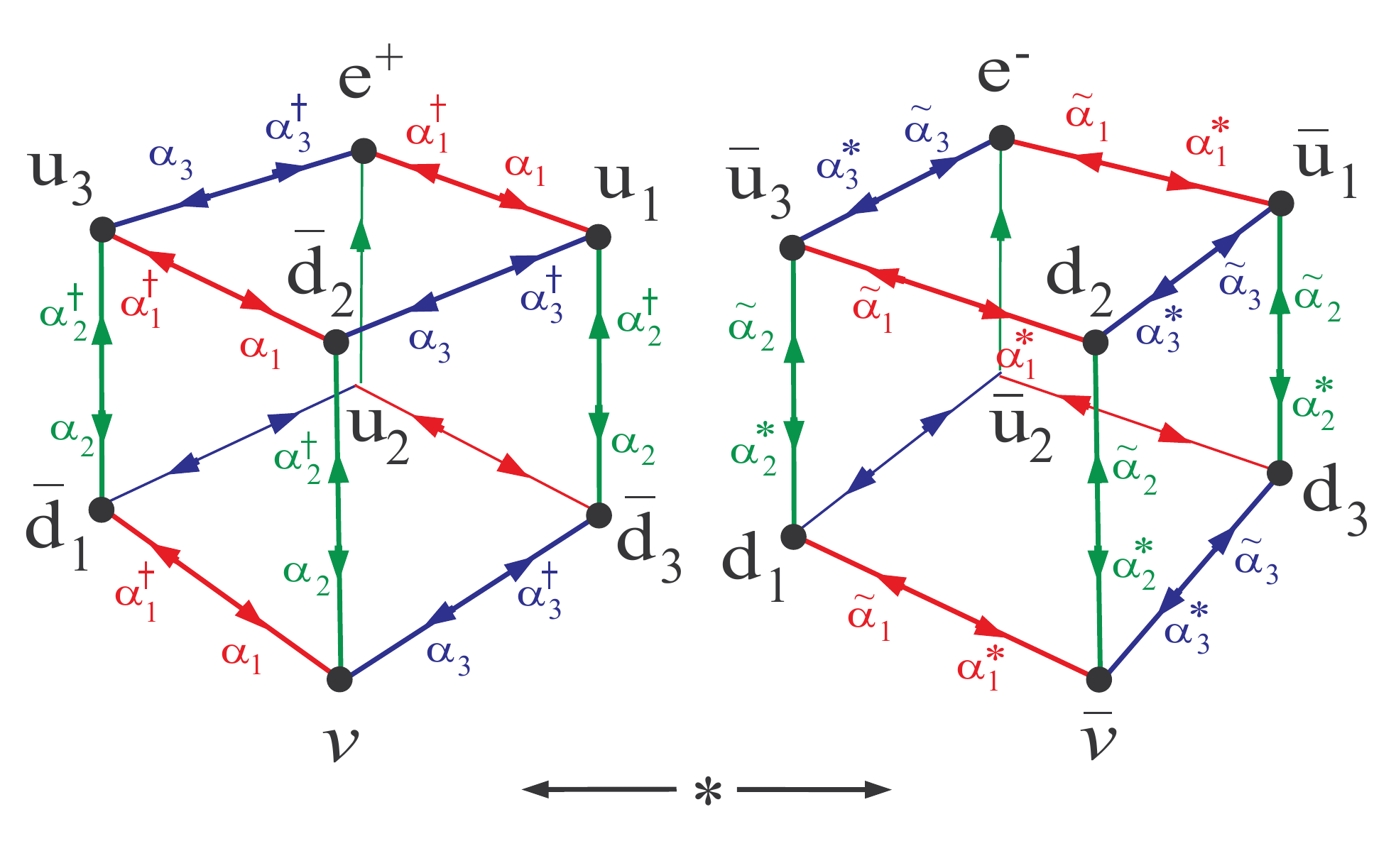}
\caption{\label{cube}  States behaving as a  full generation of standard model fermions, represented by cubes $S^u$ (left) and $S^d$ (right).  Quark and electron states may be viewed as excitations from the neutrino or anti-neutrino.  As the ``vacuum" represents the neutrino, and not the zero particle state, this model does not constitute a composite model in the usual sense. }
\end{center}\end{figure}

We have now shown a pair of conjugate ideals, which behave under $SU(3)_c$ and $U(1)_{em}$ as does a full generation of  standard model fermions.   These are summarized in Figure (\ref{cube}).

%\begin{figure}[h!]
%\includegraphics[width=9.1cm]{cubicalconjugates3.eps}
%\caption{\label{cube}  A full generation represented by cubes $S^u$ (left) and $S^d$ (right).  Quark and electron states may be viewed as excitations from the neutrino or anti-neutrino.  As the ``vacuum" represents the neutrino, and not the zero particle state, this model does not constitute a composite model in the usual sense. }
%\end{figure}

\section{Towards the weak force\label{toweak} }

Perhaps unexpectedly, it turns out that $S^u$  packages all of the  isospin up-type states together,  and $S^d$ packages all of the  down-type states together.  This is of course, if one goes ahead and makes an assumption about the placement of $\nu$ into $S^u$ and $\bar{\nu}$ into $S^d$.

We point out that $\omega$ is (automatically) negatively charged, and converts isospin up particles into isospin down, via \it right \rm multiplication on $S^u$.   It thereby exhibits features of the $W^-$ boson.  Similarly, $\omega^{\dagger}$ is positively charged, and converts isospin down particles into isospin up, via \it right \rm multiplication on $S^d$.  In doing so, it exhibits features of the $W^+$ boson.  

Other characteristics of the W bosons do not appear at the level of this chapter.   For example, there is nothing to specify that these candidate bosons act only on left-handed particles.  We also have no description here for the polarization states of these would-be bosons.

Readers may notice that right multiplication by $\omega$ and $\omega^{\dagger}$ generate a representation of $\CLtwo$, so that their inclusion would mean that we are then interested in a representation of $\CLsix\otimes_{\C}\CLtwo$.  This is clearly the analogue of the extra $\CLtwo$ which came from right multiplication  of $\CH$, which related left- and right-handed Weyl spinors with each other.   Please see Section~\ref{rar}.

%An obvious first step in this direction is to consider the Clifford algebra, $\CLeight$, which comes from the algebra $\RCHO$ acting on itself.  This is in analogy to how we found that $\CLsix$ results from $\CO$ acting on itself via left multiplication, as is explained in \cite{Gen}.  The idea, then, is to find minimal left ideals in $\CLeight$, and follow the same procedure as was introduced here.

\section{Summary  }

Using only the complex octonions acting on themselves, we were able to recover a number of aspects of the standard model's structure.

First of all, we introduced unitary MTIS symmetries, which led \it uniquely \rm to the two unbroken gauge symmetries of the standard model, $SU(3)_c$ and $U(1)_{em}$.  This new $U(1)_{em}$ generator, $Q$, happens to be proportional to a number operator, thereby suggesting an unexpected resolution to the question:  Why is electric charge quantized?

Then, using octonionic ladder operators, we have built a pair of minimal left ideals, which is found to transform under these unbroken symmetries  as does a full generation of quarks and leptons.  

%In analogy to how a pair of Weyl spinors can be constructed, we show how to build a pair of ideals behaving as a full generation of quarks and leptons.  

%Under G\"{u}naydin and G\"{u}rsey's $SU(3)_c$ generators, these ideals were found to transform as does a full generation.  We then introduced a new electric charge operator, $Q$, which distributes charges also in agreement with the standard model.  Furthermore, those charges were found to fit in perfectly with the $SU(3)_c$ representations.  There was no reason to have expected this.  Our new $Q$ is seen to be quantized simply because it is proportional to a number operator.

%Finally, we note that this model relates particles and antiparticles using only the complex conjugate $i\mapsto -i$, and in addition, seems to exhibit some rudimentary weak isospin structure. 

\medskip

%If nature truly has omitted the octonions from fundamental physics all together, it is then a striking coincidence that the extent of particle structure, mentioned above, appears to bear the mark of this algebra. 

 If the algebra of the complex octonions
is \it not \rm behind the structure of the standard model, it is
then a striking coincidence that $SU(3)_c$ and $U(1)_{em}$ both
follow readily from its ladder operators.

\chapter{Why does $SU(2)_L$ act on only left-handed fermions?\label{su2L}}

\section{A spotlight on right multiplication  }

In Section~\ref{da}, we found a four-complex-dimensional Dirac spinor, $\Psi_D\equiv\Psi_L+\Psi_R$, as an irreducible representation of  $\CLtwo \otimes_{\C} \CLtwo$.  Here, the second factor of $\CLtwo$ came from right-multiplying $\CH$ on itself, and it effected transitions between L and R states.

Likewise, in Section~\ref{susd}, we found a 16-complex-dimensional spinor, $S\equiv S^u+S^d$, as an irreducible representation of $\CLsix \otimes_{\C} \CLtwo$.  Again, we see that the factor of $\CLtwo$  came from right multiplication, but this time it effected transitions between isospin up-type and isospin down-type states.

We can now combine these spinors by taking the tensor product of the spaces $\Psi_D$ and $S$, as  in $\Upsilon\equiv \Psi_D \otimes_{\C} S$, resulting in a 64-complex-dimensional spinor space.  The Clifford algebra associated with spinors in $\Upsilon$  can easily be seen to be $\CLsix\otimes_{\C}\CLtwo\otimes_{\C}\CLtwo\otimes_{\C}\CLtwo\simeq \C l(12)$.

In this chapter, we will be interested in only the $\CLtwo\otimes_{\C}\CLtwo \simeq \CLfour$ sector of this Clifford algebra, which comes exclusively from right multiplication.  This sector encodes transitions between L and R states, and also between isospin up-type and down-type states.  

In doing so, we will be able to catch a glimpse of how $SU(2)_L$ symmetries are expected to come about in this model.  Focussing on the leptonic sector, we will again make use of MTIS symmetries, which were already introduced in Section~\ref{MTIS}.  

It turns out that the MTIS symmetries for $\CLfour$ are generated by $su(2)$ and $u(1)$.  This $su(2)$, applied to the minimal right ideal which we will now construct, automatically acts on only one of the fermions' chiralities.  In other words, MTIS symmetries seem to be able to offer an explanation for the curious favouritism exhibited by particle physics, for fermions of a particular handedness.  Such an explanation is absent in the standard model.

\section{Leptonic subspace of $\CLfour$  }

We will now build a minimal \it right \rm ideal, using the procedure laid out in Section~\ref{proc}.  Readers should note that when multiplying elements constructed from $\RCHO$, the quaternionic and octonionic imaginary units always commute with each other.

\vspace{1cm}

1.  Let us consider a Clifford algebra, $\CLfour$, generated by the vectors, $\{\tau_1 i\epsilon_x, \hspace{1mm} \tau_2 i\epsilon_x, \hspace{1mm}\tau_3 i\epsilon_x, \hspace{1mm} i\epsilon_y\}$,  where 

\begin{equation}
\tau_1 \equiv \omega + \omega^{\dagger}, \hspace{5mm} \tau_2 \equiv i\omega -i \omega^{\dagger}, \hspace{5mm} \tau_3 \equiv \oot - \oto.
\end{equation}

 \noindent Notice, that these four generators involve the right-multiplied octonionic object, $\omega$, of Section~\ref{toweak}, and also the right-multiplied quaternionic objects of Section~\ref{da}.

\vspace{7mm}

2.  Within this generating space, we may now identify an MTIS spanned by the objects $\beta_1$ and $\beta_2$ over $\C$, and another MTIS spanned by the objects $\beta_1^{\ddag}$ and $\beta_2^{\ddag}$ over $\C$.  Here, the conjugate, $\ddag$, maps the complex $i\mapsto-i$, the quaternionic $\epsilon_i\mapsto -\epsilon_i$ and the octonionic $e_j \mapsto -e_j$.  It also reverses the order of multiplication.  The lowering operators, $\beta_j$, will be defined here as 

\begin{equation}
\beta_1\equiv \frac{1}{2}\left(-\epsilon_y +i\epsilon_x\tau_3\right) \hspace{5mm} \textup{and} \hspace{5mm}\beta_2\equiv \omega^{\dagger}i\epsilon_x.  
\end{equation}

\noindent Readers may confirm that $\{\beta_i, \beta_j\} = \{\beta_i^{\ddag}, \beta_j^{\ddag}\} =0$, and that $\{\beta_i, \beta_j^{\ddag}\} \equiv \delta_{ij}$, on the leptonic subspace.

%It should be noted here, that the $\beta_j$ used in this chapter have been chosen so as to demonstrate $SU(2)_L$ structure for leptons only.  Other choices for the $\beta_j$ will need to be employed when all fermionic degrees of freedom are considered.

\vspace{7mm}

3.  The nilpotent object, $\Omega$, may now be constructed as $\Omega = \beta_2 \beta_1$.

\vspace{7mm}

4.  The projector, $\Omega^{\ddag}\Omega$, is then found to be $\Omega^{\ddag}\Omega = \oot \epsilon_{\downarrow\downarrow}$.  This idempotent will act as our vacuum state, and can be identified with the spin-down right-handed neutrino, in keeping with the vacua of previous chapters.

\vspace{7mm}

5.  Finally, right-multiplying $\CLfour$ onto our projector, we obtain the minimal right ideal, 

\begin{equation}\label{spin4_v2} \begin{array}{lll} 

\Omega^{\ddag}\Omega \hspace{1mm}\CLfour &=& \mathcal{V}_{R}^{\downarrow}\hspace{1mm} \oot\epsilon_{\downarrow\downarrow} + \mathcal{V}_{L}^{\downarrow} \hspace{1mm}\oot\epsilon_{\downarrow\downarrow} \beta_1^{\ddag}  +\hspace{1mm} \mathcal{E}_{L}^{-\downarrow}\hspace{1mm} \oot\epsilon_{\downarrow\downarrow} \beta_2^{\ddag} + \mathcal{E}_{R}^{-\downarrow}\hspace{1mm}\oot\epsilon_{\downarrow\downarrow} \beta_1^{\ddag}\beta_2^{\ddag}   \vspace{4mm} \\

&=&\mathcal{V}_{R}^{\downarrow}\hspace{1mm} \oot\epsilon_{\downarrow\downarrow}+ \mathcal{V}_{L}^{\downarrow} \hspace{1mm}\oot\epsilon_{\downarrow\uparrow} + \mathcal{E}_{L}^{-\downarrow}\hspace{1mm}\omega \epsilon_{\downarrow\uparrow} + \mathcal{E}_{R}^{-\downarrow}\hspace{1mm}\omega\epsilon_{\downarrow\downarrow},
\end{array}\end{equation}  

\noindent where the basis elements have been labelled consistently with previous chapters.  

The above minimal right ideal clearly corresponds to spin-down leptons;  the spin-up states can easily be found by acting with a $\CH$ raising operator, as was shown in Section~\ref{fock}.

\section{MTIS symmetries:  $su(2)_L$ and $u(1)_Y$   }

Now that we have constructed a minimal right ideal, we would like to see how it transforms under the MTIS symmetries for this Clifford algebra, as defined in Section~\ref{MTIS}.

Again, MTIS symmetries are those which act on the generating space as $e^{i\phi_kg_k}\hspace{1mm}|\hspace{1mm}e^{-i\phi_kg_k}$, which map non-trivially the lowering operators to themselves, and the raising operators to themselves, and which commute with hermitian conjugation, here, $\ddag$.  The generators $g_k$ this time will be elements of $(\CH)\otimes_{\C}(\COOO) \equiv \RCHOOO$.

Readers may confirm that there are four solutions to these constraints,  given by 

\begin{equation} T_1\equiv \tau_1 \hspace{1mm}\epsilon_{\uparrow\uparrow}, \hspace{4mm}T_2\equiv \tau_2 \hspace{1mm}\epsilon_{\uparrow\uparrow}, \hspace{4mm}T_3\equiv \tau_3 \hspace{1mm}\epsilon_{\uparrow\uparrow}, \hspace{4mm}N'\equiv\frac{1}{2}\left(1+i\epsilon_z\tau_3\right)+\oto.
\end{equation}

\begin{figure}[h!]\begin{center}
\includegraphics[width=6cm]{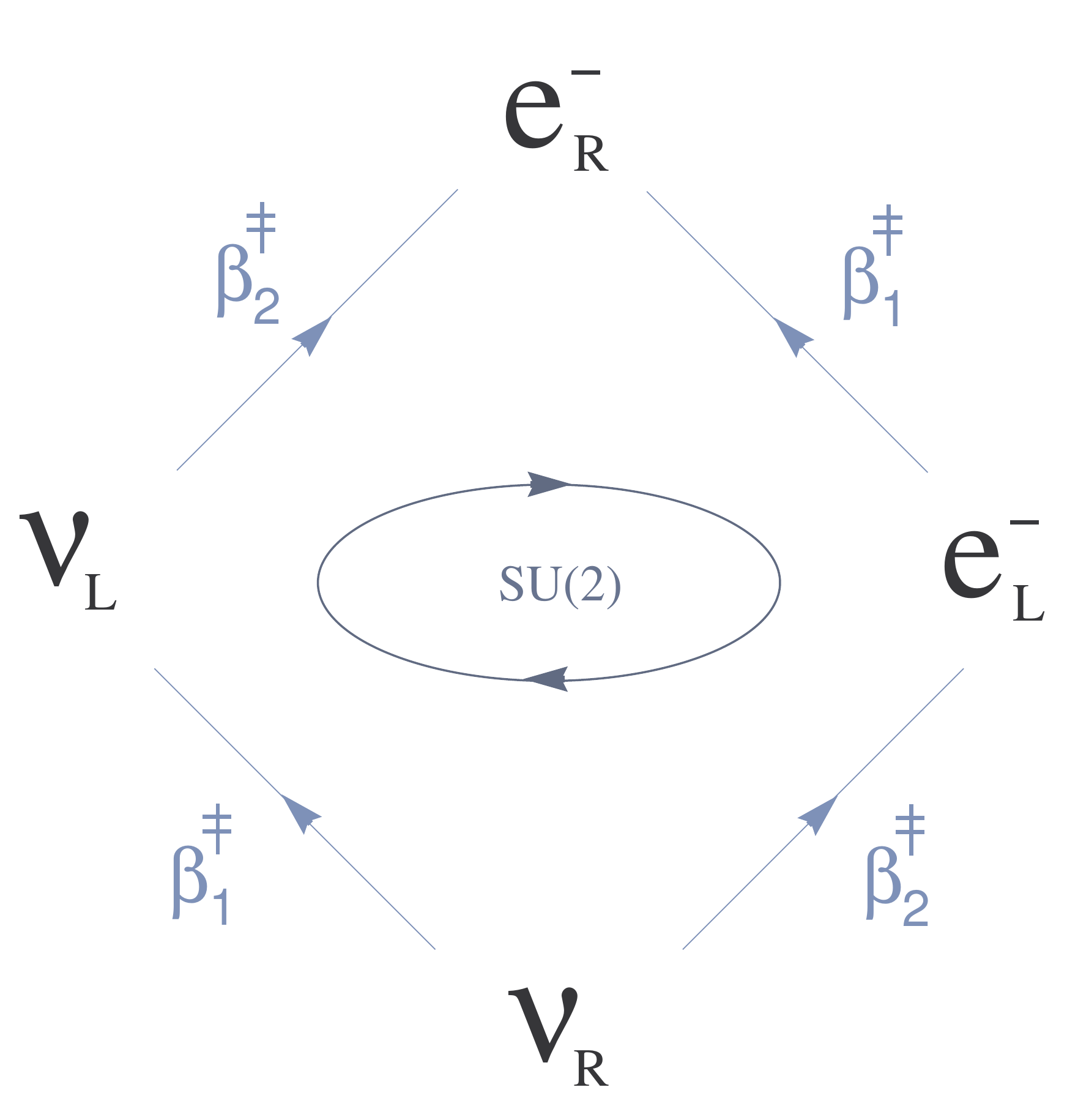}
\caption{\label{lepton}  This figure may be thought of as the analogue of Figure~\ref{cube}.  A minimal right ideal from the Clifford algebra $\CLfour$, spanned by the states $\nu_{\textup{L}}, \hspace{1mm} \nu_{\textup{R}}, \hspace{1mm}e^-_{\textup{L}},\hspace{1mm} e^-_{\textup{R}}$.  Unitary MTIS symmetries here are given by $su(2)$ and $u(1)$.  Under $su(2)$, we find that the minimal right ideal breaks down into two singlets, $\nu_{\textup{R}}$ and $e^-_{\textup{R}}$, and a doublet, $\left(\hspace{.5mm}\nu_{\textup{L}}    \hspace{1mm}e^-_{\textup{L}}\hspace{.5mm}\right)$.  This model then demonstrates a representation of $su(2)$ which acts automatically on states of only one chirality.  On this leptonic subspace, the remaining $u(1)$ turns out to be weak hypercharge.    }
\end{center}\end{figure}

\noindent The three $T_i$ can be seen to generate $SU(2)$, whereas $N'$ commutes with the $T_i$, and itself generates $U(1)$.  $N'$ may be viewed as a number operator for the $\CLfour$ system; it is nothing more than the weak isospin operator, $Y$, on this leptonic subspace.  Or, more accurately, $N'=-2Y$.  On the other hand, the $T_i$ can be seen to annihilate the right-handed neutrino and right-handed electron states, while transforming the left-handed leptons together as a doublet.  Please see Figure~\ref{lepton}.

 \newpage

\section{Summary  and outlook}

In this chapter, we have focussed our attention on $\CLtwo\otimes_{\C}\CLtwo\simeq\CLfour$ structure, coming from the right action on our previously constructed minimal left ideals.  From this Clifford algebra, $\CLfour$, we have then constructed a 4-complex-dimensional minimal right ideal, corresponding to a leptonic subspace of a generation of standard model particles.  We then found the unitary MTIS symmetries corresponding to this Clifford algebra.  It turns out that the four MTIS symmetry generators correspond to  $su(2)_L$ and $Y$ on this leptonic subspace, and furthermore provide an explanation as to why left-handed particles in this model interact via $SU(2)$, whereas right-handed particles do not.  

%As a final note, we point out that this model produced 4-dimensional spinors, and a 4-dimensional space of symmetries.  Likewise, one could pair our 8-dimensional spinors, e.g. $S^u$ of Chapter~\ref{CO}, with the 8-dimensional space of $SU(3)_c$ symmetries.  It would then be interesting to investigate whether or not a connection to triality could be made in these models.

As a final note, we point out that the dimensions of the spinors introduced in this thesis suggest  a possible connection to triality.

\chapter{Group representation structure of some grand unified theories \label{GUT}}

% guts may reduce number of coupling constants
% can explain electric charge quantization JJ
% why these reps?
% why these hyper charges?  JJ
% why q and l structure so similar?JJ
% why L and R structure different?  JJ

%% rather generically suffer from magnetic monopoles - inflation used to fix
%% rather generically suffer from proton decay

We mentioned in the previous chapter, that combing our results for $\CH$ and $\CO$ will bring us to consider the Clifford algebra $\CLtwelve$.  That is, our Dirac spinor, $\Psi_D = \Psi_L+\Psi_R$ was found to be an irreducible representation of $\CLtwo \otimes_{\C} \CLtwo$, and our generation of quarks and leptons of Chapter~\ref{CO} was found to be an irreducible representation of $\CLsix \otimes_{\C} \CLtwo$.  Taking the tensor product of the two Clifford algebras gives $\CLtwo \otimes_{\C} \CLtwo \otimes_{\C} \CLsix \otimes_{\C} \CLtwo\simeq\CLtwelve$.

%In upcoming work, we will be analyzing minimal left ideals within $\CLtwelve$, whose MTIS symmetries are naturally broken.  This breaking occurs, not fundamentally from implementing an additional Higgs field, but instead due to the fact that rotations between the quarternionic and octonionic sectors are disallowed.  One might refer to this as \it algebraic symmetry breaking. \rm  

Now, the Clifford algebra, $\CLtwelve$, can be rewritten as $\CLten\otimes_{\C} \CLtwo$, where we have separated out a factor of $\CLtwo$, corresponding to spin.  The remaining $\CLten$ algebra is tied in closely with some well-known grand unified theories, which we will now introduce.

%Grand unified theories are appealing because they can reduce the number of coupling constants in the theory, and can offer explanations to certain mysteries, such as the curious distribution of hypercharge values within a generation of particles. 

\section{$SU(5)$ unification  }

% 24 generators
% rank 4 like SM  Lang
% H. Georgi and  S. Glashow 1974 (JJ)

% proton decays too quickly (JJ)  lang
% lifetime 10^31pm2 yrs lang mann
% Superkamiodande lang
% gets running of couplings wrong?  check (JJ)

\subsection{Introduction  }

The $SU(5)$ grand unified theory offers a classic example of what it means to be a grand unified theory.  It was proposed in 1974 by H. Georgi and S. Glashow, \cite{su5}, and can be seen to successfully unify the gauge groups of the standard model.  However, the $SU(5)$ model only partially unifies a generation of quarks and leptons.  

The group $SU(5)$ has rank 4, as with the standard model's gauge group, $G_{SM}$; that is, its Lie algebra has a 4-dimensional Cartan subalgebra, \cite{Lang}.  The Lie algebra of $SU(5)$ essentially doubles the gauge symmetry of the standard model by subsuming the standard model's 12 symmetry generators, and then incorporating another 12, for a total of 24. 

Irreducible representations for $SU(5)$ are listed as 

$$\underline{1}, \hspace{2mm} \underline{5}, \hspace{2mm} \underline{10}, \hspace{2mm} \underline{15}, \hspace{2mm} \underline{24}, \hspace{2mm}  \underline{35}, \hspace{2mm} \underline{40}, \hspace{2mm} \underline{45},\hspace{2mm} \underline{50},  \dots
$$

\noindent Most descriptions of the $SU(5)$ model give a generation of quarks and leptons, which stretches across the $\underline{5}^*$ and the $\underline{10}$, and optionally, the singlet, $\underline{1}$.  This singlet can be included so as to represent the right-handed neutrino.  In contrast, gauge bosons lie in the $\underline{24}$, while the Higgs fields typically are represented by the $\underline{5}$ and the $\underline{24}$.

Despite its anticipated candidacy as a suitable fit to nature, the $SU(5)$ grand unified theory has run into trouble with  experiment.  For example, the theory predicts proton decay, with a lifetime of  $10^{31\pm2}$ years, depending on the details and source of the calculation, \cite{Mann}, \cite{Lang}.    This is generally seen to be at odds with the findings of the Super-kamiokande experiment, \cite{skk1}, \cite{skk2}, which gives a proton lifetime of $\geq5.9\times10^{33}$ years at the $90\%$ confidence level.

\subsection{$G_{SM}$ inside  $SU(5)$  }

% actually, literature says 5* and 10, i think
% SU(5)
%  split to SU(3) and SU(2)  JJ
% embed GSM into SU(5)  JJ  lang

Here, we detail the embedding of the standard model's gauge group into $SU(5)$, as given in \cite{agut}.

Elements of $SU(5)$ can be written down as $5\times5$ complex matrices, wherein we may embed the standard model's gauge group, \cite{agut}, \cite{Lang}.  Starting with $SU(2)$ and $SU(3)$, we have

\begin{equation}
\left( \begin{matrix}h_2&0\\0&h_3 \end{matrix}  \right) \hspace{4mm} \in \hspace{4mm} SU(5),
\end{equation}

\noindent where $h_2$ is a $2\times2$ matrix representing elements of $SU(2)$, and $h_3$ is a $3\times3$ matrix, representing elements of $SU(3)$.   Written in this way, the $SU(2)$ and $SU(3)$ subgroups clearly commute with each other.  We might then ask, what groups remain in $SU(5)$, which commute with both of these $SU(2)$ and $SU(3)$ subgroups?  

It is straightforward to see, \cite{agut}, \cite{Robinson}, that any $5\times 5$ matrix commuting with 

$$ \left( \begin{matrix}h_2&0\\0&h_3 \end{matrix}  \right)
$$

\noindent must be of the form 

\begin{equation}\begin{array}{ccc}\label{u1}
H_1 & = &   \left( \begin{matrix}c_2\hspace{1mm} \mathbb{I}_{2\times2}&0\\0&c_3\hspace{1mm} \mathbb{I}_{3\times 3} \end{matrix}  \right),
\end{array}\end{equation}

\noindent for $c_2, c_3\in\C$.  Furthermore, for $H_1$ to be an element of $SU(5)$, it must have a determinant of one, so that

\begin{equation}\begin{array}{ccc}\label{u1}
H_1 & = &   \left( \begin{matrix}(h_1)^{3}\hspace{1mm} \mathbb{I}_{2\times2}&0\\0&(h_1)^{-2}\hspace{1mm} \mathbb{I}_{3\times 3} \end{matrix}  \right),
\end{array}\end{equation}

\noindent where $h_1\in\mathbb{C}$.

Putting these all together, we see that it is possible to map any element of $SU(3)\times SU(2) \times U(1)$ into $SU(5)$,

\begin{equation}
   \left( \begin{matrix}(h_1)^3\hspace{1mm} h_2&0\\0&(h_1)^{-2}\hspace{1mm}h_3 \end{matrix}  \right) \hspace{4mm}\in \hspace{4mm} SU(5),
\end{equation}

\noindent where $h_1\in U(1)$, $h_2\in SU(2)$, and $h_3\in SU(3)$.

Now, it turns out that after considering the action of the $SU(2)$ and $SU(3)$ subgroups on the $\underline{5}^*$ and the $\underline{10}$, we see that this $U(1)$ of equation~(\ref{u1}) is none other than $U(1)_Y$, weak hypercharge.  That is, $SU(5)$ theory is able to elegantly explain the strikingly inelegant arrangement of hypercharges asserted by the standard model.

\medskip

A question now arises, as to whether or not every element of $SU(3)\times SU(2) \times U(1)$ is mapped to a \it unique \rm element of $SU(5)$'s subgroup.  That is, we might ask, is this mapping invertible?  It turns out that the answer is no, \cite{agut}, as it is possible to find distinct elements of $SU(3)\times SU(2) \times U(1)$, given by

\begin{equation}\left(h_3 = z_n^{2}\hspace{1mm}\mathbb{I}_{3\times3}, \hspace{2mm} h_2 = z_n^{-3}\hspace{1mm}\mathbb{I}_{2\times2}, \hspace{2mm}h_1 = z_n   \right)  \hspace{4mm}\in\hspace{4mm}SU(3)\times SU(2) \times U(1),
\end{equation}

%\begin{equation}\begin{array}{ccccc}

%U(1)\times SU(2)\times SU(3) & \rightarrow &  SU(5) & & \vspace{2mm}\\

%\left(z_n, \hspace{1mm} z_n^{-3}\hspace{1mm}\mathbb{I}_{2\times2}, \hspace{1mm}  z_n^{2}\hspace{1mm}\mathbb{I}_{3\times3}\right) & 
%\mapsto &   
%\left( \begin{matrix}     z_n^3\hspace{1mm}  z_n^{-3}\hspace{1mm}\mathbb{I}_{2\times2}     &0\\0&z_n^{-2}\hspace{1mm} z_n^{2}\hspace{1mm}\mathbb{I}_{3\times3}\end{matrix}  \right)  & 
%=&  \mathbb{I}_{5\times5},
%\end{array}\end{equation}

\noindent which each map to the same identity element of our $SU(5)$ subgroup,

\begin{equation}
\left( \begin{matrix}     
z_n^3\hspace{1mm}  z_n^{-3}\hspace{1mm}\mathbb{I}_{2\times 2}     &0\\0&z_n^{-2}\hspace{1mm} z_n^{2}\hspace{1mm}\mathbb{I}_{3\times 3}
\end{matrix}  
\right)  \hspace{2mm} 
= \hspace{2mm}  {\mathbb{I}}_{5\times 5} \hspace{4mm} \in \hspace{4mm} SU(5).
\end{equation}

The above elements, $z_n$, are restricted to be sixth roots  of unity, $z_n=e^{2\pi i n/6}$,   since $z_n^{-3}\hspace{1mm}\mathbb{I}_{2\times2}\in SU(2)$ and $z_n^{2}\hspace{1mm}\mathbb{I}_{3\times3}\in SU(3)$ must each have a determinant of one.  This generates the finite group $\mathbb{Z}_6$.

So it turns out that our subgroup of $SU(5)$ is not $SU(3)\times SU(2) \times U(1)$, but rather,  $SU(3)\times SU(2) \times U(1)/\mathbb{Z}_6$.  This could be problematic for $SU(5)$ theory, if the standard model representations were indeed able to distinguish between these six values of $z_n$.  

However, by checking explicitly one can show that the standard model representations are insensitive to these distinct values of $z_n$.  Furthermore, $\mathbb{Z}_6$ is said, \cite{agut}, to constitute the entire kernel of this representation.   This leaves us with the conclusion that the standard model's gauge group is not $SU(3)\times SU(2) \times U(1)$, but rather,  $SU(3)\times SU(2) \times U(1)/\mathbb{Z}_6$.

% 24 goes to (810) + (130) +(110) + (3,2^*,-5/6)+(3^*, 2, 5/6)  lang
% 5^* and 10 go to...  lang  robinson
% extra bosons mediate proton decay Lang  Mann - leptoquark bosons

%Symmetry breaking via two Higgs fields 5 and 24 Lang
% GUT hierarchy problem  p531 Lang
% Doublet triplet splitting problem Lang

%\newpage

\subsection{Some notes on symmetry breaking  }

The adjoint Higgs field  is responsible for breaking $SU(5)$ symmetry down to $SU(3)\times SU(2) \times U(1)/\mathbb{Z}_6$, \cite{Robinson}.  In the process, the fermionic $\underline{5}^*$ and $\underline{10}$ break into $\left(\hspace{1mm}SU(3)_c, \hspace{1mm}SU(2)_L, \hspace{1mm}U(1)_Y\hspace{1mm}\right)$ irreps as

\begin{equation}\begin{array}{rcccccc}
\underline{5}^* & \mapsto & \left(\hspace{1mm} \underline{3}^*, \hspace{1mm}\underline{1}, \hspace{1mm}\frac{1}{3} \hspace{1mm}\right) &+&\left(\hspace{1mm} \underline{1}, \hspace{1mm}\underline{2}^*, \hspace{1mm}-\frac{1}{2} \hspace{1mm}\right), & & \vspace{4mm}\\

& & \bar{d}_L &&\ell_L&  &   \vspace{8mm}\\

\underline{10} & \mapsto &  \left( \hspace{1mm}\underline{3}^*, \hspace{1mm}\underline{1}, \hspace{1mm}-\frac{2}{3} \hspace{1mm}\right)&+&\left( \hspace{1mm}\underline{3}, \hspace{1mm}\underline{2}, \hspace{1mm}\frac{1}{6}\hspace{1mm} \right)&+&\left( \hspace{1mm}\underline{1}, \hspace{1mm}\underline{1}, \hspace{1mm}1\hspace{1mm} \right). \vspace{4mm}\\

& & \bar{u}_L &&q_L&  &   e^+_L

\end{array}
\end{equation}

\noindent On the other hand, gauge generators break down as
\begin{equation}\begin{array}{rcl}
\underline{24} & \mapsto &  \left(\hspace{1mm}\underline{8},\hspace{1mm}\underline{1},\hspace{1mm}0\hspace{1mm}\right) + \left(\hspace{1mm}\underline{1},\hspace{1mm}\underline{3},\hspace{1mm}0\hspace{1mm}\right) +\left(\hspace{1mm}\underline{1},\hspace{1mm}\underline{1},\hspace{1mm}0\hspace{1mm}\right) + \left(\hspace{1mm}\underline{3},\hspace{1mm}\underline{2}^*,\hspace{1mm}-\frac{5}{6}\hspace{1mm}\right)+\left(\hspace{1mm}\underline{3}^*, \hspace{1mm}\underline{2},\hspace{1mm} \frac{5}{6}\hspace{1mm}\right),
\end{array}
\end{equation}
\noindent where the $ \left(\hspace{1mm}\underline{8},\hspace{1mm}\underline{1},\hspace{1mm}0\hspace{1mm}\right) $,  $\left(\hspace{1mm}\underline{1},\hspace{1mm}\underline{3},\hspace{1mm}0\hspace{1mm}\right)$, and  $\left(\hspace{1mm}\underline{1},\hspace{1mm}\underline{1},\hspace{1mm}0\hspace{1mm}\right)$ generate $SU(3)_c$, $SU(2)_L$, and $U(1)_Y$, respectively, and the  $\left(\hspace{1mm}\underline{3},\hspace{1mm}\underline{2}^*,\hspace{1mm}-\frac{5}{6}\hspace{1mm}\right)$ and $\left(\hspace{1mm}\underline{3}^*,\hspace{1mm} \underline{2}, \hspace{1mm}\frac{5}{6}\hspace{1mm}\right)$ generators give rise to 12 new gauge bosons which can be seen to mediate proton decay.  

The familiar Higgs doublet, $\phi$, can be seen to emerge from the $\underline{5}$ as
\begin{equation}\begin{array}{rcccc}
\underline{5} & \mapsto & \left(\hspace{1mm} \underline{3}, \hspace{1mm}\underline{1}, \hspace{1mm}-\frac{1}{3} \hspace{1mm}\right) &+&\left(\hspace{1mm} \underline{1}, \hspace{1mm}\underline{2}, \hspace{1mm}\frac{1}{2} \hspace{1mm}\right), \vspace{4mm} \\

 &&\mathcal{H}&&\phi
\end{array}
\end{equation}
\noindent where the $\left(\hspace{1mm} \underline{3}, \hspace{1mm}\underline{1}, \hspace{1mm}-\frac{1}{3} \hspace{1mm}\right)$ describes a new triplet Higgs field, $\mathcal{H}$.  

The Higgs sector, mentioned here, is commonly thought to be the source of two additional outstanding problems for $SU(5)$ theory, \cite{Lang}.  Namely, there is no clear explanation for the large differences between the GUT scale, where $SU(5)$ breaks, and the weak scale, where $G_{SM}$ breaks (GUT hierarchy problem).  Secondly, as the Higgs $\underline{5}$ breaks into a doublet under $SU(2)_L$ and a triplet under $SU(3)_c$, both pieces must be accounted for.  The Higgs boson, as we now know, has a relatively small mass of about 125 GeV.  However, in order to evade proton decay, the remaining triplet must acquire a large mass $\geq 10^{14}$ GeV, and there is no obvious reason for these two mass scales to be so far apart (doublet triplet splitting problem)~\cite{Lang}.

\section{A fermionic binary code \label{binary}  }
%5 bit code 32 dof    JJ
% iso up?  iso down?  r? g? b?  JJ

It turns out that there is a very efficient way of describing the standard model's fermions, which is helpful in understanding $SU(5)$, $Spin(10)$, Pati-Salam theories, and incidentally, also upcoming $\RCHO$ models based on $\C l(12)$.   This is in terms of a binary code, \cite{agut}, \cite{Robinson}.  For a thorough explanation of this topic, the reader is encouraged to consult \cite{agut}.  

This fermionic binary code can be thought of as a sequence of answers to five different yes or no questions:

%\vspace{ 1cm}
\newpage

\hspace{1cm} Is the fermion red?

\hspace{1cm} Is the fermion green?

\hspace{1cm} Is the fermion blue?

\hspace{1cm}  Does the fermion have isospin up?

\hspace{1cm}  Does the fermion have isospin down?

\vspace{ 0.7cm}

% get exterior alg Lambda C5  JJ

\begin{figure}[h!]\begin{center}
\includegraphics[width=15cm]{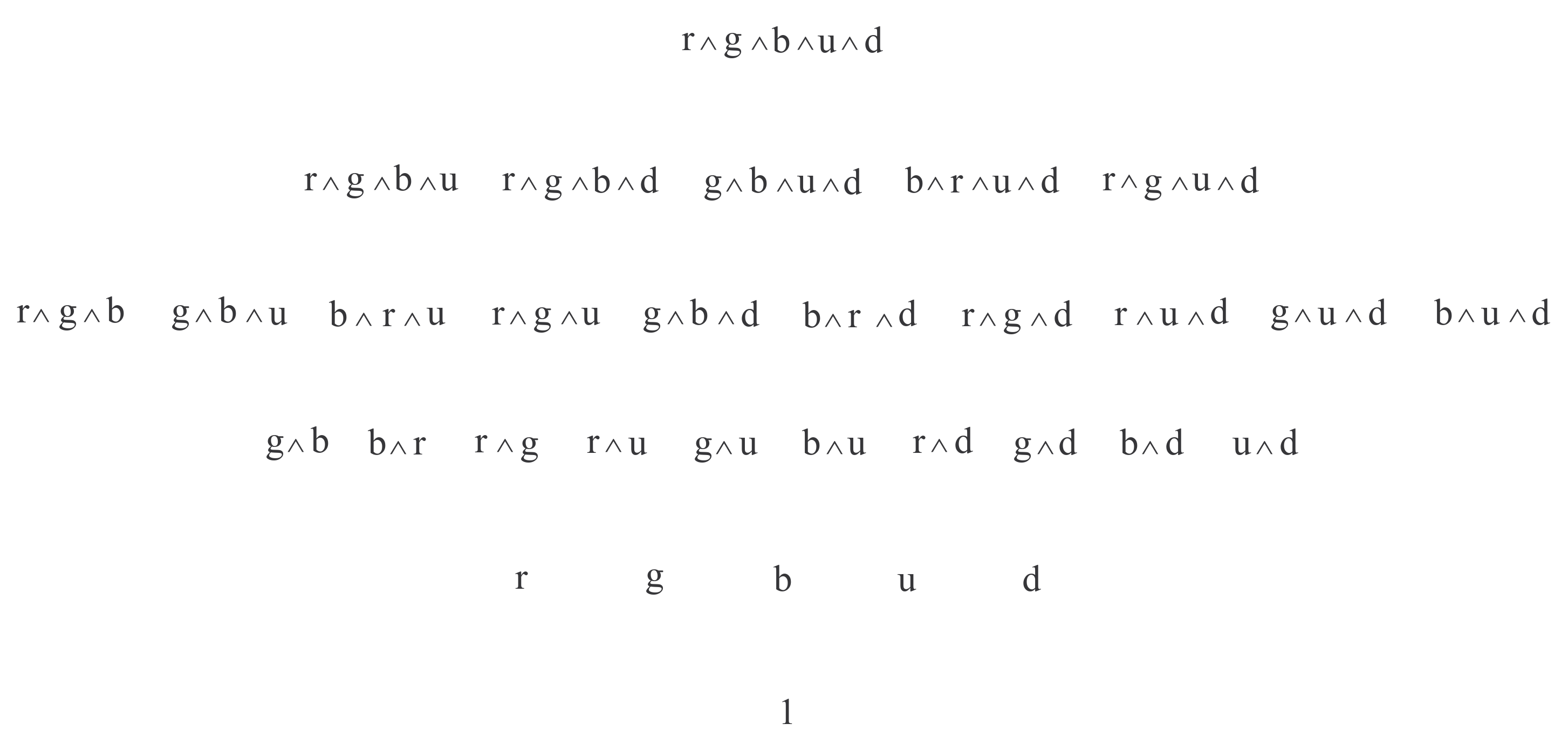}
\caption{\label{c5} The exterior algebra $\Lambda \C^5$, representing a fermionic binary code. }
\end{center}\end{figure}

The set of answers to these five questions can be represented by an exterior algebra, $\Lambda \C^5$, which is $2^5=32$-complex dimensional.  For example, a fermion which is  red, $r$, and has isospin up, $u$, would be represented by $r\wedge u$.  A fermion which is said to be both red, $r$, and blue, $b$, and has both isospin up, $u$, and isospin down, $d$, would be represented by $r\wedge b \wedge u \wedge d$, and so on.  (It may seem odd for a particle to be both red and blue, or to have both isospin up and isospin down, but this should become clear shortly.)  Please see Figure~\ref{c5}.

% now can label the binary code spinor  JJ
%diagram

\begin{figure}[h!]\begin{center}
\includegraphics[width=16.5cm]{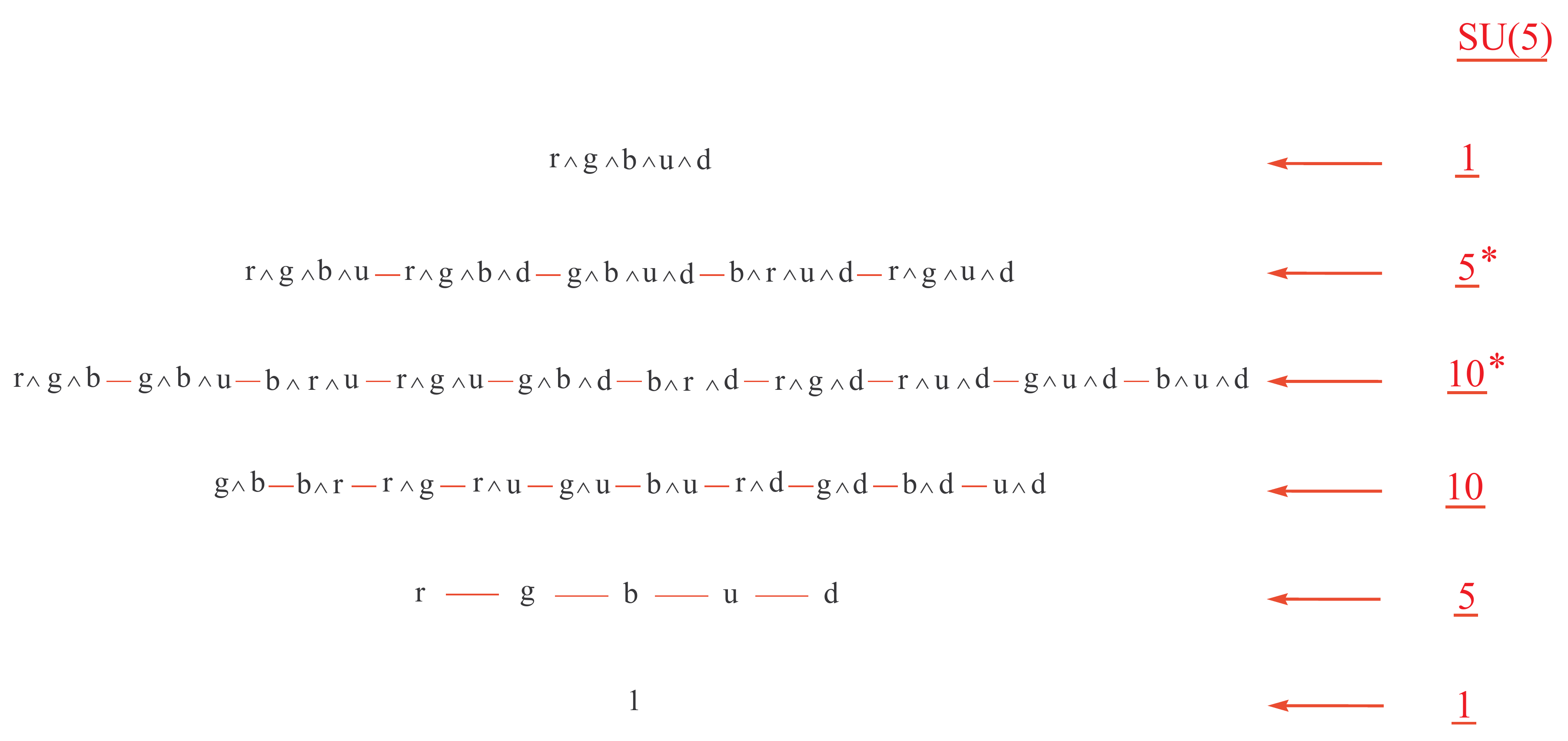}
\caption{\label{c5su5} The exterior algebra $\Lambda \C^5$, representing a fermionic binary code, broken down into irreps of $SU(5)$.  Here, basis elements which are part of the same irreducible representations are connected by red lines.}
\end{center}\end{figure}

Now, complex linear combinations of the basis vectors, $r, \hspace{1mm}g,  \hspace{1mm}b,  \hspace{1mm}u,  \hspace{1mm}d,$ may be acted upon by $SU(5)$, where they form the irreducible representation, $\underline{5}$.  By extension, the bivectors of $\Lambda \C^5$ form the $\underline{10}$, the three-vectors form the $\underline{10}^*$, the four-vectors form the $\underline{5}^*$, whereas both the unit, 1, and the five-vector form singlets.  We then see that we have exactly the right $SU(5)$ representations to describe one generation of standard model particles, together with the right-handed neutrino, including all corresponding antiparticles.  Please see Figure~\ref{c5su5}.   

\begin{figure}[h!]\begin{center}
\includegraphics[width=16cm]{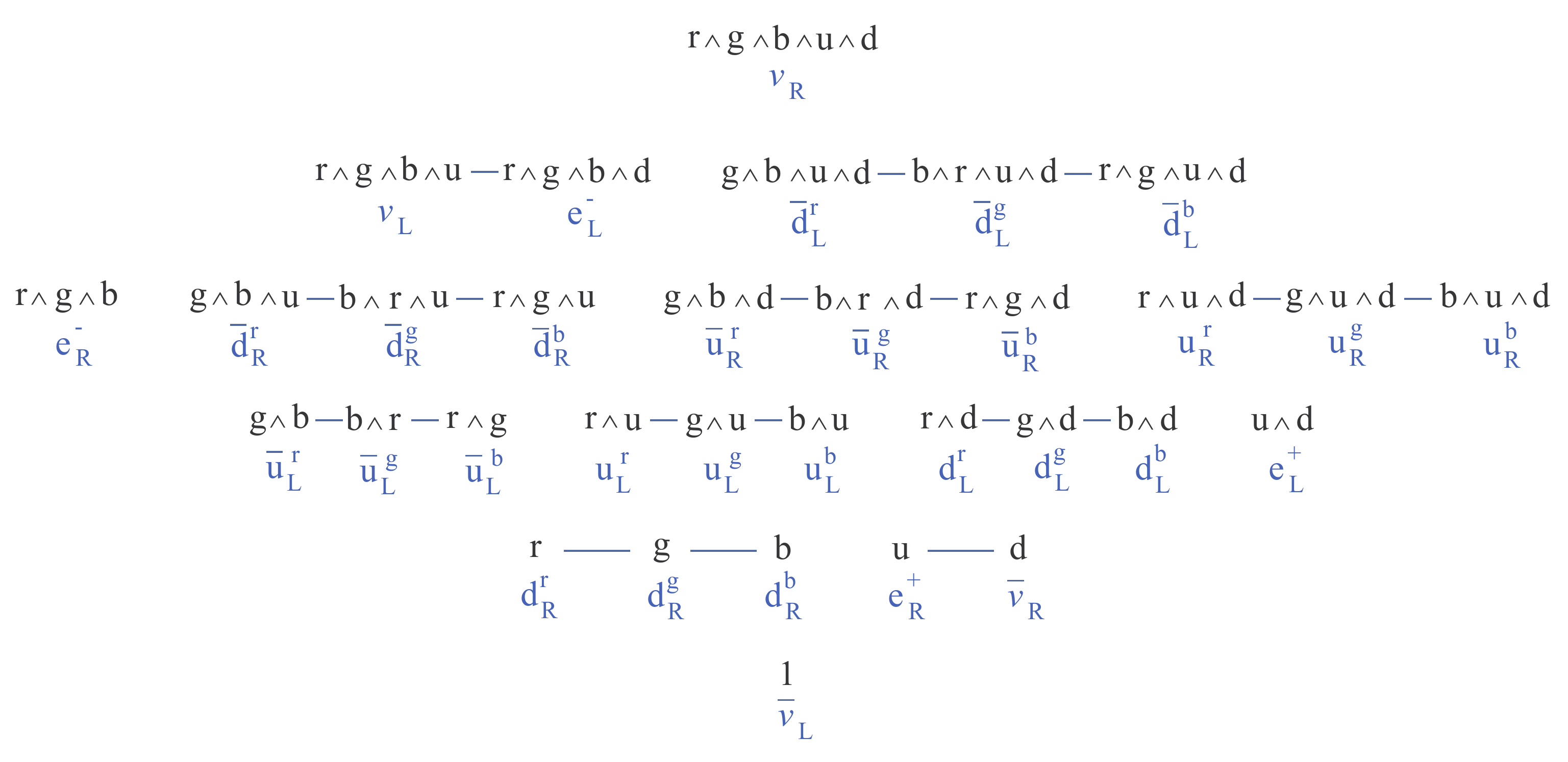}
\caption{\label{c5sm} The exterior algebra $\Lambda \C^5$, representing a fermionic binary code, with the corresponding standard model particles written underneath.  Here, basis elements which are part of the same irreps under $G_{SM}$ are connected by blue lines.}
\end{center}\end{figure}

Upon spontaneous symmetry breaking, the $\underline{5}$ breaks into  $\left(\hspace{1mm} \underline{3}, \hspace{1mm}\underline{1}, \hspace{1mm}-\frac{1}{3} \hspace{1mm}\right)$  and   $\left(\hspace{1mm} \underline{1}, \hspace{1mm}\underline{2}, \hspace{1mm}\frac{1}{2} \hspace{1mm}\right)$, and the rest of $\Lambda \C^5$ can then be seen to follow suit.  This then allows us to identify each element of $\Lambda \C^5$ with one of the standard model's fermions, as shown in Figure~\ref{c5sm}.

Readers may notice that $SU(5)$ respects the grading of $\Lambda \C^5$, and that all left-handed particles reside in the even grades, while all right-handed particles reside in the odd grades.

\section{$Spin(10)$ unification  }

%Problem with SU(5) is doesn't unify all fermions.  this can be done
Although the $SU(5)$ model was able to draw together the three gauge groups of the standard model into a single group, it only partially unified the standard model's fermions into the $\underline{5}^*$ and the $\underline{10}$.  One might wonder if an alternate Lie group might exist, which could fully unify not only the standard model's gauge bosons, but also its fermions. 

% rank 5 Lang
%45 generators
% H. Georgi 1974 (JJ) and Mink??
%particles form spinor of spin 10 JJ
%existence of RH v required (JJ)

One Lie group which does hit the mark is $Spin(10)$.  The $Spin(10)$ grand unified theory is more commonly known in the literature as the ``$SO(10)$" model, and was proposed by H. Georgi in 1974/1975, \cite{georgi10}, and independently by H. Fritzsch  and P. Minkowski in 1975, \cite{mink}.  

$Spin(10)$ is the double cover of the group $SO(10)$, and is of rank 5, \cite{Lang}.  It unifies the standard model's gauge group, and extends it from 12 dimensions to 45.  A generation of quarks and leptons in this model is unified into a 16-dimensional spinor, with a conjugate spinor accounting for the required anti-particles.  This spinor representation can be seen to naturally account for a right-handed neutrino, \cite{agut}.

% spinor can be thought of as Lambda C5   JJ
%Spin 10 breaks Lambda C5 into two irreps - Weyl spinors even and odd   16 + 16   JJ
%diagram
%SU(5) can be embedded in Spin(10)  JJ

\begin{figure}[h!]\begin{center}
\includegraphics[width=16.5cm]{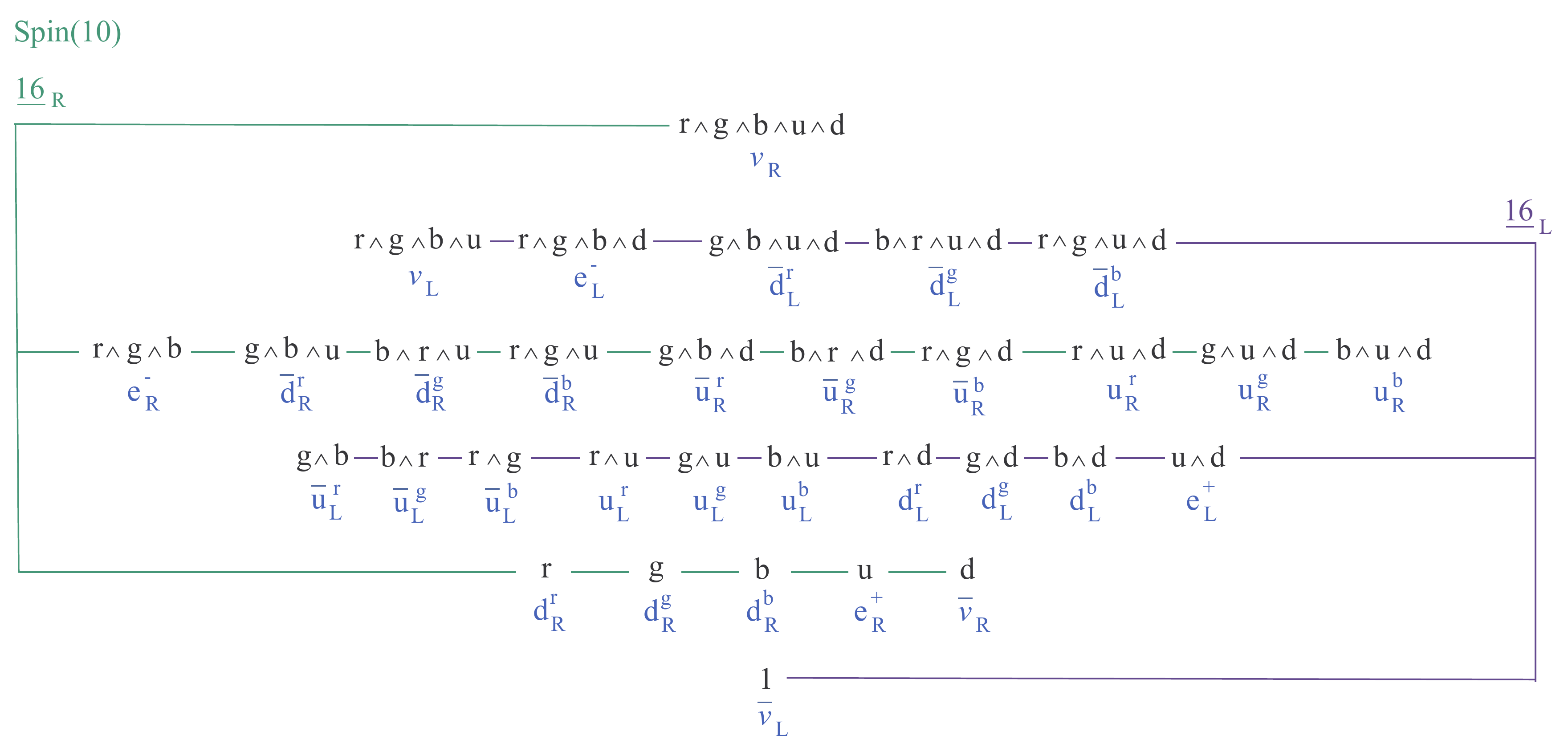}
\caption{\label{c5sp10} The exterior algebra $\Lambda \C^5$, representing a fermionic binary code, with the corresponding standard model particles written underneath.  Here, basis elements connected by lines in the even part of the algebra form one irreducible representation under $Spin(10)$, with left-handed chirality.  Similarly, those basis elements in the odd part of the algebra form another irreducible representation under $Spin(10)$, with right-handed chirality.}
\end{center}\end{figure}

The group, $Spin(10)$, is associated with the Clifford algebra $\C l (10)$, which has an irreducible representation given by a 32-dimensional Dirac spinor.  Under $Spin(10)$, this Dirac spinor breaks down into two irreducible representations, known as Weyl spinors, which are each 16-dimensional. Now, it can be shown, \cite{agut}, that Dirac spinors of the group $Spin(2n)$ can be represented by the exterior algebra, $\Lambda \C ^n$.  Hence, we see that the exterior algebra, $\Lambda \C^5$, which we just described in the previous section, is capable of acting not only as a representation of $SU(5)$, but also as a representation of $Spin(10)$.  Please see Figure~\ref{c5sp10}.

Readers may recognize that this same pattern has appeared more than once in this thesis.  That is, minimal left ideals, such as $\Psi _L$ and $S^u$, have been built using raising operators of the Clifford algebras $\CLtwo$ and $\CLsix$, respectively.  These raising operators may be viewed as generating exterior algebras.  For example, by looking at equation~(\ref{su}) of Section~\ref{syst}, it is easy to see how the Dirac spinor, $S^u$, is built from the exterior algebra, $\Lambda \C^3$, generated by $\alpha_1^{\dagger}$, $\alpha_2^{\dagger}$, and $\alpha_3^{\dagger}$. 

Now, it can be shown that our representation of $SU(5)$ resides inside that of $Spin(10)$, \cite{agut}.  As with $SU(5)$, the group $Spin(10)$ can be seen to respect the chirality of particles;  it does not mix left- and right-handed species.  However, it does not respect the grading of $\Lambda \C^5$, as do $SU(5)$ and $G_{SM}$.

%STILL VIABLE?

%PROTON DECAY?

%SYMMETRY BREAKING SECTOR?

%OUTSTANDING PROBLEMS?

\section{Pati-Salam model  }

Before the introduction of $SU(5)$- and $Spin(10)$-theories came an extension to the standard model, now known as the Pati-Salam model.  The model is named after its inventors, J. Pati and A. Salam, who published the proposal in 1974, \cite{ps}.  

The Pati-Salam model may be considered to be a more conservative approach to unification.  It does not unify the gauge groups of the standard model into a simple Lie group, as with $SU(5)$ and $Spin (10)$.  Instead, it is based on the gauge group, $SU(2)\times SU(2) \times SU(4) \simeq Spin(4)\times Spin(6)$, \cite{agut}, which extends the dimension of $G_{SM}$ from 12 to $3+3+15=21$.  

One main goal of the Pati-Salam model is to capitalize on the fact that the standard model treats quarks and leptons similarly under weak isospin.  Hence, the authors proposed that leptons might be considered as just another type of quark, which happen to be of a fourth colour, beyond the usual red, green and blue of $SU(3)_c$.  This fourth colour would then be identified with  lepton number, thereby extending $SU(3)_c$ to $SU(4)$, \cite{ps}.

Perhaps perplexed by the standard model's apparent preference for left-handed particles, Pati and Salam further extended $G_{SM}$ by incorporating an extra factor of $SU(2)$, meant for right-handed particles.  With $SU(3)_c$ extended to $SU(4)$ and $SU(2)_L$ extended to $SU(2)_L\times SU(2)_R$, we then arrive at $SU(2)\times SU(2) \times SU(4)$.  It turns out that the standard model's $U(1)_Y$ stretches across the right-handed $SU(2)$, and also the $SU(4)$ factor, \cite{agut}.  

The standard model's gauge group may be mapped into $SU(2)\times SU(2) \times SU(4)$ as shown in \cite{agut},

\begin{equation}
\left( h_2, \hspace{2mm} \left(\begin{matrix}(h_1)^3&0\\0&(h_1)^{-3} \end{matrix}\right), \hspace{2mm} \left(\begin{matrix}h_1\hspace{1mm}h_3&0\\0&(h_1)^{-3} \end{matrix}\right) \right) \hspace{4mm}\in\hspace{4mm}SU(2)\times SU(2) \times SU(4) ,
\end{equation}

%\noindent Furthermore, it may be shown, \cite{agut}, that $SU(2)\times SU(2) \times SU(4)$ lies inside $Spin(10)$, CHECK, however, it does not lie inside $SU(5)$.  Conversely, $SU(5)$ also cannot be embedded in $SU(2)\times SU(2) \times SU(4)$.

\noindent for $h_1\in U(1)$, $h_2\in SU(2)$, and $h_3 \in SU(3)$.

With the gauge groups in hand, it is now fairly straightforward to build up the fermionic vector spaces proposed in this model.  The group $SU(4)$ can be made to act on a four-complex-dimensional vector, $\C^4$, and its conjugate, ${\C^4}^*$.  For the $SU(2)_L\times SU(2)_R$ sector, we choose a left-handed doublet, $\C^2 \otimes_{\C} \C$, and a right-handed doublet, $\C\otimes_{\C} \C^2$.  Putting these all together gives the Pati-Salam representation, \cite{agut}, 

\begin{equation} \begin{array}{lll}f& \equiv& \left(\left(    \C^2 \otimes_{\C} \C      \right)  \oplus   \left(   \C\otimes_{\C} \C^2    \right)    \right)  \otimes_{\C}  \left( \C^4 \oplus {\C^4}^* \right)\vspace{2mm} \\

&\simeq& \left(  \C^2 \otimes_{\C} \C  \otimes_{\C}\C^4\right)   \oplus    \left(\C \otimes_{\C} \C^2  \otimes_{\C}\C^4 \right)\oplus      \vspace{2mm} \\

& & \left(\C^2 \otimes_{\C} \C  \otimes_{\C}{\C^4}^* \right)   \oplus  \left(  \C \otimes_{\C} \C^2  \otimes_{\C}{\C^4}^*\right),
\end{array}\end{equation}

\noindent which is 32-complex dimensional. 
%and is made up of four irreducible representations under $SU(2)\times SU(2) \times SU(4)$.

In keeping with the fact that $SU(2)\times SU(2) \times SU(4) \simeq Spin(4)\times Spin(6)$, it turns out that there is another way to describe the fermionic space, $f$.  This is in terms of the Dirac spinors, which form representations of $Spin(4)$ and $Spin(6)$.  These Dirac spinors may be expressed as the exterior algebras, $\Lambda \C^2$ and $\Lambda \C^3$, respectively, so that  

\begin{equation}f \simeq \Lambda \C^2 \otimes_{\C} \Lambda \C^3.
\end{equation}

%  PS binary code  JJ

\begin{figure}[h!]\begin{center}
\includegraphics[width=16.5cm]{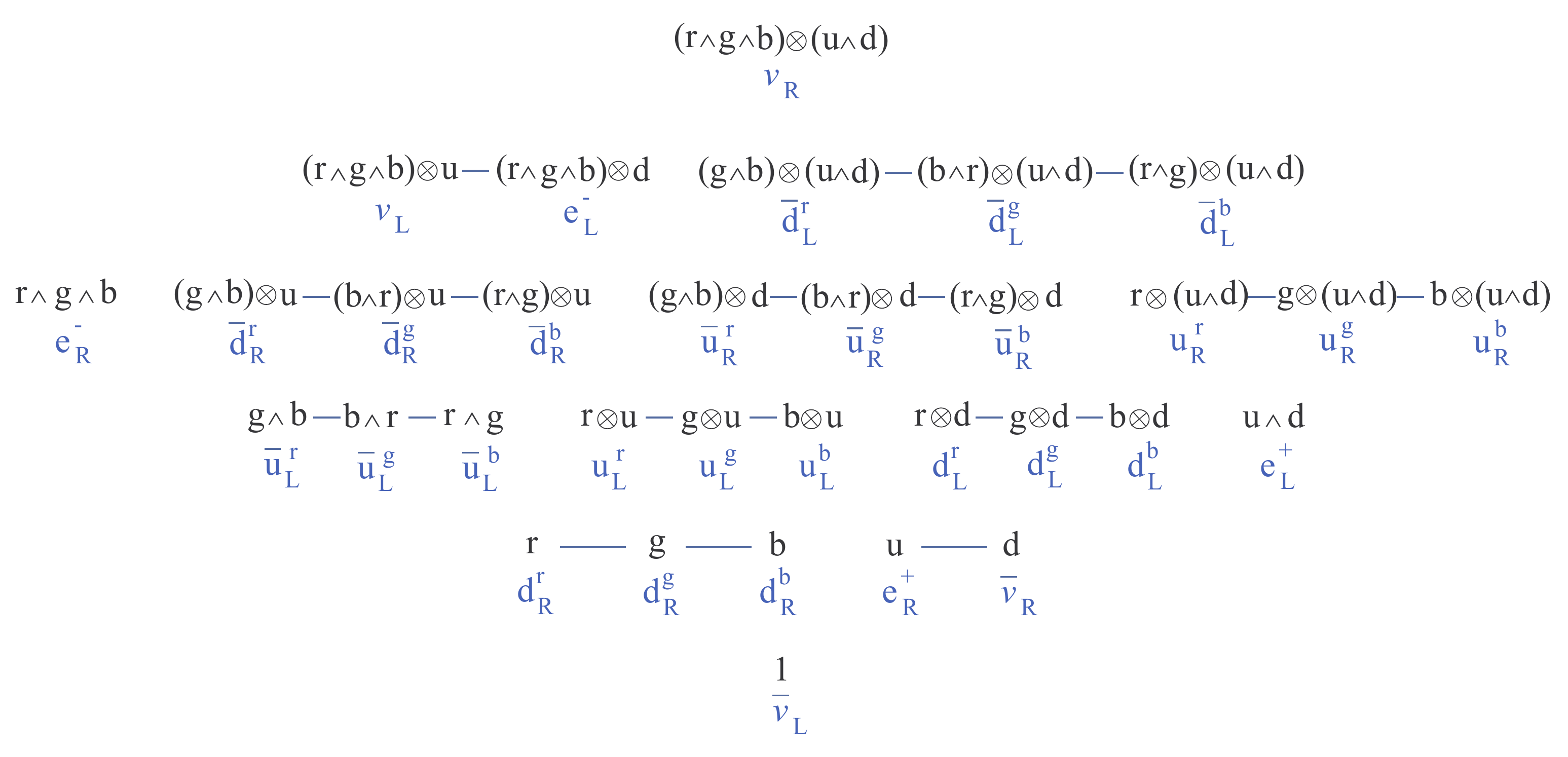}
\caption{\label{psbinary} The tensor product of the exterior algebras $\Lambda \C^2$ and $\Lambda \C^3$, arranged so as to demonstrate its similarity to $\Lambda \C^5$ from Section~\ref{binary}.  Here, standard model irreducible representations are identified within the diagram.}
\end{center}\end{figure}

Let us now consider $\Lambda \C^2$ to be generated by $u$ and $d$, and $\Lambda \C^3$ to be generated by $r$, $g$, and $b$, which are familiar from Section~\ref{binary}.  We can now take the tensor product between these two exterior algebras, and arrange them so as to see the similarity between $\Lambda \C^2 \otimes_{\C} \Lambda \C^3$ and $\Lambda \C^5$ from Section~\ref{binary}.  Please see Figure~\ref{psbinary}.    Clearly, going from the $ \Lambda \C^2 \otimes_{\C} \Lambda \C^3$ fermionic space to the $\Lambda \C^5$ fermionic space, requires simply replacing the tensor product in objects such as $r\otimes u$ with the wedge product, giving $r \wedge u$,~\cite{agut}. 

%Please see Figure~(\ref{split}).

%\begin{figure}[h!]\begin{center}
%\includegraphics[width=16.5cm]{extalgs10.eps}
%\caption{\label{split} The exterior algebras $\Lambda \C^2$ and $\Lambda \C^3$, generated by the binary code elements, $u$, $d$, and $r$, $g$, $b$, respectively.}
%\end{center}\end{figure}

%  PS can also be embedded in Spin 10   JJ

%  diagram

\chapter{Towards a three-generation model\label{GEN}}

% three- gen model
\section{Introduction }

Despite the wide range of proposals to simplify the standard model, most schemes tend to share the same impedances.  Few unified models naturally offer more than a single generation of particles, and few are able to evade proton decay without repercussion.

In previous chapters, we were also concerned with finding just a one-generation model of particle physics, based on $\CO$.  However, this begs the question:  could  $\CO$ provide room for three?  

\medskip

In this chapter, we point out a somewhat mysterious appearance of $SU(3)_c$ representations, which exhibit the behaviour of three full generations of standard model particles.  These representations are found in the Clifford algebra $\CLsix$, arising from the complex octonions.  Back in Section~\ref{chain}, we explained how this 64-complex-dimensional space came about.  With the algebra in place, we will now identify new generators of $SU(3)$ within it.  These $SU(3)$ generators then act to partition the remaining part of the 64-dimensional Clifford algebra into six triplets, six singlets, and their antiparticles.  That is, the algebra mirrors the chromodynamic structure of exactly three generations of the standard model's quarks and leptons.  

Passing from particle to antiparticle, or \it vice versa, \rm requires nothing more than effecting the complex conjugate, $*$: $i \mapsto -i$. The entire result is achieved using only the eight-dimensional complex octonions as a single ingredient.

\newpage

The purpose of  this chapter is \it not \rm to offer a completed unified gauge theory, or even a completed description of QCD.  Instead, we propose a gateway from which such a theory might be found.

To the best of this author's knowledge, \cite{Gen} was the first account of these three-generation results found either within the octonions, or $\CLsix$.

%We come forward with some early blueprints, hinting at  an unusually efficient chromodynamic model.  The $SU(3)_c$ generators here are drafted from the same algebra as the fermions that they act on.  Better still, this algebra readily supports multiple generations, despite being built from nothing more than the complex octonions: an eight-complex-dimensional algebra.  Paradoxically, it is in fact the \it non-associativity \rm of the octonions that enables a larger \it associative \rm algebra to arise, as peculiar as this may initially sound.  (Section~\ref{chain})

%This discovery is expected to strengthen several lines of research.  It may prompt investigators to reinvest in early theories, \cite{GGquarks}-\cite{dixon_fam}, which are based on the idea of division algebras acting on themselves.  It may provide important clues for those working on novel constructions of particle physics~\cite{conlott}-\cite{e6md}.   It also opens up a full arena for study to $G_2$ gauge theory enthusiasts, \cite{rastogi} - \cite{cas}.  Furthermore, this finding releases~\cite{UTI}, from the confines of a single generation, and finally grants anti-particles a space all to their own, which was not a luxury of the original algebra.

\section{Generators of $SU(3)_c$  }  

As mentioned earlier, the automorphism group of the octonions is $G_2$, which is a 14-dimensional exceptional Lie group.  Within $G_2$, we may find a subgroup, $SU(3)$, which is defined as that subgroup of $G_2$ which keeps the imaginary unit, $e_7$, invariant.  The Lie algebra of this $SU(3)$,  acting on $f\in\mathbb{C}\otimes\mathbb{O}$ may be expressed as

\begin{equation}\begin{array}{l}\label{su3gens}

\Lambda_1f = \frac{i}{2}\big(e_1\left(e_5\hspace{2mm} f\hspace{1mm}\right)-e_3\left(e_4 \hspace{2mm}f\hspace{1mm}\right) \big),  \vspace{3mm} \\

\Lambda_2f = -\frac{i}{2}\big(e_1\left(e_4\hspace{2mm} f\hspace{1mm}\right)+e_3\left(e_5 \hspace{2mm}f\hspace{1mm}\right) \big),  \vspace{3mm} \\

\Lambda_3f = \frac{i}{2}\big(e_4\left(e_5\hspace{2mm} f\hspace{1mm}\right)-e_1\left(e_3 \hspace{2mm}f\hspace{1mm}\right) \big) , \vspace{3mm} \\

 \Lambda_4f = \frac{i}{2}\big(e_2\left(e_5\hspace{2mm} f\hspace{1mm}\right)+e_4\left(e_6 \hspace{2mm}f\hspace{1mm}\right) \big),  
 \vspace{3mm} \\

\Lambda_5f = \frac{i}{2}\big(e_5\left(e_6\hspace{2mm} f\hspace{1mm}\right)-e_2\left(e_4 \hspace{2mm}f\hspace{1mm}\right) \big),  \vspace{3mm} \\

\Lambda_6f = \frac{i}{2}\big(e_1\left(e_6\hspace{2mm} f\hspace{1mm}\right)+e_2\left(e_3 \hspace{2mm}f\hspace{1mm}\right) \big),  \vspace{3mm} \\

\Lambda_7f = \frac{i}{2}\big(e_1\left(e_2\hspace{2mm} f\hspace{1mm}\right)+e_3\left(e_6 \hspace{2mm}f\hspace{1mm}\right) \big),  \vspace{3mm} \\

\Lambda_8f = \frac{i}{2\sqrt{3}}\big(e_1\left(e_3\hspace{2mm} f\hspace{1mm}\right)+e_4\left(e_5 \hspace{2mm}f\hspace{1mm}\right)   -    2e_2\left(e_6 \hspace{2mm}f\hspace{1mm}\right) \big).

\end{array}\end{equation}

\noindent  The Lie algebra's commutation relations take the form

\begin{equation}\label{comm}
\Bigg[ \frac{\Lambda_a}{2} \hspace{0.5mm} ,\hspace{0.5mm} \frac{\Lambda_b}{2} \Bigg] f \equiv \left[ \frac{\Lambda_a}{2} \frac{\Lambda_b}{2} -\frac{\Lambda_b}{2} \frac{\Lambda_a}{2} \right]f= ic_{abc} \frac{\Lambda_c}{2} f,
\end{equation}

\noindent $\forall f \in \mathbb{C}\otimes\mathbb{O}$, with the usual $SU(3)$ structure constants, $c_{abc}$.  

Clearly, the $\Lambda_i$, as expressed above, constitute elements of $\mathbb{C}\otimes\overleftarrow{\mathbb{O}}$.  In earlier references, ~\cite{GGquarks}, \cite{GGstats}, \cite{dixon},  these $\Lambda_i$ are shown to act on quark and lepton representations in the eight-dimensional $\mathbb{C}\otimes\mathbb{O}$, or multiple copies thereof.  In contrast, here we introduce the $\Lambda_i$ acting on quark and lepton representations within the 64-dimensional $\mathbb{C}\otimes\overleftarrow{\mathbb{O}}$.

Taking a hint from~\cite{UTI}, let us now introduce a related representation of $su(3)$, which will draw out structure in $\mathbb{C}\otimes\overleftarrow{\mathbb{O}}$, familiar from the behaviour of quarks and leptons.

Consider a resolution of the identity in $\mathbb{C}\otimes\overleftarrow{\mathbb{O}}$

\begin{equation} 1f = \left[\nu + \nu^*\right]f,
\end{equation}

\noindent where $\nu \equiv \frac{1}{2}(1+i\overleftarrow{e_7})$.  Both $\nu$ and $\nu^*$ act as projectors, whereby $\nu  \nu = \nu$, $\hspace{2mm}\nu^*  \nu^* = \nu^*$, and $\hspace{1mm}\nu  \nu^*=\nu^*  \nu=0$.

As \begin{equation} \big[\Lambda_n \hspace{0.5mm} ,\hspace{0.5mm} \nu \big] = 0 \hspace{1cm} \forall n=1\dots8,
\end{equation}

\noindent equation~(\ref{comm}) then leads to

\begin{equation}\label{commnu}
\Bigg[ \frac{\Lambda_a}{2}\nu \hspace{0.5mm} ,\hspace{0.5mm} \frac{\Lambda_b}{2}\nu \Bigg] = ic_{abc} \frac{\Lambda_c}{2}\nu.
\end{equation}

\noindent That is, the eight $\frac{1}{2}\Lambda_n\nu$ form a representation of su(3).  Taking the complex conjugate of~(\ref{commnu}) gives

\begin{equation}\label{commnustar}
\Bigg[ -\frac{\Lambda_a^*}{2}\nu^* \hspace{0.5mm} ,\hspace{0.5mm} -\frac{\Lambda_b^*}{2}\nu^* \Bigg] = ic_{abc} \left[-\frac{\Lambda_c^*}{2}\nu^*\right],
\end{equation}

\noindent so that the $-\frac{1}{2}\Lambda_n^*\nu^*$ give a further representation.

\section{Three generations under $SU(3)_c$ }

Knowing that the $\Lambda_n\nu$ behave as an eight dimensional representation under the action of $\left[\Lambda_m\nu \hspace{0.5mm} , \hspace{1.5mm}\cdot\hspace{1.5mm}\right]$, one might wonder how objects of the more general form $a\nu$ behave under $\left[\Lambda_m\nu \hspace{0.5mm} ,\hspace{1.5mm}\cdot\hspace{1.5mm}\right]$.

Obeying $\left[\Lambda_m\nu\hspace{0.5mm} ,\hspace{0.5mm} \ell_j\nu\right]=0$ $\forall m=1\dots8$, we find six $SU(3)$ singlets, whose basis vectors are given by

\begin{equation}\begin{array}{l}\label{lept}
\ell_{a}\equiv \left(1+ie_{13}+ie_{26}+ie_{45}\right)\nu, \vspace{\spacer}\\
\ell_{b}\equiv\left(3-ie_{13}-ie_{26}-ie_{45}\right)\nu, \vspace{\spacer}\\
\ell_{c}\equiv \left(-ie_{124}-e_{125}+e_{146}-ie_{156}\right)\nu, \vspace{\spacer} \\
\ell_{d}\equiv \left(-ie_{1}-e_{3}+e_{126}+e_{145}\right)\nu, \vspace{\spacer} \\
\ell_{e}\equiv \left(ie_{2}+e_{6}+e_{123}+ie_{136}\right)\nu, \vspace{\spacer} \\
\ell_{f}\equiv \left(ie_{4}+e_{5}-e_{134}+ie_{135}\right)\nu, \vspace{\spacer} \\
\end{array}\end{equation}

\noindent where the left-pointing arrows were dropped throughout for notational simplicity, and right-to-left multiplication is still meant to occur.  The notation $e_{ab}$ is meant here to be shorthand for $e_a(e_b \hspace{1mm}\cdot\hspace{1mm})$, etc.  

The set of basis vectors

\begin{equation}\begin{array}{l}\label{triplet}
q^R_{1}\equiv\left(-ie_{12}-e_{16}+e_{23}+ie_{36}\right)\nu   \vspace{\spacer} \\
q^G_{1}\equiv\left(-ie_{24}-e_{25}+e_{46}-ie_{56}\right)\nu \vspace{\spacer} \\
q^B_{1}\equiv\left(ie_{14}+e_{15}+e_{34}-ie_{35}\right)\nu \vspace{\spacer} \\
\end{array}\end{equation}

\noindent acts as a triplet under commutation with the $\Lambda_m\nu$.  Next, we find five anti-triplets given by

\begin{equation}\begin{array}{l}\label{triplet2}
\bar{q}^R_{2}\equiv\left(ie_{12}-e_{16}+e_{23}-ie_{36}\right)\nu   \vspace{\spacer} \\
\bar{q}^G_{2}\equiv\left(ie_{24}-e_{25}+e_{46}+ie_{56}\right)\nu \vspace{\spacer} \\
\bar{q}^B_{2}\equiv\left(-ie_{14}+e_{15}+e_{34}+ie_{35}\right)\nu, \vspace{\spacer} \\
\end{array}\end{equation}
\begin{equation}\begin{array}{l}\label{triplet3}
\bar{q}^R_{3}\equiv\left(ie_{4}+e_{5}+e_{134}-ie_{135}\right)\nu   \vspace{\spacer} \\
\bar{q}^G_{3}\equiv\left(ie_{1}+e_{3}+e_{126}+e_{145}\right)\nu   \vspace{\spacer} \\
\bar{q}^B_{3}\equiv\left(ie_{2}+e_{6}-e_{123}-ie_{136}\right)\nu,   \vspace{\spacer} \\
\end{array}\end{equation}
\begin{equation}\begin{array}{l}\label{triplet4}
\bar{q}^R_{4}\equiv\left(ie_{1}-e_{3}+e_{126}-e_{145}\right)\nu   \vspace{\spacer} \\
\bar{q}^G_{4}\equiv\left(-ie_{4}+e_{5}+e_{134}+ie_{135}\right)\nu   \vspace{\spacer} \\
\bar{q}^B_{4}\equiv\left(ie_{124}-e_{125}-e_{146}-ie_{156}\right)\nu,   \vspace{\spacer} \\
\end{array}\end{equation}
\begin{equation}\begin{array}{l}\label{triplet5}
\bar{q}^R_{5}\equiv\left(-ie_{2}+e_{6}+e_{123}-ie_{136}\right)\nu   \vspace{\spacer} \\
\bar{q}^G_{5}\equiv\left(ie_{124}-e_{125}+e_{146}+ie_{156}\right)\nu   \vspace{\spacer} \\
\bar{q}^B_{5}\equiv\left(ie_{4}-e_{5}+e_{134}+ie_{135}\right)\nu,   \vspace{\spacer} \\
\end{array}\end{equation}
\begin{equation}\begin{array}{l}\label{triplet6}
\bar{q}^R_{6}\equiv\left(ie_{124}+e_{125}+e_{146}-ie_{156}\right)\nu   \vspace{\spacer} \\
\bar{q}^G_{6}\equiv\left(ie_{2}-e_{6}+e_{123}-ie_{136}\right)\nu   \vspace{\spacer} \\
\bar{q}^B_{6}\equiv\left(-ie_{1}+e_{3}+e_{126}-e_{145}\right)\nu.   \vspace{\spacer} \\
\end{array}\end{equation}

Taking the complex conjugate, $*$: $i\mapsto-i$, of these 32 basis vectors gives 32 new linearly independent basis vectors. Under commutation with $-\Lambda_m^* \nu^*$,

\begin{equation}\begin{array}{l}\label{antilept}
\ell_{a}^*=\left(1-ie_{13}-ie_{26}-ie_{45}\right)\nu^*, \vspace{\spacer}\\
  \ell_{b}^*=\left(3+ie_{13}+ie_{26}+ie_{45}\right)\nu^*, \vspace{\spacer}\\
\ell_{c}^*= \left(ie_{124}-e_{125}+e_{146}+ie_{156}\right)\nu^*, \vspace{\spacer} \\
\ell_{d}^*= \left(ie_{1}-e_{3}+e_{126}+e_{145}\right)\nu^*, \vspace{\spacer} \\
\ell_{e}^*= \left(-ie_{2}+e_{6}+e_{123}-ie_{136}\right)\nu^*, \vspace{\spacer} \\
\ell_{f}^*= \left(-ie_{4}+e_{5}-e_{134}-ie_{135}\right)\nu^* \vspace{\spacer} \\
\end{array}\end{equation}

\noindent transform as $SU(3)$ singlets,

\begin{equation}\begin{array}{l}\label{antitriplet}
q^{R*}_{1}=\left(ie_{12}-e_{16}+e_{23}-ie_{36}\right)\nu^* \equiv\bar{q}^R_1   \vspace{\spacer} \\
q^{G*}_{1}=\left(ie_{24}-e_{25}+e_{46}+ie_{56}\right)\nu^* \equiv\bar{q}^G_1 \vspace{\spacer} \\
q^{B*}_{1}=\left(-ie_{14}+e_{15}+e_{34}+ie_{35}\right)\nu^* \equiv\bar{q}^B_1 \vspace{\spacer} \\
\end{array}\end{equation}

\noindent behaves as an anti-triplet,

\begin{equation}\begin{array}{l}\label{antitriplet2}
\bar{q}^{R^*}_{2}=\left(-ie_{12}-e_{16}+e_{23}+ie_{36}\right)\nu^*\equiv q_2^R   \vspace{\spacer} \\
\bar{q}^{G^*}_{2}=\left(-ie_{24}-e_{25}+e_{46}-ie_{56}\right)\nu^*\equiv q_2^G \vspace{\spacer} \\
\bar{q}^{B^*}_{2}=\left(ie_{14}+e_{15}+e_{34}-ie_{35}\right)\nu^*\equiv q_2^B \vspace{\spacer} \\
\end{array}\end{equation}

\noindent behaves as a triplet, and so on.

That is, unlike the standard model, we are able to pass back and forth between particle and anti-particle using \it only \rm the complex conjugate $i\mapsto-i$.   This feature appeared early on in the work of~\cite{GGquarks} for some  internal degrees of freedom, and also in~\cite{UTI} and Section~\ref{lr} when passing between left- and right-handed Weyl spinors.

\section{A sample calculation }

We introduce to the reader how calculations are carried out in $\mathbb{C}\otimes\overleftarrow{\mathbb{O}}$  by working through an example.  Let us find the action of the first $SU(3)$ generator of the form $\Lambda \nu$, which we will define as $\Lambda_1\nu\equiv\frac{i}{2}\left(e_{15}-e_{34}\right)\nu$, in accordance with equation~(\ref{liealg}).  Let $\Lambda_1\nu$ act on $q^R_{1}$, as defined in equations~(\ref{triplet}):
\begin{equation}\begin{array}{ll}\label{calc} & \Big[ \Lambda_1\nu \hspace{1mm},\hspace{1mm} q^R_{1} \Big]\vspace{\spacer}   \\
=& \Big[\frac{i}{2}\left(e_{15}-e_{34}\right)\nu\hspace{0.5mm} ,\hspace{1mm}\left(-ie_{12}-e_{16}+e_{23}+ie_{36}\right)\nu \Big]\vspace{\spacer} \\
=&\frac{i}{2}\Big( \left(e_{15}-e_{34}\right)\left(-ie_{12}-e_{16}+e_{23}+ie_{36}\right)\vspace{\spacer} \\
 &- \left(-ie_{12}-e_{16}+e_{23}+ie_{36}\right)\left(e_{15}-e_{34}\right) \Big)\hspace{0.2mm}\nu\vspace{\spacer} \\
%\end{array}\end{equation}
%\begin{equation}\begin{array}{ll}
=&\frac{i}{2}\big(-ie_{1512}-e_{1516}+e_{1523}+ie_{1536}\vspace{\spacer} \\
&\hspace{0.85cm}+ie_{3412}+e_{3416}-e_{3423}-ie_{3436}\vspace{\spacer} \\
&\hspace{0.85cm}+ie_{1215}+e_{1615}-e_{2315}-ie_{3615}\vspace{\spacer} \\
&\hspace{0.85cm}-ie_{1234}-e_{1634}+e_{2334}+ie_{3634}\big)\hspace{0.2mm}\nu\vspace{\spacer} \\
%\end{array}
%\end{equation}
%\begin{equation}
%\begin{array}{ll}
=&\frac{i}{2}\big(-ie_{52}-e_{56}+e_{1235}-ie_{1356}\vspace{\spacer} \\
&\hspace{0.85cm}+ie_{1234}+e_{1346}+e_{42}-ie_{46}\vspace{\spacer} \\
&\hspace{0.85cm}+ie_{25}+e_{65}-e_{1235}+ie_{1356}\vspace{\spacer} \\
&\hspace{0.85cm}-ie_{1234}-e_{1346}-e_{24}+ie_{64}\big)\hspace{0.2mm}\nu\vspace{\spacer} \\
=&i\big(ie_{25}-e_{56}-e_{24}-ie_{46}\big)\hspace{0.2mm}\nu=q^G_{1}.\vspace{\spacer} \\
\end{array}\end{equation}

\noindent This is the result we would expect for the first of the $su(3)$ Gell-Mann matrices, $\Lambda^{GM}_1$, from the standard model, acting to convert a red basis vector, $\underline{R}\equiv (1,0,0)^\top$, into a green basis vector, $\underline{G}\equiv (0,1,0)^\top$.
\begin{equation} \Lambda^{GM}_1\underline{R} = \left(
\begin{array}{ccc}
0 & 1 & 0 \\
1 & 0 & 0 \\
0 & 0 & 0
\end{array} \right)\left(
\begin{array}{c}
1  \\
0  \\
0
\end{array} \right) = \left(
\begin{array}{c}
0  \\
1  \\
0
\end{array} \right) = \underline{G}.
\end{equation}

\section{Summary for $SU(3)_c$ }

\begin{figure}[h!]\begin{center}
\includegraphics[width=10cm]{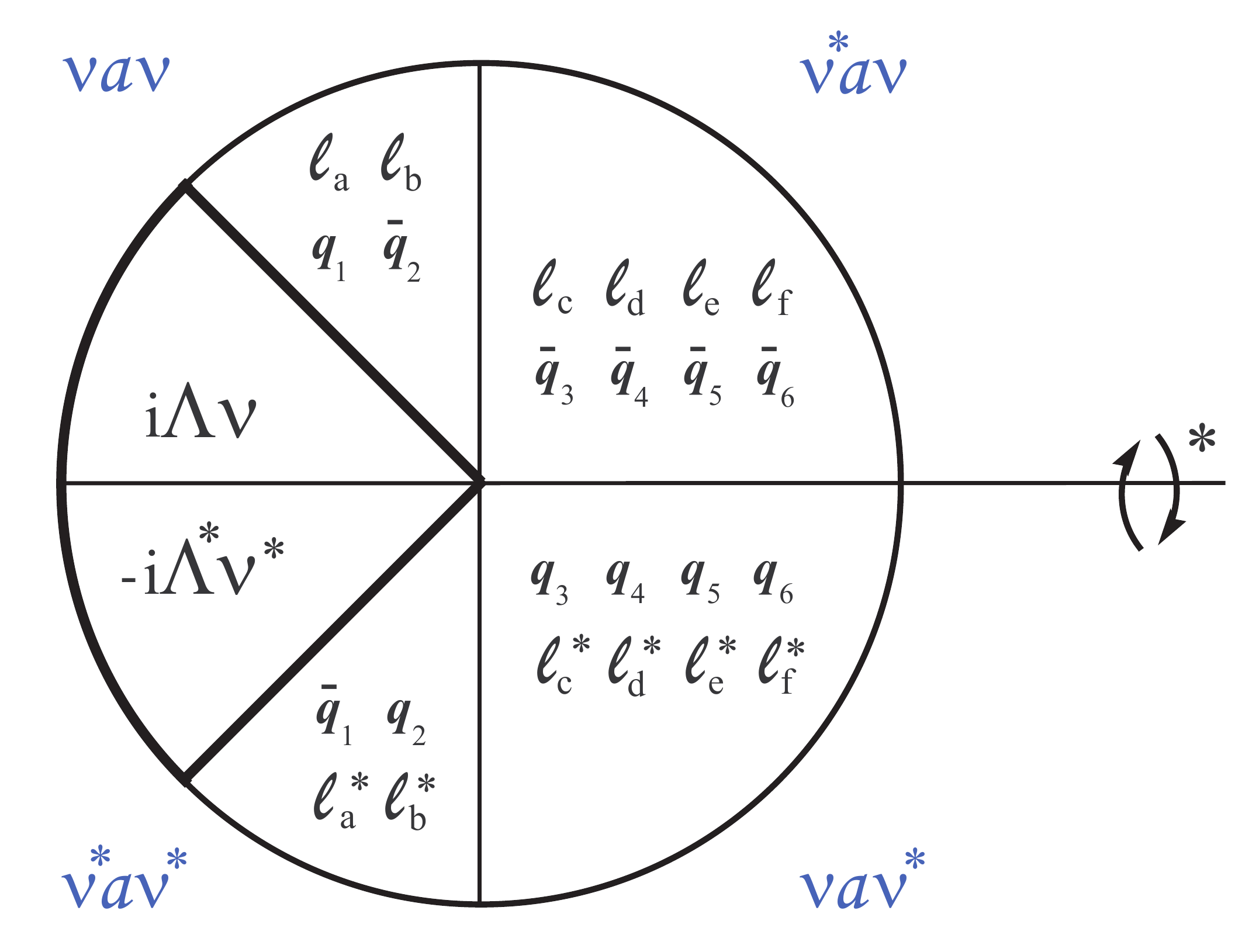}
\caption{\label{64} The 64-dimensional octonionic chain algebra splits into two sets of $SU(3)$ generators of the form $i\Lambda\nu$ and $-i\Lambda^*\nu^*$, six $SU(3)$ singlets $\ell_j$, six triplets $q_k$, and their complex conjugates.  These objects are sectioned off above into four quadrants according to their forms: $\nu a \nu$, $\nu^* a \nu$, $\nu a \nu^*$ and $\nu^* a \nu^*$ for $a$ in the chain algebra. Transforming particles into anti-particles, and \it vice versa, \rm requires only the complex conjugate $*$: $i\mapsto-i$ in our formalism.}
\end{center}\end{figure}

Finally, we conclude by summarizing the main result of this chapter in Figure~\ref{64}: the breakdown of the 64-dimensional $\mathbb{C}\otimes\overleftarrow{\mathbb{O}}$ into irreducible representations of $SU(3)$.

Using only the eight-dimensional complex octonions, $\mathbb{C}\otimes\mathbb{O}$, we have explained how to build up a 64-complex-dimensional associative algebra.  The $SU(3)$ generators identified within this algebra then break down the remaining space into six singlets, six triplets, and their antiparticles, with no extra particles beyond these.  

These representations are a curious finding.  They effortlessly suggest the existence of exactly three generations, they relate particles to antiparticles by using only the complex conjugate $i\mapsto -i$, and finally, they fill these tall orders while working from but a modest eight-complex-dimensional algebra.

%\section{Jordan algebras}

% intro to JAs
% defn
% why useful

% is you complete JA of the Lamda s you get Q
%curious result.

\vspace{1cm}

\section{Three generations under $SU(3)_c$ and $U(1)_{em}$  \label{ass}}

% the question is now - can you find a way for Q to act on 3 generations and give the right results?
%describe 3 gen with charges

%\noindent \bf Outlook. \rm

%\subsection{Electric charge assignments\label{ass}  }

Since the time that these irreducible representations of $SU(3)_c$ were identified within $\COOO$, more standard model structure has been uncovered in the algebra.  It turns out that we can extend our three-generation results for $SU(3)_c$ to include $U(1)_{em}$ as well (presented for the first time here).  

That is, our $SU(3)_c$ action on $\COOO$ given by 

\begin{equation}\begin{array}{lll}\label{simpleaction}\large[\hspace{1mm}\Lambda_j \nu,\hspace{1mm} a \nu \hspace{1mm}\large], & \hspace{1cm}&\large[\hspace{1mm}-\Lambda_j^* \nu^*, \hspace{1mm}a \nu^* \hspace{1mm}\large] 
\end{array}\end{equation}

\noindent can be generalized, so as to include electric charges.  Furthermore, we will be able to incorporate $U(1)_{em}$ by making use of the $Q$ we had already introduced in our one-generation model of Section~\ref{ladders}, 
\begin{equation}Q=\frac{N}{3} = \sum_{i=1}^3 \alpha^{\dagger}_i\alpha_i.
\end{equation}

Consider the following action for the $SU(3)_c$ generators,

\begin{equation}\begin{array}{lll} \label{newaction} \left[\hspace{1mm}\Lambda_j \nu,\hspace{1mm} S^*a \nu \hspace{1mm}\right], & \hspace{1cm}&\left[\hspace{1mm}-\Lambda_j^* \nu^*, \hspace{1mm}Sa \nu^* \hspace{1mm}\right],  \vspace{4mm}\\

\left[\hspace{1mm}\Lambda_j \nu,\hspace{1mm} Sa \nu \hspace{1mm}\right], & \hspace{1cm}&\left[\hspace{1mm}-\Lambda_j^* \nu^*, \hspace{1mm}S^*a \nu^* \hspace{1mm}\right],
\end{array}\end{equation}

\noindent where $S$ is the projector given by
\begin{equation} S\equiv \frac{1}{2}-\frac{ie_7}{4}+\frac{i}{4}\left(e_{13}+e_{26}+e_{45}\right).
\end{equation}

 \noindent Clearly, $S+S^* = 1$, so that we see that this new action~(\ref{newaction}) is no different from the original action~(\ref{simpleaction}).  Although it has not been mentioned before, $S$ is a projector which has come up frequently in complex-octonionic work;  it is nothing more than the right-multiplication analogue of $\nu = \frac{1}{2}\left(1+ie_7\right)$.  Or in other words, 
 
 \begin{equation}
 f \hspace{1mm}\frac{1}{2}\left(1+ie_7\right) = Sf \hspace{1cm} \forall f\in \CO.
 \end{equation}
 
 %It is the projector one finds by right multiplying $\frac{1}{2}\left(1+ie_7\right)$ onto any $f\in\CO$.  That is,
 %\begin{equation} f \frac{1}{2}\left(1+ie_7\right) = Sf \hspace{1cm} \forall f\in\CO.
 %\end{equation}
 
 It so happens that $-\Lambda_j^* = \Lambda_j$ $\forall j = 1 \dots 8$, so that the action~(\ref{newaction}) may be rewritten as 
 
 \begin{equation}\begin{array}{lll} \label{newnewaction} \left[\hspace{1mm}\Lambda_j \nu,\hspace{1mm} S^*a \nu \hspace{1mm}\right], & \hspace{1cm}&\left[\hspace{1mm}-\Lambda_j^* \nu^*, \hspace{1mm}Sa \nu^* \hspace{1mm}\right],  \vspace{4mm}\\

\left[\hspace{1mm}-\Lambda_j^* \nu,\hspace{1mm} Sa \nu \hspace{1mm}\right], & \hspace{1cm}&\left[\hspace{1mm}\Lambda_j \nu^*, \hspace{1mm}S^*a \nu^* \hspace{1mm}\right].
\end{array}\end{equation}
 
Now, an action for electric charge may be found, which matches the $SU(3)_c$ action of (\ref{newnewaction}).  Namely, 

\begin{equation}\begin{array}{lll} \label{elecaction} \left[\hspace{1mm}Q \nu,\hspace{1mm} S^*a \nu \hspace{1mm}\right], & \hspace{1cm}&\left[\hspace{1mm}-Q^* \nu^*, \hspace{1mm}Sa \nu^* \hspace{1mm}\right],  \vspace{4mm}\\

\left[\hspace{1mm}-Q^* \nu,\hspace{1mm} Sa \nu \hspace{1mm}\right], & \hspace{1cm}&\left[\hspace{1mm}Q \nu^*, \hspace{1mm}S^*a \nu^* \hspace{1mm}\right].
\end{array}\end{equation}

\noindent Under these actions (\ref{newnewaction}) and (\ref{elecaction}), we find $SU(3)_c$ and $U(1)_{em}$ charge assignments, which are consistent with three generations of standard model particles.  Below, we relabel the states given earlier in equations~(\ref{lept})-(\ref{antitriplet2}), so as to now specify their electric charges.  

At this level, we are not specifying which generation each state belongs to, so for $i=1,2,3$, the three states with electric charge of -1 will be labelled $e^-_i$;  the three states with electric charge of +1 will be labelled $e^+_i$; the three states with electric charge of $2/3$ will be labelled $u_i$;  the three states with electric charge of $-2/3$ will be labelled $\bar{u}_i$;  the three states with electric charge $-1/3$ will be labelled $d_i$, and  the three states with electric charge $1/3$ will be labelled $\bar{d}_i$.  Since $SU(3)_c$ and $U(1)_{em}$ do not distinguish between neutrinos, $\nu_i$ and anti-neutrinos, $\bar{\nu}_i$, we will then label the six states with electric charge of zero as $n _i$ and $\bar{n} _i$, where the symbol $n_i$ could represent either a neutrino or anti-neutrino.

\begin{equation}\begin{array}{l}\label{newlept}
n_1\longleftarrow\ell_{a}\equiv\left(1+ie_{13}+ie_{26}+ie_{45}\right)\nu = S\hspace{1mm}\ell_{a}\hspace{1mm}\nu,   \vspace{\spacer}\\
n_2\longleftarrow\ell_{b}\equiv\left(3-ie_{13}-ie_{26}-ie_{45}\right)\nu = S^*\hspace{1mm}\ell_{b}\hspace{1mm}\nu, \vspace{\spacer}\\
n_3\longleftarrow\ell_{c}\equiv \left(-ie_{124}-e_{125}+e_{146}-ie_{156}\right)\nu = S^*\hspace{1mm}\ell_{c}\hspace{1mm}\nu, \vspace{\spacer} \\
e^+_1\longleftarrow\ell_{d}\equiv \left(-ie_{1}-e_{3}+e_{126}+e_{145}\right)\nu = S\hspace{1mm}\ell_{d}\hspace{1mm}\nu, \vspace{\spacer} \\
e^+_2\longleftarrow\ell_{e}\equiv \left(ie_{2}+e_{6}+e_{123}+ie_{136}\right)\nu = S\hspace{1mm}\ell_{e}\hspace{1mm}\nu, \vspace{\spacer} \\
e^+_3\longleftarrow\ell_{f}\equiv \left(ie_{4}+e_{5}-e_{134}+ie_{135}\right)\nu = S\hspace{1mm}\ell_{f}\hspace{1mm}\nu, \vspace{\spacer} \\
\end{array}\end{equation}

\begin{equation}\begin{array}{l}\label{newtriplet}
u_1^R\longleftarrow q^R_{1}\equiv\left(-ie_{12}-e_{16}+e_{23}+ie_{36}\right)\nu  = S^*\hspace{1mm} q^R_{1}\hspace{1mm}\nu  \vspace{\spacer} \\
u_1^G\longleftarrow q^G_{1}\equiv\left(-ie_{24}-e_{25}+e_{46}-ie_{56}\right)\nu = S^*\hspace{1mm} q^G_{1}\hspace{1mm}\nu\vspace{\spacer} \\
u_1^B\longleftarrow q^B_{1}\equiv\left(ie_{14}+e_{15}+e_{34}-ie_{35}\right)\nu = S^*\hspace{1mm} q^B_{1}\hspace{1mm}\nu\vspace{\spacer} \\
\end{array}\end{equation}

\begin{equation}\begin{array}{l}\label{newtriplet2}
\bar{u}_2^R\longleftarrow \bar{q}^R_{2}\equiv\left(ie_{12}-e_{16}+e_{23}-ie_{36}\right)\nu  =  S \hspace{1mm} \bar{q}^R_{2} \hspace{1mm}\nu  \vspace{\spacer} \\
\bar{u}_2^G\longleftarrow \bar{q}^G_{2}\equiv\left(ie_{24}-e_{25}+e_{46}+ie_{56}\right)\nu =  S \hspace{1mm} \bar{q}^G_{2} \hspace{1mm}\nu \vspace{\spacer} \\
\bar{u}_2^B\longleftarrow \bar{q}^B_{2}\equiv\left(-ie_{14}+e_{15}+e_{34}+ie_{35}\right)\nu=  S \hspace{1mm} \bar{q}^B_{2} \hspace{1mm}\nu, \vspace{\spacer} \\
\end{array}\end{equation}
\begin{equation}\begin{array}{l}\label{newtriplet3}
\bar{u}_3^R\longleftarrow \bar{q}^R_{3}\equiv\left(ie_{4}+e_{5}+e_{134}-ie_{135}\right)\nu  = S^*\hspace{1mm}\bar{q}^R_{3}\hspace{1mm}\nu  \vspace{\spacer} \\
\bar{u}_3^G\longleftarrow \bar{q}^G_{3}\equiv\left(ie_{1}+e_{3}+e_{126}+e_{145}\right)\nu= S^*\hspace{1mm}\bar{q}^G_{3}\hspace{1mm}\nu   \vspace{\spacer} \\
\bar{u}_3^B\longleftarrow \bar{q}^B_{3}\equiv\left(ie_{2}+e_{6}-e_{123}-ie_{136}\right)\nu= S^*\hspace{1mm}\bar{q}^B_{3}\hspace{1mm}\nu,   \vspace{\spacer} \\
\end{array}\end{equation}
\begin{equation}\begin{array}{l}\label{newtriplet4}
\bar{d}_1^R\longleftarrow \bar{q}^R_{4}\equiv\left(ie_{1}-e_{3}+e_{126}-e_{145}\right)\nu  = S\hspace{1mm}\bar{q}^R_{4}\hspace{1mm}\nu \vspace{\spacer} \\
\bar{d}_1^G\longleftarrow \bar{q}^G_{4}\equiv\left(-ie_{4}+e_{5}+e_{134}+ie_{135}\right)\nu = S\hspace{1mm}\bar{q}^G_{4}\hspace{1mm}\nu   \vspace{\spacer} \\
\bar{d}_1^B\longleftarrow \bar{q}^B_{4}\equiv\left(ie_{124}-e_{125}-e_{146}-ie_{156}\right)\nu = S\hspace{1mm}\bar{q}^B_{4}\hspace{1mm}\nu,   \vspace{\spacer} \\
\end{array}\end{equation}
\begin{equation}\begin{array}{l}\label{newtriplet5}
\bar{d}_2^R\longleftarrow \bar{q}^R_{5}\equiv\left(-ie_{2}+e_{6}+e_{123}-ie_{136}\right)\nu =S\hspace{1mm}\bar{q}^R_{5}\hspace{1mm}\nu  \vspace{\spacer} \\
\bar{d}_2^G\longleftarrow \bar{q}^G_{5}\equiv\left(ie_{124}-e_{125}+e_{146}+ie_{156}\right)\nu =S\hspace{1mm}\bar{q}^G_{5}\hspace{1mm}\nu   \vspace{\spacer} \\
\bar{d}_2^B\longleftarrow \bar{q}^B_{5}\equiv\left(ie_{4}-e_{5}+e_{134}+ie_{135}\right)\nu=S\hspace{1mm}\bar{q}^B_{5}\hspace{1mm}\nu ,   \vspace{\spacer} \\
\end{array}\end{equation}
\begin{equation}\begin{array}{l}\label{newtriplet6}
\bar{d}_3^R\longleftarrow \bar{q}^R_{6}\equiv\left(ie_{124}+e_{125}+e_{146}-ie_{156}\right)\nu =S\hspace{1mm}\bar{q}^R_{6}\hspace{1mm}\nu  \vspace{\spacer} \\
\bar{d}_3^G\longleftarrow \bar{q}^G_{6}\equiv\left(ie_{2}-e_{6}+e_{123}-ie_{136}\right)\nu=S\hspace{1mm}\bar{q}^G_{6}\hspace{1mm}\nu    \vspace{\spacer} \\
\bar{d}_3^B\longleftarrow \bar{q}^B_{6}\equiv\left(-ie_{1}+e_{3}+e_{126}-e_{145}\right)\nu=S\hspace{1mm}\bar{q}^B_{6}\hspace{1mm}\nu .   \vspace{\spacer} \\
\end{array}\end{equation}

Taking the complex conjugate, $*$: $i\mapsto-i$, of these 32 basis vectors gives 32 new linearly independent basis vectors:

\begin{equation}\begin{array}{l}\label{newantilept}
\bar{n}_1\longleftarrow\ell_{a}^*=\left(1-ie_{13}-ie_{26}-ie_{45}\right)\nu^* = S^*\hspace{1mm}\ell_{a}^*\hspace{1mm}\nu^*,  \vspace{\spacer}\\
 \bar{n}_2\longleftarrow \ell_{b}^*=\left(3+ie_{13}+ie_{26}+ie_{45}\right)\nu^*= S\hspace{1mm}\ell_{b}^*\hspace{1mm}\nu^*, \vspace{\spacer}\\
\bar{n}_3\longleftarrow\ell_{c}^*= \left(ie_{124}-e_{125}+e_{146}+ie_{156}\right)\nu^*= S\hspace{1mm}\ell_{c}^*\hspace{1mm}\nu^*, \vspace{\spacer} \\
e^-_1\longleftarrow\ell_{d}^*= \left(ie_{1}-e_{3}+e_{126}+e_{145}\right)\nu^*= S^*\hspace{1mm}\ell_{d}^*\hspace{1mm}\nu^*, \vspace{\spacer} \\
e^-_2\longleftarrow\ell_{e}^*= \left(-ie_{2}+e_{6}+e_{123}-ie_{136}\right)\nu^*= S^*\hspace{1mm}\ell_{e}^*\hspace{1mm}\nu^*, \vspace{\spacer} \\
e^-_3\longleftarrow\ell_{f}^*= \left(-ie_{4}+e_{5}-e_{134}-ie_{135}\right)\nu^*= S^*\hspace{1mm}\ell_{f}^*\hspace{1mm}\nu^*, \vspace{\spacer} \\
\end{array}\end{equation}

\begin{equation}\begin{array}{l}\label{newantitriplet}
\bar{u}_1^R\longleftarrow q^{R*}_{1}=\left(ie_{12}-e_{16}+e_{23}-ie_{36}\right)\nu^* \equiv\bar{q}^R_1= S\hspace{1mm}\bar{q}^R_1\hspace{1mm}\nu^*   \vspace{\spacer} \\
\bar{u}_1^G\longleftarrow q^{G*}_{1}=\left(ie_{24}-e_{25}+e_{46}+ie_{56}\right)\nu^* \equiv\bar{q}^G_1= S\hspace{1mm}\bar{q}^G_1\hspace{1mm}\nu^* \vspace{\spacer} \\
\bar{u}_1^B\longleftarrow q^{B*}_{1}=\left(-ie_{14}+e_{15}+e_{34}+ie_{35}\right)\nu^* \equiv\bar{q}^B_1= S\hspace{1mm}\bar{q}^B_1\hspace{1mm}\nu^* \vspace{\spacer} \\
\end{array}\end{equation}

\begin{equation}\begin{array}{l}\label{newantitriplet2}
u_2^R\longleftarrow \bar{q}^{R^*}_{2}=\left(-ie_{12}-e_{16}+e_{23}+ie_{36}\right)\nu^*\equiv q_2^R  = S^*\hspace{1mm}q_2^R\hspace{1mm}\nu^*  \vspace{\spacer} \\
u_2^G\longleftarrow \bar{q}^{G^*}_{2}=\left(-ie_{24}-e_{25}+e_{46}-ie_{56}\right)\nu^*\equiv q_2^G= S^*\hspace{1mm}q_2^G\hspace{1mm}\nu^* \vspace{\spacer} \\
u_2^B\longleftarrow \bar{q}^{B^*}_{2}=\left(ie_{14}+e_{15}+e_{34}-ie_{35}\right)\nu^*\equiv q_2^B= S^*\hspace{1mm}q_2^B\hspace{1mm}\nu^* \vspace{\spacer} \\
\end{array}\end{equation}
\noindent and so on.

%\subsection{Jordan Algebras and the unbroken symmetries  }

%It may have been noticed that the generators we are using here, $\Lambda_i \nu$ and $Q\nu$, and their conjugates, bear a striking resemblance to those used in our one-generation model of Section~\ref{ladders}, namely, $\Lambda_i$ and $Q$.  However, we have introduced an extra factor of $\nu = \frac{1}{2}(1+ie_7)$ in this chapter, and it is natural to wonder why.

%One possible explanation for $\nu$ could lie in a requirement that these symmetry generators need be elements of a Jordan algebra.  A Jordan algebra is an algebra over a field, $\mathbb{F}$, such that multiplication is commutative,

%\begin{equation} x* y = y* x, \hspace{1.5cm} \forall x, y
%\end{equation}

%\noindent and 

%\begin{equation} (x* y)*(x* x) = x* (y* (x* x))\hspace{1.5cm} \forall x, y.
%\end{equation}

%defn checked mccrimmon
%Under multiplication given by the anti-commutator, $x* y\equiv xy +yx$, we quicky find that the original $\Lambda_i$ and $Q$ operators do not form a closed (Jordan) algebra.  However, the set of $\Lambda_i\nu$ and $Q\nu$ do.

%Furthermore, it is interesting to note that the $\Lambda_i\nu$ alone are not closed under anti-commutation, but that they require $Q\nu$ so as to complete the algebra.  Incidentally, $Q\nu$ is able to form a 1-dimensional Jordan algebra of its own under the anti-commutator.  These results are expected to appear in future work, where we will consider Jordan algebras appearing in standard model representations.

\section{Outlook:  From one generation to three}

% two extra DOF - how to get it?
% possibly from AA^T + A^TA form

In Section~\ref{ladders}, we described a one-generation model in $\COOO$, transforming under the symmetry generators $\Lambda_j$ and $Q$.  
In Section~\ref{GEN} we described a  three-generation model in $\COOO$, transforming under the symmetry generators $\Lambda_j\nu$ and $Q\nu$. 
Given that they make use of the same algebras, and very similar symmetry generators, one might wonder if the one-generation and three-generation models could be connected.  

Direct verification shows that the one-generation representation and its symmetries do not fit directly into the three-generation model in any obvious way.  However, we suspect that the similarity between these two models is no coincidence and there could be a way to go from one to the other.  \footnote{For example, one might consider sums of objects of the form $S^u{S^u}^{\dagger}$ and $S^d{S^d}^{\dagger}$ in order to move from states $S^u$ and $S^d$ in the one-generation model to states in the three-generation model.}.

\chapter{Conclusion}

As a non-associative algebra, one might have naturally expected that $\RCHO$ would not be fit to describe the action of groups.  And with no more than 32 complex dimensions, one might have further anticipated that $\RCHO$ would not have the capacity to describe much of the standard model, whose particle content spans hundreds of degrees of freedom.  

Upon closer inspection, though, this algebra can be seen to exhibit a surprising amount of the standard model's structure.  Over the years, numerous authors have pointed out various Lorentz representations, within $\CH$.  In this thesis, we have then gone on to consolidate all of the standard model's Lorentz representations in terms of generalized ideals of $\CH$.  

In the early seventies, G\"{u}naydin and G\"{u}rsey  showed  $SU(3)_c$ quark structure within the octonions, \cite{GGquarks}.  
%Subsequently, Casalbuoni \it et al., \rm \cite{Grass}, \cite{wow}, demonstrated further features of the standard model, by considering the tensor products of different Clifford algebras, which are close relatives of $\RCHO$.
This thesis then subsequently provided a way to extend this octonionic quark model so as to include leptons and the electromagnetic charge operator.  This completes one full generation of quarks and leptons, and describes their behaviour under the unbroken gauge symmetries of the standard model.  Our use of minimal left ideals from \it only \rm $\CO$ allowed us to provide a straightforward explanation for the quantization of electric charge.

We also  demonstrated a rudimentary leptonic model with this $\RCHO$ algebra, whereby $SU(2)_L$ acts automatically on only left-handed states.  We have repeatedly shown the generators of standard model gauge symmetries appearing, uniquely, as particular symmetries of the algebra's ladder operators.  

Finally, within the octonionic sector of $\RCHO$, we found the  $SU(3)_c$ and $U(1)_{em}$ representations corresponding to three full generations of quarks and leptons.  Given that most unified theories are based on a single generation, this may be viewed as an unusual finding.

Although evidence is accumulating in support of a connection between the standard model and $\RCHO$, we certainly do not have a complete model, at the moment.  However, with every new discovery, it becomes a little more clear that this unlikely algebra is not going away.

\bibliographystyle{plain}
% This specifies the location of the file containing the bibliographic information.  
% It assumes you're using BibTeX (if not, why not?).
\cleardoublepage % This is needed if the book class is used, to place the anchor in the correct page,
                 % because the bibliography will start on its own page.
                 % Use \clearpage instead if the document class uses the "oneside" argument
\phantomsection  % With hyperref package, enables hyperlinking from the table of contents to bibliography             
% The following statement causes the title "References" to be used for the bibliography section:
\renewcommand*{\bibname}{References}

% Add the References to the Table of Contents
\addcontentsline{toc}{chapter}{\textbf{References}}

\bibliography{uw-ethesis}

\begin{thebibliography}{10}

\bibitem{ablam}
R.~Ab\l amowicz.
\newblock {\em Construction of spinors via Witt decomposition and primitive
  idempotents: a review}.
\newblock Clifford algebras and spinor structures. Kluwer Academic Publishers,
  1995.

\bibitem{magic}
A.~Anastasiou, L.~Borsten, M.~J. Duff, L.J. Hughes, and S.~Nagy.
\newblock A magic pyramid of supergravities.
\newblock {\em JHEP}, 04:178, 2014.

\bibitem{mth}
A.~Anastasiou, L.~Borsten, M.~J. Duff, L.J. Hughes, and S.~Nagy.
\newblock An octonionic formulation of the {M-theory} algebra.
\newblock {\em JHEP}, 11:022, 2014.

\bibitem{baez}
J.~Baez.
\newblock The octonions.
\newblock {\em Bull.Am.Math.Soc.}, 39:145--205, 2002.

\bibitem{jjI}
J.~Baez and J.~Huerta.
\newblock Division algebras and supersymmetry {I}.
\newblock {\em Superstrings, Geometry, Topology, and C*-algebras, eds. R.
  Doran, G. Friedman and J. Rosenberg, Proc. Symp. Pure Math.}, 81:65--80,
  2010.

\bibitem{jjII}
J.~Baez and J.~Huerta.
\newblock Division algebras and supersymmetry {II}.
\newblock {\em Adv. Math. Theor. Phys.}, 15:1373--1410, 2011.

\bibitem{agut}
J.~C. Baez and J.~Huerta.
\newblock The algebra of grand unified theories.
\newblock {\em Bull.Am.Math.Soc.}, 47:483--552, 2010.

\bibitem{Grass}
A.~Barducci, F.~Buccella, R.~Casalbuoni, L.~Lusanna, and E.~Sorace.
\newblock Quantized grassmann variables and unified theories.
\newblock {\em Phys. Letters B}, 67(344), 1977.

\bibitem{barrett}
J.W. Barrett.
\newblock Matrix geometries and fuzzy spaces as finite spectral triples.
\newblock {\em arXiv:1502.05383 [math-ph]}, 2015.

\bibitem{rafael}
L.~Bombelli, J.~Lee, D.~Meyer, and R.~D. Sorkin.
\newblock Space-time as a causal set.
\newblock {\em Phys. Rev. Lett.}, 59(521), 1987.

\bibitem{boylefar}
L.~Boyle and S.~Farnsworth.
\newblock Non-commutative geometry, non-associative geometry and the standard
  model of particle physics.
\newblock {\em New J. Phys.}, 16:123027, 2014.

\bibitem{BM}
C.~Burgess and G.~Moore.
\newblock {\em The standard model: a primer}.
\newblock Cambridge University Press, 2007.

\bibitem{wow}
R.~Casalbuoni and R.~Gatto.
\newblock Unified description of quarks and leptons.
\newblock {\em Phys. Letters B}, 88(306), 1979.

\bibitem{wowagain}
R.~Casalbuoni and R.~Gatto.
\newblock Unified theories for quarks and leptons based on {Clifford} algebras.
\newblock {\em Phys. Letters}, 90B(1,2), 1979.

\bibitem{cliff}
W.~K. Clifford.
\newblock Applications of grassmann's extensive algebra.
\newblock {\em American Journal of Mathematics}, 1(4):350--358, 1878.

\bibitem{ATLAS}
ATLAS Collaboration.
\newblock Observation of a new particle in the search for the standard model
  {Higgs boson with the ATLAS detector at the LHC}.
\newblock {\em Phys. Lett. B}, 716:1--29, 2012.

\bibitem{CMSnat2}
CMS Collaboration.
\newblock Observation of a new boson at a mass of {125 GeV with the CMS
  experiment at the LHC}.
\newblock {\em Phys. Lett. B}, 716:30--61, 2012.

\bibitem{CMSnat1}
CMS Collaboration.
\newblock Evidence for the direct decay of the {125 GeV Higgs} boson to
  fermions.
\newblock {\em Nature Physics}, 10:557--560, 2014.

\bibitem{CMSjhep}
CMS Collaboration.
\newblock Search for the associated production of the {H}iggs boson with a
  top-quark pair.
\newblock {\em JHEP}, 09:087, 2014.

\bibitem{skk1}
The Super-Kamiokande Collaboration.
\newblock Search for proton decay via {$p \rightarrow e^+ \pi^0$ and $p
  \rightarrow \mu^+ \pi^0$} in a large water {C}herenkov detector.
\newblock {\em Phys.Rev.Lett.}, 102:141801, 2009.

\bibitem{skk2}
The Super-Kamiokande Collaboration.
\newblock Search for proton decay via {$p\rightarrow\nu \hspace{1mm}K^+$ using
  260 kiloton-year data of Super-Kamiokande}.
\newblock {\em Phys. Rev. D}, 90:072005, 2014.

\bibitem{conneslott}
A.~Connes and J.~Lott.
\newblock Particle models and non-commutative geometry.
\newblock {\em Nuc.Phys.B}, 18B:29--47, 1990.

\bibitem{conway}
J.H. Conway and D.A. Smith.
\newblock {\em On quaternions and octonions, their geometry, arithmetic, and
  symmetry}.
\newblock Peters, 2003.

\bibitem{QECS}
M.~Cort\^{e}s and L.~Smolin.
\newblock Quantum energetic causal sets, {See note added in proof, Phys.Rev.D
  version, and updated abstract, arXiv version 2}.
\newblock {\em Phys.Rev.D}, 90(044035), 2014.

\bibitem{Universe}
M.~Cort\^{e}s and L.~Smolin.
\newblock The universe as a process of unique events, {See updated abstract,
  arXiv version 3}.
\newblock {\em Phys.Rev.D}, 90(084007), 2014.

\bibitem{dixon}
G.~Dixon.
\newblock {\em Division algebras: octonions, quaternions, complex numbers and
  the algebraic design of physics}.
\newblock Kluwer Academic Publishers, 1994.

\bibitem{matter}
G.~Dixon.
\newblock Seeable matter; unseeable antimatter.
\newblock {\em Conference Proceedings, in press}, 2015.

\bibitem{dowker}
F.~Dowker.
\newblock Causal sets and the deep structure of spacetime.
\newblock {\em Article contributed to ``100 Years of Relativity - Space-time
  Structure: Einstein and Beyond" ed Abhay Ashtekar (World Scientific)}, 2005.

\bibitem{bill}
W.~Edwards.
\newblock Non-embeddable relational configurations.
\newblock {\em arXiv:1409.0271 [gr-qc]}, 2014.

\bibitem{fqxi}
FQXi.
\newblock The physics of what happens,
  {http://fqxi.org/community/forum/topic/2469}.

\bibitem{mink}
H.~Fritzsch and P.~Minkowski.
\newblock Unified interactions of leptons and hadrons.
\newblock {\em Annals Phys.}, 93:193--266, 1975.

\bibitem{cmnotes}
C.~Furey.
\newblock An algebraic causal set constructed at the fundamental level out of
  particles (not space-time points). {H}and-written notes emailed {L. Smolin on
  Sept 2, 2011, per his request. } {Plea}se see amendment to abstracts of
  \cite{QECS} and \cite{Universe}, {arXiv versions}.

\bibitem{cm}
C.~Furey.
\newblock {An algebraic causal set constructed at the fundamental level out of
  particles (not space-time points). Cambridge University Part III Mathematics
  research essay, later circulated at Perimeter Institute}.
\newblock 2006.

\bibitem{UTI}
C.~Furey.
\newblock A unified theory of ideals.
\newblock {\em Phys. Rev. D}, 86(025024), 2012.

\bibitem{Gen}
C.~Furey.
\newblock Generations: three prints, in colour.
\newblock {\em JHEP}, 10(046), 2014.

\bibitem{qq}
C.~Furey.
\newblock Charge quantization from a number operator.
\newblock {\em Phys. Lett. B}, 742:195--199, 2015.

\bibitem{georgi10}
H.~Georgi.
\newblock The state of the art - gauge theories.
\newblock {\em AIP Conf. Proc.}, 23:575--582, 1975.

\bibitem{su5}
H.~Georgi and S.~Glashow.
\newblock Unity of all elementary-particle forces.
\newblock {\em Phys.Rev.Lett.}, 32:438, 1974.

\bibitem{paddy}
P.~Girard.
\newblock {\em Quaternions, Clifford algebras and relativistic physics}.
\newblock Birkhauser Velag, 2007.

\bibitem{gs}
M.~Greiter and D.~Schuricht.
\newblock Imaginary in all directions.
\newblock {\em Eur.J.Phys.}, 24(397), 2003.

\bibitem{GGquarks}
M.~G\"{u}naydin and F.~G\"{u}rsey.
\newblock Quark structure and the octonions.
\newblock {\em J. Math. Phys.}, 14, 1973.

\bibitem{GGstats}
M.~G\"{u}naydin and F.~G\"{u}rsey.
\newblock Quark statistics and octonions.
\newblock {\em Phys. Rev. D}, 9, 1974.

\bibitem{jjIII}
J.~Huerta.
\newblock Division algebras and supersymmetry {III}.
\newblock {\em arXiv:1109.3574 [hep-th]}, 2011.

\bibitem{jjIV}
J.~Huerta.
\newblock Division algebras and supersymmetry {IV}.
\newblock {\em arXiv:1409.4361 [hep-th]}, 2014.

\bibitem{achim}
A.~Kempf.
\newblock Quantum gravity on a quantum computer?
\newblock {\em Found Phys}, pages 472--482, 2014.

\bibitem{kibble}
T.~Kibble.
\newblock The standard model of particle physics.
\newblock {\em Invited talk at 25 Anniversary Meeting of Academia Europaea,
  {arXiv:1412.4094 [physics.hist-ph]}}, 2013.

\bibitem{km}
D.~W. Kribs and F.~Markopoulou.
\newblock Geometry from quantum particles.
\newblock {\em arXiv:gr-qc/0510052}, 2005.

\bibitem{Lang}
P.~Langacker.
\newblock {\em The standard model and beyond}.
\newblock CRC Press, 2010.

\bibitem{andras}
A.~L\'{a}szl\'{o}.
\newblock A natural extension of the conformal lorentz group in a field theory
  context.
\newblock {\em Proceedings of Gribov-85 Memorial Workshop, arXiv:1507.08039
  [math-ph]}, 2015.

\bibitem{deleoGUT}
S.~De Leo.
\newblock Quaternions for {GUTs}.
\newblock {\em Int.J.Theor.Phys.}, 35:1821, 1996.

\bibitem{loun}
P.~Lounesto.
\newblock {\em Clifford algebras and spinors}.
\newblock Cambridge University Press, 1997.

\bibitem{sak}
A.~Kempf M.~Saravani, S.~Aslanbeigi.
\newblock Spacetime curvature in terms of scalar field propagators.
\newblock {\em arXiv:1510.02725 [gr-qc]}, 2015.

\bibitem{Mann}
R.~Mann.
\newblock {\em An introduction to particle physics and the standard model}.
\newblock CRC Press, 2010.

\bibitem{e6md}
C.~A. Manogue and T.~Dray.
\newblock Octonions, {E6,} and particle physics.
\newblock {\em J. Phys. Conf. Ser.}, 254:012005, 2010.

\bibitem{fotini}
F.~Markopoulou.
\newblock The internal description of a causal set: {W}hat the universe looks
  like from the inside.
\newblock {\em Commun. Math. Phys.}, 211:559--583, 2000.

\bibitem{okubo}
S.~Okubo.
\newblock {\em Introduction to octonion and other non- associative algebras in
  physics}.
\newblock Cambridge University Press, 1995.

\bibitem{ps}
J.~Pati and A.~Salam.
\newblock Lepton number as the fourth `color'.
\newblock {\em Phys.Rev. D}, 10:275--289, 1974.

\bibitem{fedo}
F.~Piazza.
\newblock Quantum degrees of freedom of a region of spacetime.
\newblock {\em AIP Conf. Proc.}, 841:566, 2006.

\bibitem{fedoglim}
F.~Piazza.
\newblock Glimmers of a pre-geometric perspective.
\newblock {\em Found. Phys.}, 40:239--266, 2010.

\bibitem{costapiazza}
F.~Piazza and F.~Costa.
\newblock Volumes of space as subsystems.
\newblock {\em arXiv:0711.3048 [gr-qc]}.

\bibitem{raam}
M.~Van Raamsdonk.
\newblock Building up spacetime with quantum entanglement.
\newblock {\em Gen.Rel.Grav.}, 42:2323--2329, 2010.

\bibitem{ramjbsm}
P.~Ramond.
\newblock {\em Journeys beyond the standard model}.
\newblock Perseus Books, 1999.

\bibitem{adreams}
P.~Ramond.
\newblock Algebraic dreams.
\newblock {\em Contribution to {Francqui Foundation Meeting in the honor of
  Marc Henneaux, arXiv:hep-th/0112261}}, 2001.

\bibitem{Robinson}
M.~Robinson.
\newblock {\em Symmetry and the standard model}.
\newblock Springer, 2011.

\bibitem{thooft}
G.~'t~Hooft.
\newblock Renormalizable lagrangians for massive yang-mills fields.
\newblock {\em Nuc. Phys. B}, 35:176--188, 1971.

\bibitem{west}
P.~West.
\newblock {\em Introduction to strings and branes}.
\newblock Cambridge University Press, 2012.

\bibitem{Wolfgang}
W.~Wieland.
\newblock New action for simplicial gravity in four dimensions.
\newblock {\em Class. Quantum Grav.}, 32:015016, 2015.

\end{thebibliography}
% Tip 5: You can create multiple .bib files to organize your references. 
% Just list them all in the \bibliogaphy command, separated by commas (no spaces).

% The following statement causes the specified references to be added to the bibliography% even if they were not 
% cited in the text. The asterisk is a wildcard that causes all entries in the bibliographic database to be included (optional).
%\nocite{*}

\end{document}